\documentclass[11pt,hyper,letterpaper]{JHEP3}
\usepackage[dvips]{epsfig}
\usepackage{graphicx}
\usepackage{dcolumn}% Align table columns on decimal point
\usepackage{bm}% bold math
\usepackage{cite}
\usepackage{amsmath,amssymb,epsf,amsfonts}
\usepackage{bbm}
\usepackage{slashed}

\newcommand{\tr}{\text{tr}}
\newcommand{\beq}{\begin{equation}}
\newcommand{\eeq}{\end{equation}}
\newcommand{\bea}{\begin{eqnarray}}
\newcommand{\eea}{\end{eqnarray}}
\newcommand{\bi}{\begin{itemize}}
\newcommand{\ei}{\end{itemize}}
\newcommand{\ben}{\begin{enumerate}}
\newcommand{\een}{\end{enumerate}}

\newcommand{\N}{{\mathcal N}}

\newcommand{\simp}{s_{\textrm{imp}}}
\newcommand{\imps}{S_{\textrm{imp}}}
\newcommand{\lirr}{\lambda_{\textrm{irr.}}}
\newcommand{\oirr}{\mathcal{O}_{\textrm{irr.}}}
\newcommand{\oirrv}{\langle \mathcal{O}_{\textrm{irr.}}\rangle}
\newcommand{\dirr}{\Delta_{\textrm{irr.}}}
\newcommand{\phinf}{\phi_{\infty}}
\newcommand{\ainf}{a_{\infty}}
\newcommand{\thetainf}{\theta_{\infty}}
\newcommand{\Oa}{\mathcal{O}_a}
\newcommand{\Ophi}{\mathcal{O}_{\phi}}

\title{ \LARGE A Holographic Model of the Kondo Effect}
\author{Johanna Erdmenger,\!$^1$\footnotemark[1]\, Carlos Hoyos,\!$^2$\footnotemark[2]\, Andy O'Bannon,\!$^{3,4}$\footnotemark[3]\, and Jackson Wu\!$^5$\footnotemark[4]
\\
$^1$Max-Planck-Institut f\"{u}r Physik (Werner-Heisenberg-Institut) \\ F\"{o}hringer Ring 6, 80805 M\"{u}nchen, Germany
\\
$^2$Raymond and Beverly Sackler School of Physics and Astronomy \\ Tel-Aviv University, Ramat-Aviv 69978, Israel
\\
$^3$Department of Applied Mathematics and Theoretical Physics \\ University of Cambridge, Cambridge CB3 0WA, United Kingdom
\\
$^4$Rudolf Peierls Centre for Theoretical Physics \\ University of Oxford, 1 Keble Road, Oxford OX1 3NP, United Kingdom
\\
$^5$National Center for Theoretical Sciences, Physics Division \\ No. 101, Section 2, Kuang Fu Road, Hsinchu, Taiwan 300, R.O.C.}

\footnotetext[1]{E-mail address: \email{jke@mppmu.mpg.de}}
\footnotetext[2]{E-mail address: \email{choyos@post.tau.ac.il}}
\footnotetext[3]{E-mail address: \email{obannon@physics.ox.ac.uk}}
\footnotetext[4]{E-mail address: \email{jknw350@yahoo.com}}

\abstract{We propose a model of the Kondo effect based on the Anti-de Sitter/Conformal Field Theory (AdS/CFT) correspondence, also known as holography. The Kondo effect is the screening of a magnetic impurity coupled anti-ferromagnetically to a bath of conduction electrons at low temperatures. In a (1+1)-dimensional CFT description, the Kondo effect is a renormalization group flow triggered by a marginally relevant (0+1)-dimensional operator between two fixed points with the same Kac-Moody current algebra. In the large-$N$ limit, with spin $SU(N)$ and charge $U(1)$ symmetries, the Kondo effect appears as a (0+1)-dimensional second-order mean-field transition in which the $U(1)$ charge symmetry is spontaneously broken. Our holographic model, which combines the CFT and large-$N$ descriptions, is a Chern-Simons gauge field in (2+1)-dimensional AdS space, $AdS_3$, dual to the Kac-Moody current, coupled to a holographic superconductor along an $AdS_2$ subspace. Our model exhibits several characteristic features of the Kondo effect, including a dynamically generated scale, a resistivity with power-law behavior in temperature at low temperatures, and a spectral flow producing a phase shift. Our holographic Kondo model may be useful for studying many open problems involving impurities, including for example the Kondo lattice problem.}

\preprint{DAMTP-2013-57\\MPP-2013-128\\OUTP-13-21P\\ TAUP-2977/13}

\begin{document}

\vspace{-1cm}

%%%%%%%%%%%%%%%%%%%%%%%%%%%%%%%%%%%%%%%%%%%%%%%%%%%%%%%%%%%%%%%%%%%%%%%%%%%%%%%%%%%%%%%%%%%%%%%%%%%%%%%%%
%%%%%%%%%%%%%%%%%%%%%%%%%%%%%%%%%%%%%%%%%%%%%%%%%%%%%%%%%%%%%%%%%%%%%%%%%%%%%%%%%%%%%%%%%%%%%%%%%%%%%%%%%
\section{Introduction and Summary}
\label{intro}
%%%%%%%%%%%%%%%%%%%%%%%%%%%%%%%%%%%%%%%%%%%%%%%%%%%%%%%%%%%%%%%%%%%%%%%%%%%%%%%%%%%%%%%%%%%%%%%%%%%%%%%%%
%%%%%%%%%%%%%%%%%%%%%%%%%%%%%%%%%%%%%%%%%%%%%%%%%%%%%%%%%%%%%%%%%%%%%%%%%%%%%%%%%%%%%%%%%%%%%%%%%%%%%%%%%

The Kondo effect~\cite{PTP.32.37} is the screening at low temperatures $T$ of a magnetic moment coupled anti-ferromagnetically to a bath of conduction electrons. The Kondo interaction involves only the spins of the magnetic impurity and the electrons, both of which are spin-$1/2$ representations of the $SU(2)$ spin symmetry. Heuristically, the screening occurs when an electron becomes bound to the impurity, forming the so-called Kondo singlet, below a characteristic, dynamically-generated scale, the Kondo temperature $T_K$. More precisely, the screening occurs when a many-body resonance forms, the Kondo resonance. The Kondo effect has been observed in many systems, the canonical examples being metals doped with magnetic ion impurities~\cite{0034-4885-37-2-001,GrŸner1978591} and quantum dots~\cite{Goldhaber1998,Cronenwett24071998,vanderWiel22092000}. A key experimental signature of the Kondo effect appears in the resistivity, $\rho$, which behaves as $-\ln \left(T /T_K\right)$ in the regime $T \gg T_K$~\cite{PTP.32.37}.

The theory of the Kondo effect employs many techniques, including Wilson's numerical renormalization group (RG)~\cite{Wilson:1974mb}, Nozi\`{e}res' Fermi liquid description~\cite{springerlink:10.1007/BF00654541,Nozieres:1975}, the Bethe Ansatz~\cite{PhysRevLett.45.379,Wiegmann:1980,RevModPhys.55.331,doi:10.1080/00018738300101581,0022-3719-19-17-017}, large-$N$ limits~\cite{0022-3719-19-17-017,PhysRevB.35.5072,RevModPhys.59.845,1998PhRvB..58.3794P}, conformal field theory (CFT)~\cite{Affleck:1990zd,Affleck:1990by,Affleck:1990iv,Affleck:1991tk,PhysRevB.48.7297,Affleck:1995ge}, and others. For reviews of many of these, see refs.~\cite{Hewson:1993,doi:10.1080/000187398243500}. Taken together, these techniques provide complete information about the spectrum, thermodynamics, and transport properties of the Kondo system at all energy scales. The single-impurity Kondo problem is thus considered a solved problem.

Many open questions remain about the Kondo effect, however. For example, an important unsolved problem is the generalization from a single impurity to multiple impurities with Ruderman-Kittel-Kasuya-Yosida (RKKY) interactions among one another, which can promote anti-ferromagnetic ordering of the impurity spins~\cite{Doniach1977231}. Many heavy fermion compounds realize a dense lattice of magnetic moments, \textit{i.e.}\ a ``Kondo lattice,'' wherein a competition between the Kondo and RKKY interactions gives rise to a quantum phase transition~\cite{Doniach1977231,RevModPhys.69.809,2006cond.mat.12006C,2010uqpt.book..193S,2008NatPh...4..186G}. The quantum critical degrees of freedom subsequently give rise to non-Fermi liquid behavior, namely the ``strange metal'' state with resistivity $\rho\propto T$~\cite{2008NatPh...4..186G}. Various obstacles have prevented a solution of the Kondo lattice problem (in more than one spatial dimension~\cite{RevModPhys.69.809}). For example, even with just two impurities, integrability is lost. Moreover, the quantum critical degrees of freedom are strongly coupled. Other important open questions involve entanglement entropy~\cite{AffleckEE} and far-from-equilibrium evolution~\cite{2006PhRvB..74x5113A,2011arXiv1102.3982L} of Kondo systems, both of which are difficult to study using the techniques mentioned above.

With an eye towards these (and other) open questions, our goal is to apply a new technique to the Kondo problem: the Anti-de Sitter/CFT Correspondence (AdS/CFT), also known as gauge-gravity duality or holography~\cite{Maldacena:1997re,Gubser:1998bc,Witten:1998qj}. AdS/CFT equates a weakly-coupled theory of gravity in (d+1)-dimensional AdS spacetime, $AdS_{d+1}$, with a strongly-coupled d-dimensional CFT ``living'' on the AdS boundary. In the best-understood examples the CFT is a non-Abelian Yang-Mills (YM) theory in the 't Hooft large-$N$ limit.

Holographic models come in two kinds: top-down and bottom-up. Top-down models are obtained from string theory constructions, while bottom-up models are \textit{ad hoc}\ toy models that may or may not descend from an ultra-violet (UV)-complete string theory. Various holographic Kondo models have been proposed, most of them top-down, with a strongly-coupled non-Abelian YM theory playing the role of the electrons, the $SU(N)$ gauge group playing the role of the spin symmetry, and a Wilson line operator playing the role of the impurity~\cite{Kachru:2009xf,Sachdev:2010um,Kachru:2010dk,Sachdev:2010uj,Mueck:2010ja,Faraggi:2011bb,Jensen:2011su,Karaiskos:2011kf,Harrison:2011fs,Benincasa:2011zu,Faraggi:2011ge,Benincasa:2012wu,Itsios:2012ev}.\footnote{See also ref.~\cite{Matsueda:2012gc} for an attempt to translate the Kondo Hamiltonian to holography directly, using a multi-scale entanglement renormalization ansatz, in the spirit of refs.~\cite{Swingle:2009bg,Swingle:2012wq}.} Some basic phenomena of the Kondo effect are missing from these top-down models, however, a prominent example being the dynamically-generated scale $T_K$.

Our goal is to find a bottom-up holographic Kondo model, with simple, generic ingredients that might be useful for further model building. We consider only a single impurity. Our strategy is in fact to consider a new top-down model with symmetries closer to those of actual Kondo systems. We then identify the minimal, essential ingredients of the top-down model to motivate a bottom-up model. We perform calculations only in our bottom-up model, to demonstrate that the model indeed captures some basic Kondo physics.

We employ the CFT and large-$N$ approaches to the Kondo effect. The CFT approach~\cite{Affleck:1990zd,Affleck:1990by,Affleck:1990iv,Affleck:1991tk,PhysRevB.48.7297,Affleck:1995ge} begins by reducing the problem to one spatial dimension. The key step is a partial wave decomposition of the electrons, retaining only the s-wave. The resulting effective theory is simply free chiral fermions in one dimension, which is a (1+1)-dimensional CFT with a Kac-Moody current algebra. Solving the Kondo problem then reduces to an exercise in Kac-Moody algebra representation theory, namely, determining how representations re-arrange between UV and infra-red (IR) fixed points. The large-$N$ approach~\cite{0022-3719-19-17-017,PhysRevB.35.5072,RevModPhys.59.845,1998PhRvB..58.3794P} begins by generalizing the $SU(2)$ spin group to $SU(N)$, followed by a standard vector-like large-$N$ limit. If we write the impurity's spin operator as a bi-linear in auxiliary ``slave'' fermions, then the Kondo coupling is double-trace with respect to $SU(N)$, and the Kondo effect appears as ``superconductivity'' at the location of the impurity~\cite{0022-3719-19-17-017,PhysRevB.35.5072,2003PhRvL..90u6403S,2004PhRvB..69c5111S}: a (0+1)-dimensional charged scalar operator condenses at a critical temperature $T_c$ near $T_K$. The charged scalar is built from an electron and a slave fermion, and its condensation represents the formation of the Kondo singlet. Crucially, the large-$N$ limit is only reliable for low $T$~\cite{0022-3719-19-17-017,PhysRevB.35.5072,RevModPhys.59.845,1998PhRvB..58.3794P,Hewson:1993}, essentially because the impurity can only affect observables when $T \leq T_c$. In particular, the characteristic $-\ln \left(T /T_K\right)$ contribution to $\rho$ at $T \gg T_K$ does not appear at leading order in the large-$N$ limit.

Our top-down model begins with (3+1)-dimensional $\N=4$ supersymmetric YM theory (SYM) with gauge group $SU(N_c)$, in the large-$N_c$ limit and with large 't Hooft coupling. We then introduce chiral fermions in the fundamental representation of $SU(N_c)$ localized to a (1+1)-dimensional defect. Our chiral fermions do not come from an s-wave reduction, but they do realize a Kac-Moody current algebra. They are also our new ingredient, compared to the earlier holographic models~\cite{Kachru:2009xf,Kachru:2010dk,Mueck:2010ja,Faraggi:2011bb,Jensen:2011su,Karaiskos:2011kf,Harrison:2011fs,Benincasa:2011zu,Faraggi:2011ge,Benincasa:2012wu,Itsios:2012ev}. We introduce the impurity as a Wilson line of $SU(N_c)$ described by (0+1)-dimensional slave fermions. The holographic dual of $\N=4$ SYM in the limits above is type IIB supergravity in $AdS_5 \times S^5$~\cite{Maldacena:1997re}. The chiral fermions are dual to D7-branes along $AdS_3 \times S^5$~\cite{Skenderis:2002vf,Harvey:2007ab,Buchbinder:2007ar,Harvey:2008zz} while the slave fermions are dual to D5-branes along $AdS_2 \times S^4$~\cite{Pawelczyk:2000hy,Camino:2001at,Skenderis:2002vf,Yamaguchi:2006tq,Gomis:2006sb}. We consider coincident D7- and D5-branes, which we treat as probes, neglecting their back-reaction on supergravity fields.

The probe D-brane actions reveal the essential ingredients for our holographic model. The D7-brane action includes a Chern-Simons (CS) gauge field in the $AdS_3$ part of its worldvolume, dual to a current obeying a Kac-Moody algebra~\cite{Gukov:2004id,Kraus:2006nb,Kraus:2006wn,Jensen:2010em,Andrade:2011sx}. The D5-brane action includes a YM gauge field in the $AdS_2$ part of the worldvolume whose electric flux encodes the representation of the Wilson line~\cite{Yamaguchi:2006tq,Gomis:2006sb}. Open strings between the D7- and D5-branes give rise to a complex scalar bi-fundamental under the two gauge fields and localized to the intersection of the D-branes, $AdS_2 \times S^4$. The dual scalar operator is built from the chiral and slave fermions, and the double-trace Kondo coupling appears in the bulk through a special boundary condition on the scalar~\cite{Witten:2001ua,Berkooz:2002ug}. We do not know the potential for the scalar in our top-down model, but having identified the essential ingredients we can proceed to our bottom-up model.

Our bottom-up model consists of an $AdS_3$ CS gauge field, an $AdS_2$ YM field, and a bi-fundamental $AdS_2$ scalar with a non-zero mass but no self-interactions. We introduce a black hole in $AdS_3$ with Hawking temperature $T$ and introduce electric flux of the $AdS_2$ YM field. We show analytically that at low $T$ the trivial solution for the scalar is unstable because of the special boundary condition. We then construct non-trivial solutions numerically, which exist only because of the special boundary condition~\cite{Faulkner:2010gj}. By computing the free energy and condensate numerically, we demonstrate that a second-order mean-field phase transition occurs from the trivial solution to a non-trivial solution as $T$ decreases. In short, our system is a holographic superconductor in $AdS_2$~\cite{Hartnoll:2008vx,Hartnoll:2008kx}. In field theory terms, a (0+1)-dimensional charged scalar operator condenses due to the double-trace Kondo coupling, as expected.

In holography the AdS radial coordinate is dual to the field theory's RG scale. Our non-trivial scalar solution is thus the holographic representation of an RG flow between two fixed points. In contrast, most previous holographic Kondo models described only fixed points~\cite{Mueck:2010ja,Faraggi:2011bb,Karaiskos:2011kf,Harrison:2011fs,Benincasa:2011zu,Faraggi:2011ge,Benincasa:2012wu,Itsios:2012ev}. To be clear, our field theory has two couplings, the single-trace 't Hooft coupling and the double-trace Kondo coupling. We work in the probe limit, so the 't Hooft coupling does not run, and is always large. As a result, any fixed points in our model are necessarily strongly-coupled. On the other hand, we will demonstrate that our Kondo coupling runs in a way similar to that in the original Kondo system.

We do not find a $-\ln \left(T /T_K\right)$ contribution to $\rho$ when $T \gg T_K$, due to the large-$N$ limit, as mentioned above. We do find many of the other ``smoking gun'' phenomena characteristic of the Kondo effect, however, including:

\begin{itemize}

\item \textbf{Dynamical scale generation:} A holographic calculation in our model reveals that our double-trace Kondo coupling diverges at a dynamically-generated scale, our $T_K$. We also find that $T_c$ is on the order of $T_K$.

\item \textbf{Power-law scalings at low $T$:} We will argue that in our model, the entropy, resistivity, and other observables exhibit power law behavior in $T$ when $T \ll T_c$, with the powers of $T$ fixed by the dimension of the leading irrelevant operator when we deform about the IR fixed point. In our model that dimension is non-integer, indicating a non-trivial IR fixed point, as expected, given the large 't Hooft coupling.

\item \textbf{Screening of the impurity:} In our model, when $T \leq T_c$ the non-trivial scalar draws electric flux away from the $AdS_2$ YM field deep in the bulk of $AdS_2$. The reduced electric flux deep in the bulk represents an impurity in a lower-dimensional representation of $SU(N_c)$ in the IR. In other words, the impurity is screened in the IR.

\item \textbf{Phase shift:} As we review in section~\ref{kondo}, the Kondo effect produces a phase shift for the electrons at the IR fixed point. In our model, when $T \leq T_c$ the non-trivial bi-fundamental scalar transfers electric flux from the $AdS_2$ YM field to the CS gauge field, generating a Wilson loop for the CS gauge field deep in the bulk of $AdS_3$, which leads to a phase shift for the chiral fermions (our electrons) at the IR fixed point~\cite{Kraus:2006nb,Kraus:2006wn}.

\end{itemize}

In short, our model captures much of the essential physics of the large-$N$, single-impurity Kondo effect. We hope that our model may be useful for studying the many open problems about the Kondo effect, especially those for which holography is particularly well-suited, such as entanglement entropy and far-from-equilibrium phenomena.

This paper is organized as follows. In section~\ref{kondo} we review the CFT and large-$N$ approaches to the Kondo problem. In section~\ref{topdown} we present our top-down model. In section~\ref{bottomup} we present our bottom-up model, and present all of the results mentioned above. We end in section~\ref{summary} with a summary and a discussion of future research directions.

%%%%%%%%%%%%%%%%%%%%%%%%%%%%%%%%%%%%%%%%%%%%%%%%%%%%%%%%%%%%%%%%%%%%%%%%%%%%%%%%%%%%%%%%%%%%%%%%%%%%%%%%%
%%%%%%%%%%%%%%%%%%%%%%%%%%%%%%%%%%%%%%%%%%%%%%%%%%%%%%%%%%%%%%%%%%%%%%%%%%%%%%%%%%%%%%%%%%%%%%%%%%%%%%%%%
\section{Review of the Kondo Effect}
\label{kondo}
%%%%%%%%%%%%%%%%%%%%%%%%%%%%%%%%%%%%%%%%%%%%%%%%%%%%%%%%%%%%%%%%%%%%%%%%%%%%%%%%%%%%%%%%%%%%%%%%%%%%%%%%%
%%%%%%%%%%%%%%%%%%%%%%%%%%%%%%%%%%%%%%%%%%%%%%%%%%%%%%%%%%%%%%%%%%%%%%%%%%%%%%%%%%%%%%%%%%%%%%%%%%%%%%%%%

The literature about the Kondo effect is enormous. In this section we will very briefly review only those subjects essential to the construction of our holographic model.

The Kondo effect occurs when a magnetic impurity is coupled to free electrons, or more precisely, a Landau Fermi liquid (LFL) of electrons. The Kondo Hamiltonian density is~\cite{PTP.32.37}
\beq
\label{eq:kondoham1}
H_K = \psi^{\dagger}_{\alpha} \frac{-\nabla^2}{2m} \psi_{\alpha} + \hat{\lambda}_K \, \delta(\vec{x}) \, \vec{S} \cdot \psi^{\dagger}_{\alpha'} \frac{1}{2} \vec{\tau}_{\alpha'\alpha} \, \psi_{\alpha},
\eeq
where the first term is the electron kinetic term and the second term represents the interaction between the electrons and the impurity. Here $\psi^{\dagger}_{\alpha}$ and $\psi_{\alpha}$ are creation and annihilation operators for an electron of spin $\alpha = \uparrow$ or $\downarrow$, \textit{i.e.}\ the electrons are in the fundamental representation of the spin $SU(2)$ symmetry, $\nabla^2$ is the Laplacian of flat $\mathbb{R}^3$, $m$ is the electron mass, $\vec{S}$ is the spin of the impurity, which is also in the fundamental representation of $SU(2)$, $\vec{\tau}$ is the vector of Pauli matrices, and $\hat{\lambda}_K$ is the Kondo coupling. Anti-ferromagnetic coupling means $\hat{\lambda}_K>0$, ferromagnetic means $\hat{\lambda}_K<0$.

The leading-order perturbative result for $\hat{\lambda}_K$'s beta function is negative. As a result, when $\hat{\lambda}_K<0$, the effective coupling goes to zero at low energy. When $\hat{\lambda}_K>0$, however, the system exhibits asymptotic freedom, a dynamically generated scale, $T_K$, and a coupling that appears to diverge at low energy, leading to the Kondo problem: what is the ground state of the Kondo Hamiltonian when $\hat{\lambda}_K>0$?

As mentioned in section~\ref{intro}, the single-impurity Kondo problem has been solved using a combination of complementary techniques. The solution is usually expressed in terms of an RG flow from a UV fixed point to an IR fixed point. Thanks to asymptotic freedom, the UV fixed point is a LFL and a decoupled spin. As we flow to lower energy, the ground state changes at the scale $T_K$. Heurstically, an electron becomes bound to the impurity, their spins combining into an anti-symmetric singlet of $SU(2)$, the Kondo singlet. More precisely, a many-body resonance, the Kondo resonance, forms, and the impurity spin is screened. The change in the ground state is a crossover, not a phase transition. At the IR fixed point the spin is absent, having been completely screened, while the remaining, unbound electrons form a LFL with a special boundary condition: the electronic wave function vanishes at the impurity's location. Intuitively, the reason is that any electron that attempts to penetrate that location must break apart the Kondo singlet, which has a very large binding energy, $\propto \hat{\lambda}_K$, making such events energetically costly and so extremely unlikely. The special boundary condition means that the IR spectrum is shifted relative to the UV, \textit{i.e.}~some spectral flow occurs. In short, although from the perspective of the UV degrees of freedom $\hat{\lambda}_K$ diverges in the IR, the IR degrees of freedom arrange themselves into a LFL with a special boundary condition.

The Kondo Hamiltonian admits several generalizations. We can promote spin $SU(2)$ to $SU(N)$, and we can consider multiple channels, or in particle physics language multiple flavors, of electrons. With $k$ channels, the total symmetry is $SU(N)\times SU(k) \times U(1)$, with $SU(k)$ and $U(1)$ the channel and charge symmetries, respectively. By definition, the electrons are in the fundamental of $SU(N) \times SU(k)$ and the impurity is a singlet of $SU(k) \times U(1)$. A Kondo Hamiltonian is thus specified by three data: $N$, $k$, and the representation of the impurity under $SU(N)$. These generalizations are relevant for real systems, including quantum dots and some alloys: in some cases the impurity has a large spin degeneracy, so $N>2$, and/or multiple conduction bands couple to the impurity, so $k>1$.

The original Kondo problem had $N=2$, $k=1$, and impurity spin $\simp = 1/2$. For any $N$ and $k$ and for various impurity representations, solutions to the single-impurity Kondo problem have been obtained that are as rigorous as those for the original Kondo problem (numerical RG, Bethe ansatz, CFT, etc.). In many cases the IR fixed point is non-trivial, \textit{i.e.}\ is an interacting CFT rather than a LFL. Indeed, Kondo Hamiltonians provide some of the few examples of exactly solvable systems exhibiting non-Fermi liquid behavior, which is one reason they have attracted so much interest.

In the rest of this section, we will leave $N$, $k$, and the impurity representation as free parameters, unless stated otherwise. We will review the solution of the Kondo problem in detail only for some special values of these parameters, however.

%%%%%%%%%%%%%%%%%%%%%%%%%%%%%%%%%%%%%%%%%%%%%%%%%%%%%%%%%%%%%%%%%%%%%%%%%%%%%%%%%%%%%%%%%%%%%%%%%%%%%%%%%
\subsection{CFT Techniques}
\label{cft}
%%%%%%%%%%%%%%%%%%%%%%%%%%%%%%%%%%%%%%%%%%%%%%%%%%%%%%%%%%%%%%%%%%%%%%%%%%%%%%%%%%%%%%%%%%%%%%%%%%%%%%%%%

The CFT approach to the Kondo problem, developed by Affleck and Ludwig in the 1990s~\cite{Affleck:1990zd,Affleck:1990by,Affleck:1990iv,Affleck:1991tk,PhysRevB.48.7297,Affleck:1995ge}, begins by reducing the problem to one spatial dimension. The Kondo interaction in eq.~\eqref{eq:kondoham1} is point-like, \textit{i.e.}\ is $\propto \delta(\vec{x})$, and hence preserves spherical symmetry. The first step is thus to perform a partial wave decomposition of the electrons $\psi_{\alpha}$, retaining only the s-wave. The next step is to linearize the dispersion relation around the Fermi momentum $k_F$, an approximation valid at energies far below the Fermi energy. The resulting effective theory is defined on the positive real axis, representing the radial distance to the impurity, with the in- and out-going s-waves appearing as left- and right-moving fermions. By extending the axis to negative values and then reflecting the right-movers about the origin and re-labeling them as left-movers, we obtain the simplest description: left-movers alone moving on the entire real line and interacting with the impurity at the origin,
\beq
\label{eq:kondoham2}
H = +\frac{v_F}{2\pi} \, \psi_L^{\dagger} i \partial_x \psi_L + v_F \, \lambda_K \, \delta(x) \, \vec{S} \cdot \psi_L^{\dagger} \, \frac{1}{2} \vec{\tau} \, \psi_L,
\eeq
where $\psi_L$ are the left-moving fermions and $v_F = k_F/m$ is the Fermi velocity. In contrast to $H_K$ in eq.~\eqref{eq:kondoham1}, in $H$ we have suppressed the $SU(N)$ spin indices and $\lambda_K = \frac{k_F^2}{2\pi^2 v_F}\hat{\lambda}_K$. Starting now, we take $v_F \equiv 1$. The (1+1)-dimensional Kondo coupling is classically marginal: $\delta(x)$ is dimension one, $\vec{S}$ is dimensionless, and $\psi_L$ is dimension $1/2$.

%%%%%%%%%%%%%%%%%%%%%%%%%%%%%%%%%%%%%%%%%%%%%%%%%%%%%%%%%%%%%%%%%%%%%%%%%%%%%%%%%%%%%%%%%%%%%%%%%%%%%%%%%
\subsubsection{The IR Fixed Point}
\label{irfixedpoint}
%%%%%%%%%%%%%%%%%%%%%%%%%%%%%%%%%%%%%%%%%%%%%%%%%%%%%%%%%%%%%%%%%%%%%%%%%%%%%%%%%%%%%%%%%%%%%%%%%%%%%%%%%

Affleck and Ludwig's key observation was that the effective Hamiltonian in eq.~\eqref{eq:kondoham2} has a much larger symmetry than the original Hamiltonian $H_K$ in eq.~\eqref{eq:kondoham1}, and that this ``accidental'' symmetry can be used to determine the IR spectrum completely, as follows.

In the UV $\lambda_K$ goes to zero and $H$ reduces to $\psi_L$'s kinetic term, which  is trivially a (1+1)-dimensional CFT, with $v_F$ playing the role of the speed of light. Using Euclidean time $\tau$, if we define $z\equiv \tau + i x$ then $\psi_L$ and the spin, channel, and charge currents are all holomorphic in $z$, and moreover each of the currents obeys a Kac-Moody algebra. For example, the spin current $J^a(z)$, with $a=1,2,\ldots,N-1$, has Laurent coefficients $J^a_n$,
\beq
J^a(z) = \sum_{n \in \mathbb{Z}} z^{-n-1} J^a_n,
\eeq
which obey the $SU(N)$ level $k$ Kac-Moody algebra, $SU(N)_k$,
\beq
[J^a_n,J^b_m]= i f^{abc} J^c_{n+m} +  k \, \frac{n}{2} \, \delta^{ab} \, \delta_{n,-m},
\eeq
where $f^{abc}$ are the $SU(N)$ structure constants. Similarly, the channel current obeys an $SU(k)_N$ Kac-Moody algebra and the charge current obeys a $U(1)$ Kac-Moody algebra, whose level can be set to any value by re-scaling the $U(1)$ current. At the UV fixed point, the eigenstates of $H$ are representations of $SU(N)_k \times SU(k)_N \times U(1)$. A Kac-Moody algebra has a finite number of highest weight states, each of which corresponds to a primary opertator of the CFT. The number of highest-weight states is determined by the level of the Kac-Moody algebra, for example, $SU(2)_k$ has highest weight states of spin $0,1/2,\ldots,k/2$. Lowering operators then generate an infinite number of descendant states, producing ``conformal towers.'' The boundary conditions determine how states in the separate $SU(N)_k$, $SU(k)_N$ and $U(1)$ conformal towers combine into eigenstates. The Kac-Moody algebra and the boundary conditions thus determine the spectrum completely.

In any (1+1)-dimensional CFT with a Kac-Moody algebra, the Hamiltonian can be written in Sugawara form, quadratic in currents (with appropriate normal ordering). The Sugawara form of the Kondo Hamiltonian density $H$ in eq.~\eqref{eq:kondoham2} is
\beq
\label{eq:kondoham3}
H = \frac{1}{2\pi(N+k)} J^aJ^a + \frac{1}{2\pi(k+N)} J^AJ^A + \frac{1}{4\pi Nk} J^2 + \lambda_K \, \delta(x) \, \vec{S} \cdot \vec{J},
\eeq
with channel currents $J^A$, where $A=1,2,\ldots,k-1$, and charge current $J$. The Sugawara form of $H$ has two advantages over eq.~\eqref{eq:kondoham2}. First, the spin, channel, and charge degrees of freedom decouple. Second, the Kondo interaction involves only the spin current. Indeed, the Sugawara form of $H$ tempts us to ``complete the square'': we define a new spin current
\beq
\label{absorb}
{\mathcal{J}}^a \equiv J^a + \pi (N+k) \lambda_K \delta(x) S^a,
\eeq
in terms of which $H$ takes the form (dropping an unimportant constant $\propto\vec{S} \cdot \vec{S}$),
\beq
\label{eq:kondoham4}
H = \frac{1}{2\pi(N+k)} {\mathcal{J}}^a{\mathcal{J}}^a + \frac{1}{2\pi(k+N)} J^AJ^A + \frac{1}{4\pi Nk} J^2,
\eeq
where, crucially, $\mathcal{J}^a$'s Laurent coefficients obey the $SU(N)_k$ current algebra if and only if
\beq
\label{eq:irlambda}
\lambda_K = \frac{2}{N+k}.
\eeq
Affleck and Ludwig interpret eqs.~\eqref{eq:kondoham4} and~\eqref{eq:irlambda} as the $H$ and $\lambda_K$ of the IR fixed point.\footnote{According to ref.~\cite{Affleck:1995ge}, the coupling $\lambda_K$ in the CFT formalism is related non-linearly to the coupling in other formalisms, and hence appears to reach a finite value at the IR fixed point rather than diverging.} Intuitively, the electrons ``absorb the spin.'' The IR fixed point will thus have the same $SU(N)_k \times SU(k)_N \times U(1)$ symmetry as the UV fixed point, as indeed must be the case, thanks to anomaly matching.\footnote{We thank J.~McGreevy for emphasizing to us the importance of anomaly matching in the CFT approach to the Kondo effect.} The eigenstates in the IR will again be representations of $SU(N)_k \times SU(k)_N \times U(1)$, so the Kondo problem reduces to finding how the representations re-arrange in going from the UV to the IR.

To solve the Kondo problem in the CFT formalism, Affleck and Ludwig propose the following ansatz: the highest weight states of $SU(N)_k$ each fuse with the impurity spin according to the fusion rules, a set of non-negative integers that count the number of ways two representations of the Kac-Moody algebra combine into a third representation. Luckily the fusion rules have already been computed using various CFT techniques, as reviewed for example in ref.~\cite{DiFrancesco:1997nk}. Given that $\vec{S}$ couples only to the spin current, Affleck and Ludwig propose that nothing happens to the channel and charge highest weight states. The boundary conditions then dictate how the new spin conformal towers combine with the channel and charge conformal towers to form eigenstates. The spectrum of IR eigenstates is then specified completely, and thus their ansatz constitutes a solution of the Kondo problem.

As a simple example, consider the original Kondo problem, $N=2$, $k=1$, and $\simp=1/2$. The current algebra is $SU(2)_1 \times U(1)$. The highest weight states of $SU(2)_1$ have spins $0$ and $1/2$, leading to two conformal towers of states with integer and half-integer spins, respectively. Upon compactifying $x$ into a circle, we impose Neveu-Schwarz (NS) boundary conditions in the UV, in which case the integer and half-integer spin eigenstates have odd and even $U(1)$ charges, respectively. Now we let the electrons absorb the spin. The $SU(2)_k$ fusion rules are, for spin $s$ combining with $s_{\textrm{imp}}$ to form spin $s'$, and assuming that $k>(s+\simp)$,
\beq
\label{eq:fusion}
|s-\simp| \leq s' \leq \textrm{min}\{ s+\simp,k-(s+\simp)\}.
\eeq
According to the fusion rules for $SU(2)_1$, the spin $0$ highest weight state becomes the spin $1/2$ highest weight state, and vice-versa. In other words, the two spin conformal towers switch. The charge conformal towers do not switch, however: in the IR the eigenstates with integer and half-integer spin have \textit{even} and \textit{odd} $U(1)$ charges, respectively. That corresponds to a Ramond (R) boundary condition\footnote{Recall that only NS and R boundary conditions are consistent with Lorentz invariance, and hence with conformal invariance.}, indicating a $\pi/2$ phase shift relative to the UV. The IR fixed point is thus free left-movers with a $\pi/2$ phase shift, and no impurity.

More generally, for $SU(2)_k$ with $k\geq 1$ the nature of the IR fixed point depends on the size of $k$ versus $2s_{\textrm{imp}}$. When $k=2s_{\textrm{imp}}$, the IR fixed point is $k$ free left-movers, each with a $\pi/2$ phase shift, and no impurity. This is ``critical screening,'' as occurs in the original Kondo problem. When $k<2s_{\textrm{imp}}$ the fusion is not with $s_{\textrm{imp}}$ but with $k/2$. The system has insufficient channels to screen the impurity completely, and the IR fixed point is $k$ free left-movers, each with a $\pi/2$ phase shift, plus a decoupled impurity of reduced spin $|s_{\textrm{imp}}-k/2|$. This is ``under-screening.'' The IR physics changes dramatically when $k>2s_{\textrm{imp}}$. In that case, upon discretizing $x$, the electrons at the sites neighboring $x=0$ attempt to screen the impurity, aligning anti-ferromagnetically with it, but with so many channels a non-zero effective spin remains, which the next layer of electrons attempts to screen, and so on, like an onion. This is ``over-screening,'' which leads to a non-trivial IR fixed point, including primary fields of non-integer dimension, which could not occur with just free fermions.

The most efficient way to describe the overscreened fixed point is to bosonize: schematically, we write each holomorphic current as a derivative of a periodic boson, $J \sim \partial \phi$, producing $SU(N)_k$, $SU(k)_N$, and $U(1)$ Wess-Zumino-Witten (WZW) models. The impurity appears as an $SU(N)_k$ Wilson line~\cite{Felder:1999cv,Bachas:2004sy,Alekseev:2007in,Monnier:2008jj,Harrison:2011fs}. A phase shift corresponds to a shift in the periodicity of the $U(1)$ charge boson. The overscreened fixed point involves non-trivial boundary conditions on the bosons which do not translate into a simple boundary condition on the original chiral fermions. For more details about the overscreened case, see refs.~\cite{Affleck:1990iv,Affleck:1995ge}.

The (1+1)-dimensional single-impurity Kondo problem is integrable for any $N$, $k$, and impurity representation~\cite{RevModPhys.55.331,0022-3719-19-17-017,ZinnJustin1998,PhysRevB.58.3814}, admitting a solution via the Bethe ansatz. The CFT and Bethe ansatz solutions always agree where they overlap, which provides a non-trivial check of the CFT approach.

Let us summarize the solution of the Kondo problem for an impurity in either a totally symmetric or anti-symmetric representation of $SU(N)$, so the corresponding Young tableau has $q$ boxes in a single row or column. The following results are valid for any $N$~\cite{0022-3719-19-17-017,ZinnJustin1998,PhysRevB.58.3814,1998PhRvB..58.3794P,PhysRevB.73.224445}. For the symmetric representation, $k< q$ or $k=q$ produces under- or critical screening, respectively, while $k>q$ produces over-screening, where the IR CFT is characterized in refs.~\cite{ZinnJustin1998,PhysRevB.58.3814,PhysRevB.73.224445}. For the anti-symmetric representation, $k=1$ produces critical screening while any $k\geq 2$ produces over-screening, where the IR CFT is characterized in refs.~\cite{0022-3719-19-17-017,1998PhRvB..58.3794P,PhysRevB.73.224445}.

%%%%%%%%%%%%%%%%%%%%%%%%%%%%%%%%%%%%%%%%%%%%%%%%%%%%%%%%%%%%%%%%%%%%%%%%%%%%%%%%%%%%%%%%%%%%%%%%%%%%%%%%%
\subsubsection{Leading Irrelevant Operator and Low-$T$ Scalings}
\label{irrop}
%%%%%%%%%%%%%%%%%%%%%%%%%%%%%%%%%%%%%%%%%%%%%%%%%%%%%%%%%%%%%%%%%%%%%%%%%%%%%%%%%%%%%%%%%%%%%%%%%%%%%%%%%

The Bethe ansatz solution provides complete information about the spectrum and thermodynamics everywhere along the RG flow. The CFT solution is only valid near the fixed points, but provides complete information not only about the spectrum and thermodynamics, but also about transport properties.

Consider for example the entropy as a function of $T$, with any $N$, $k$, and impurity representation. Specifically, consider the impurity's contribution to the entropy, $\imps$, defined as the contribution that is non-extensive, \textit{i.e.}\ independent of the system size. When $T \to \infty$ the system approaches the UV fixed point, free fermions and a decoupled spin, and $\imps$ simply counts the spin states. More precisely, at any fixed point we can express $\imps$ as the logarithm of a ratio of elements of the modular $S$-matrix~\cite{Affleck:1991tk}. At the UV fixed point $\imps$ is the logarithm of the dimension of the impurity's representation. As $T \to 0$ the system approaches the IR fixed point. With under-screening, $\imps$ approaches the logarithm of the dimension of the impurity's representation in the IR, which is smaller than that in the UV. With critical screening, no impurity remains in the IR, and so $\imps \to 0$. With over-screening, in general $\imps$ is the logarithm of a non-integer number~\cite{Affleck:1991tk}, providing another sign of a non-trivial fixed point. In all cases $\lim_{T \to 0} \imps < \lim_{T \to \infty} \imps$, as expected: the number of degrees of freedom is smaller in the IR than in the UV~\cite{Affleck:1991tk}.

Now consider a transport property, namely the resistivity, $\rho$, as a function of $T$, with any $N$, $k$, and impurity representation. The imaginary part of the $(3+1)$-dimensional electrons' retarded Green's function determines the conductivity, and hence $\rho$, via the Kubo formula. The retarded Green's function of the $(3+1)$-dimensional electrons can be expressed in terms of the $(1+1)$-dimensional fermions' self-energy, $\Sigma$, and so $\rho$ can be expressed in terms of $\textrm{Im}\,\Sigma$~\cite{PhysRevB.48.7297}. In the $T \to \infty$ limit, the impurity decouples and so has no effect on $\rho$. As $T \to 0$, $\rho$ approaches a non-zero constant fixed by the $(3+1)$-dimensional electrons' density of states  at the Fermi energy and the matrix element for an electron to scatter off the impurity and into another electron ($1 \to 1$ scattering off the impurity). With under- or critical screening, that matrix element is fixed by the $(1+1)$-dimensional fermions' phase shift, which is the maximum allowed by unitarity, $\pi/2$, and the resulting $\lim_{T \to 0} \rho$ is called the ``unitary limit resistivity''~\cite{PhysRevB.48.7297}, $\rho_u$. With over-screening, in general $\lim_{T \to 0} \rho < \rho_u$~\cite{PhysRevB.48.7297}. 

For $T$ finite but large, $T \gg T_K$, perturbation theory in $\lambda_K$ reliably predicts the corrections to the $T \to \infty$ results for all observables. The corrections to $\rho$ include, at order $\lambda_K^3$, the impurity's characteristic $-\ln(T/T_K)$ contribution~\cite{PTP.32.37}.

For $T$ finite but small, $T \ll T_K$, the low-$T$ scalings of observables are determined by the leading irrelevant deformation about the IR fixed point. For example, suppose we want the low-$T$ scaling of $\imps$. First, we write the thermodynamic partition function as a path integral over $e^{-S_E}$, with $S_E$ the Euclidean action of the IR CFT. Second, we add to $S_E$ the perturbation $\int d^2x \, \delta(x) \, \lirr \oirr$, with leading irrelevant operator $\oirr$ of dimension $\dirr>1$ and irrelevant coupling $\lirr$ of dimension $1 - \dirr$. Since $T_K$ is the only intrinsic scale, $\lirr \propto T_K^{1-\dirr}$. Third, we expand the partition function in $\lirr$, producing a sum of finite-$T$ correlators of $\oirr$. When $T\ll T_K$, each such correlator will be a numerical coefficient times a power of $T$ fixed by dimensional analysis. The first non-vanishing correlator determines the leading non-trivial power of $\lirr$ in $\imps$. For example, suppose the first non-vanishing correlator of $\oirr$ at finite $T$ is the one-point function: $\oirrv \neq 0$. The leading correction to $\imps$ at low $T$ would then be linear in $\lirr$: $\imps \propto \lirr T^{\dirr -1} \propto (T/T_K)^{\dirr-1}$. Similar arguments apply for all other thermodynamic quantities. For transport quantities, such as $\rho$, the low-$T$ scalings are fixed by the leading correction in $\lirr$ to $\textrm{Im}\,\Sigma$.

Crucially, the symmetries of the IR CFT are typically sufficient to fix $\oirr$. For example, consider the cases reviewed at the end of subsection~\ref{irfixedpoint}: an impurity in a totally symmetric or anti-symmetric representation, with any $N$ and $k$. With under- or critical screening, the irrelevant operator must be built from the spin current, because the channel and charge currents do not meaningfully participate in the RG flow and the spin itself either decouples or is absent. The spin, channel, and charge symmetries, and $(3+1)$-dimensional rotational symmetry, are all unbroken at the IR fixed point. The leading irrelevant operator invariant under all these symmetries is $\mathcal{J}^a \mathcal{J}^a$, with $\dirr=2$~\cite{Affleck:1990zd}, where $\mathcal{J}^a$ is the spin current of the IR fixed point, \textit{i.e.}\ after absorbing the spin as in eq.~\eqref{absorb}. That operator is precisely the spin current's contribution to the Sugawara Hamiltonian density which, being the energy density, has non-zero one-point function at finite $T$, so $\oirrv \neq 0$ at finite $T$. Moreover, the leading correction to $\textrm{Im}\,\Sigma$ is order $\lirr^2$. As a result, at low $T$~\cite{PhysRevB.48.7297}
\begin{subequations}
\label{eq:criticalscalings}
\bea
&\imps &\propto \lirr T^{\dirr-1} \propto T/T_K, \\ & \rho &\propto \rho_u \lirr^2 T^{2(\dirr-1)} \propto \rho_u (T/T_K)^{2}.
\eea
\end{subequations}
With over-screening, $\oirr$ is obtained by contracting the spin current with the adjoint primary of $SU(N)_k$, and has $\dirr=1+\frac{N}{N+k}$~\cite{Affleck:1990iv,1998PhRvB..58.3794P}. That operator is a Virasoro primary, so $\oirrv =0$ at finite $T$. The two-point function of $\oirr$ does not vanish at finite $T$, however. The leading correction to $\textrm{Im}\,\Sigma$ is order $\lirr$~\cite{PhysRevB.48.7297,1998PhRvB..58.3794P}. As a result, at low $T$,
\begin{subequations}
\label{eq:overscreenedscalings}
\bea
&\imps & \propto \lirr^2 T^{2(\dirr-1)} \propto (T/T_K)^{2N/(N+k)}, \\ &\rho & \propto \rho_u \lirr T^{\dirr-1} \propto \rho_u (T/T_K)^{N/(N+k)}.
\eea
\end{subequations}

%%%%%%%%%%%%%%%%%%%%%%%%%%%%%%%%%%%%%%%%%%%%%%%%%%%%%%%%%%%%%%%%%%%%%%%%%%%%%%%%%%%%%%%%%%%%%%%%%%%%%%%%%
\subsection{Large-$N$ Techniques}
\label{largen}
%%%%%%%%%%%%%%%%%%%%%%%%%%%%%%%%%%%%%%%%%%%%%%%%%%%%%%%%%%%%%%%%%%%%%%%%%%%%%%%%%%%%%%%%%%%%%%%%%%%%%%%%%

For reviews of large-$N$ approaches to the Kondo problem, see refs.~\cite{RevModPhys.59.845,Hewson:1993,doi:10.1080/000187398243500}. We will review only a few points that will be useful to us later.

With $SU(N)$ spin symmetry, the Kondo Hamiltonian admits a standard large-$N$ limit: $N\to\infty$ and $\lambda_K \to 0$ with $N \lambda_K$ fixed, with any $k$. The theory is vector-like, hence standard techniques provide an exact saddle-point solution. The large-$N$ limit thus provides complete information about the spectrum, thermodynamics, and transport properties everywhere along the RG flow, with (in principle) calculable corrections suppressed by powers of $1/N$.

Of use to us will be a large-$N$ solution in which the Kondo effect appears as (0+1)-dimensional superconductivity~\cite{0022-3719-19-17-017,PhysRevB.35.5072,2003PhRvL..90u6403S,2004PhRvB..69c5111S}. Let us consider an impurity in a totally anti-symmetric representation of $SU(N)$, so the associated Young tableau has a single column with $q<N$ boxes. Let us also represent $\vec{S}$ in terms of slave fermions $\chi$, also known as Abrikosov pseudo-fermions, transforming in the fundamental representation of $SU(N)$,
\beq
\label{eq:abrikosov}
S^a = \chi^{\dagger} \, T^a \, \chi, \qquad a=1,2,\ldots,N^2-1,
\eeq
where $T^a$ are the generators of $SU(N)$, in the fundamental representation. Clearly phase rotations of $\chi$ leave $S^a$ invariant, so the system acquires an ``extra'' $U(1)$ symmetry. Introducing $\chi$ also enlarges the Hilbert space. As with any redundant degrees of freedom, we must impose a constraint to project onto the subspace of physical states. Here physical states have $q$ units of $\chi$'s $U(1)$ charge: the constraint is $\chi^{\dagger} \chi = q$. We can also reverse the logic: fixing the $U(1)$ charge $\chi^{\dagger} \chi = q$ fixes the impurity's representation to be totally anti-symmetric, where the corresponding Young tableau is a single column with $q$ boxes.

To describe an impurity in an arbitrary representation of $SU(N)$ requires multiple ``flavors'' of $\chi$, say $N_f$ flavors, in which case the extra symmetry is $U(N_f)$. In a path integral formalism, the physical state constraint is enforced via the insertion of a Wilson line in some representation of $U(N_f)$, which is equivalent to the insertion of an $SU(N)$ Wilson line in the conjugate representation~\cite{Gomis:2006sb}, with the flavor $U(1)$ charge $\chi^{\dagger} \chi= q$ being the total number of boxes in the corresponding Young tableau.

Let us work in (1+1) dimensions, with chiral fermions $\psi_L$. Consider the operator ${\mathcal{O}} \equiv \psi_L^{\dagger} \chi$, which is a function of time $t$ only, ${\mathcal{O}}(t)$, because $\chi$ cannot propagate away from $x=0$. This operator is a singlet of $SU(N)$, is bi-fundamental under $SU(k) \times U(N_f)$, and has minus the $U(1)$ charge of the electron. Recall that $\psi_L$ and $\chi$ have engineering dimensions $1/2$ and zero, respectively, so classically ${\mathcal{O}}$ has dimension $1/2$. Using the $SU(N)$ identity
\beq
\label{eq:identity}
T^a_{\alpha\beta} \, T^a_{\gamma\delta} = \frac{1}{2} \left ( \delta_{\alpha\delta} \delta_{\beta\gamma} - \frac{1}{N} \delta_{\alpha\beta} \delta_{\gamma\delta} \right), \qquad \alpha,\beta,\gamma,\delta = 1, \ldots, N,
\eeq
we can write the Kondo coupling in terms of ${\mathcal{O}}$:
\beq
\label{eq:doubletrace}
\lambda_K \, \delta(x) \, J^a S^a = \lambda_K \, \delta(x) \, \left(\psi^{\dagger}_L T^a \psi_L\right) \, \left(\chi^{\dagger} T^a \chi\right) = \frac{1}{2} \lambda_K \, \delta(x) \left [  {\mathcal{O}} {\mathcal{O}}^{\dagger} - \frac{q}{N} \left(\psi^{\dagger}_L \psi_L \right)\right],
\eeq
where we used $\chi^{\dagger} \chi = q$. The Kondo coupling is thus the coupling of a classically-marginal ``double-trace'' deformation ${\mathcal{O}}{\mathcal{O}}^{\dagger}$, where we use quotes because in terms of $SU(N)$ indices ${\psi^{\dagger}_L \chi}$ is not a trace of a matrix, but a contraction of a column vector with a row vector. A Hubbard-Stratonovich transformation can then linearize the Kondo coupling, with an auxiliary field whose on-shell value is $\propto \mathcal{O}$~\cite{0022-3719-19-17-017,PhysRevB.35.5072,2006cond.mat.12006C,2003PhRvL..90u6403S,2004PhRvB..69c5111S}.

In the large-$N$ limit, with $k=1$ and $q/N$ of order one, the solution of the saddle-point equations reveals a second-order mean-field phase transition in which ${\mathcal{O}}(t)$ condenses at low $T$: when $T>T_c$, $\langle {\mathcal{O}}(t)\rangle =0$, whereas when $T\leq T_c$, $\langle {\mathcal{O}}(t)\rangle \neq 0$, where $T_c$ is on the order of $T_K$~\cite{0022-3719-19-17-017,PhysRevB.35.5072,2003PhRvL..90u6403S,2004PhRvB..69c5111S}. More precisely, in the low-$T$ phase, in the $t \to \infty$ limit $\langle {\mathcal{O}}(t){\mathcal{O}}(0)^{\dagger} \rangle \sim 1/t^{1/N}$, indicating that when $N\to \infty$ the correlation \textit{time} diverges, and the $SU(k)\times U(N_f)$ and charge $U(1)$ symmetries are spontaneously broken to the diagonal. Of course, symmetry breaking is impossible in (0+1) dimensions. In fact, the large-$N$ limit suppresses the long-\textit{time} fluctuations that would destroy the apparent order. The $1/N$ corrections then change the Kondo effect from a sharp phase transition at $T_c$ to a smooth crossover around $T_c$~\cite{0022-3719-19-17-017}. Intuitively, the condensation of $\mathcal{O}(t) = \psi^{\dagger}_L \chi$ represents the formation of the Kondo singlet: an electron $\psi^{\dagger}_L$ gets stuck to $\chi$.

Crucially, the large-$N$ saddle-point approximation is only reliable at \textit{low} temperatures, $T \lesssim T_c$~\cite{RevModPhys.59.845,Hewson:1993}. For any $T>T_c$, where $\langle {\mathcal{O}}(t)\rangle =0$, all physics reduces to that of free chiral fermions $\psi_L$. Accessing high-temperature phenomena of the Kondo effect, such as the $- \ln \left(T/T_K\right)$ behavior of the resistivity in the $T \gg T_K$ regime, requires calculating $1/N$ corrections to the saddle-point approximation.

The upshot is that at large $N$ and low $T$ the Kondo effect can be viewed as (0+1)-dimensional superconductivity triggered by a marginally relevant double-trace coupling. This description of the Kondo effect will be essential to our holographic model.

%%%%%%%%%%%%%%%%%%%%%%%%%%%%%%%%%%%%%%%%%%%%%%%%%%%%%%%%%%%%%%%%%%%%%%%%%%%%%%%%%%%%%%%%%%%%%%%%%%%%%%%%%
%%%%%%%%%%%%%%%%%%%%%%%%%%%%%%%%%%%%%%%%%%%%%%%%%%%%%%%%%%%%%%%%%%%%%%%%%%%%%%%%%%%%%%%%%%%%%%%%%%%%%%%%%
\section{Top-Down Holographic Model}
\label{topdown}
%%%%%%%%%%%%%%%%%%%%%%%%%%%%%%%%%%%%%%%%%%%%%%%%%%%%%%%%%%%%%%%%%%%%%%%%%%%%%%%%%%%%%%%%%%%%%%%%%%%%%%%%%
%%%%%%%%%%%%%%%%%%%%%%%%%%%%%%%%%%%%%%%%%%%%%%%%%%%%%%%%%%%%%%%%%%%%%%%%%%%%%%%%%%%%%%%%%%%%%%%%%%%%%%%%%

Consider the following intersection of D-branes in type IIB string theory:
\begin{center}
        \begin{tabular}{|c|cccccccccc|}\hline
                &$x^0$&$x^1$&$x^2$&$x^3$&$x^4$&$x^5$&$x^6$&$x^7$&$x^8$&$x^9$\\ \hline
$N_c$ D3&$\bullet$&$\bullet$&$\bullet$&$\bullet$&--&--&--&--&--&--\\
$N_7$ D7&$\bullet$&$\bullet$&--&--&$\bullet$&$\bullet$&$\bullet$&$\bullet$&$\bullet$&$\bullet$\\
$N_5$ D5&$\bullet$&--&--&--&$\bullet$&$\bullet$&$\bullet$&$\bullet$&$\bullet$&--\\ \hline
        \end{tabular}
\end{center}
We are interested in the open string sector, and specifically in the field theory on the worldvolume of the D3-branes. The system above has many kinds of open strings: strings with both ends on the D3-branes, the 3-3 strings, as well as 5-5 and 7-7 strings, 3-7 and 7-3 strings, 3-5 and 5-3 strings, and 5-7 and 7-5 strings.

For $N_q$ Dq-branes, the $q$-$q$ strings give rise at low energy to (q+1)-dimensional maximally supersymmetric YM with gauge group $U(N_q)$. The YM coupling $g_q$ of that theory is related to the string coupling $g_s$ and the string length squared $\alpha'$ as $g_q^2 \propto g_s \alpha'^{(q-3)/2}$, and the corresponding 't Hooft coupling is $\lambda_q \equiv N_q \, g_q^2$.

The 3-3 strings give rise at low energy to (3+1)-dimensional $\N=4$ SYM with YM coupling $g_{YM}^2 = 4 \pi g_s$. Unless stated otherwise, we will work in the Maldacena limits, $N_c \rightarrow \infty$ and $g_s \rightarrow 0$ with $\lambda \equiv 4 \pi g_s N_c$ fixed, followed by $\lambda >> 1$. The $\N=4$ SYM theory in these limits is dual to type IIB supergravity in the near-horizon geometry of the D3-branes, $AdS_5 \times S^5$, with $N_c$ units of flux of the Ramond-Ramond (RR) five-form $F_5 = dC_4$ on the $S^5$,
\beq
\label{eq:fiveform}
\int_{S^5} F_5 = g_s (2\pi)^2 (2\pi\alpha')^2 \, N_c.
\eeq

In all that follows we will treat the D7- and D5-branes in the probe limit: we keep $N_7$ and $N_5$ fixed as $N_c \to \infty$, and then expand all physical quantities in the small parameters $N_7/N_c$ and $N_5/N_c$, retaining only the leading terms. In that limit both of the D7- and D5-brane 't Hooft couplings are suppressed by powers of $1/N_c$, and vanish to leading order: $\lambda_7 \propto N_7 /N_c$ and $\lambda_5 \propto g_{YM} N_5 /\sqrt{N_c}$. The 7-7 and 5-5 strings thus decouple from the field theory, so that the $U(N_7)$ and $U(N_5)$ gauge groups become global symmetry groups.

The 3-7 and 7-3 strings will give rise to (1+1)-dimensional chiral fermions and the associated Kac-Moody current algebra. The 3-5 and 5-3 strings will give rise to slave fermions describing an $SU(N_c)$ Wilson line. The 7-5 and 5-7 strings will describe the Kondo interaction. We will now discuss each of these open string sectors separately.

%%%%%%%%%%%%%%%%%%%%%%%%%%%%%%%%%%%%%%%%%%%%%%%%%%%%%%%%%%%%%%%%%%%%%%%%%%%%%%%%%%%%%%%%%%%%%%%%%%%%%%%%%
\subsection{The D7-branes}
\label{d7branes}
%%%%%%%%%%%%%%%%%%%%%%%%%%%%%%%%%%%%%%%%%%%%%%%%%%%%%%%%%%%%%%%%%%%%%%%%%%%%%%%%%%%%%%%%%%%%%%%%%%%%%%%%%

The D3/D7 intersection above has been studied in detail refs.~\cite{Skenderis:2002vf,Harvey:2007ab,Buchbinder:2007ar,Harvey:2008zz}, whose results we now review. We will review the field theory side first, and then the gravity side.

The D3/D7 intersection above preserves eight real supercharges. The 3-7 and 7-3 strings will have eight directions with mixed Neumann-Dirichlet (ND) boundary conditions, hence the ground state is in the Ramond sector. After the GSO projection we obtain $N_7$ Weyl fermions confined to the (1+1)-dimensional intersection with the D3-branes, the $x^0$ and $x^1$ directions, or equivalently the $x^{\pm} \equiv x^0 \pm x^1$ light-cone directions. We will choose our Weyl fermions to be left-handed, so they depend only on $x^-$, and denote them $\psi_L$. All of the preserved supercharges are then \textit{right-handed}: the theory has $\N=(0,8)$ supersymmetry. The R-symmetry is $SU(4)\simeq SO(6)$, consistent with the the fact that the D7-branes preserve the $SO(6)$ rotations in the directions transverse to the D3-branes. The $\psi_L$ are bi-fundamental under $SU(N_c) \times U(N_7)$. The symmetries constrain the action for the $\psi_L$'s to be~\cite{Buchbinder:2007ar,Harvey:2008zz}
\beq
\label{d7flavoraction}
S_7 = \frac{1}{\pi}\int d^2x \, \psi^{\dagger}_L \left( i\partial_- - A_- \right) \psi_L.
\eeq
The $\psi_L$ couple only to the restriction of the $\N=4$ SYM gauge field to the defect, and indeed only to the component $A_-$ that is inert under $\N=(0,8)$ supersymmetry.

Integrating out the $\psi_L$ in the path integral produces an $SU(N_c)$ WZW model at level $N_7$ and a $U(N_7)$ WZW model at level $N_c$, realizing an $SU(N_c)_{N_7} \times SU(N_7)_{N_c} \times U(1)$ Kac-Moody algebra~\cite{Buchbinder:2007ar}. More precisely, these WZW terms are added to the Lagrangian of $\N=4$ SYM. The $U(N_7)$ symmetry is global, so the $U(N_7)$ WZW model is non-dynamical.

The $SU(N_c)$ WZW model is not invariant under $SU(N_c)$ gauge transformations: the chiral fermions $\psi_L$ produce a gauge anomaly. The anomaly can be eliminated in several different ways, for example by a careful treatment of anomaly inflow~\cite{Buchbinder:2007ar} or by adding $O7$-planes to cancel the D7-brane charge (\textit{i.e.}\ changing the theory)~\cite{Harvey:2007ab,Harvey:2008zz}. Luckily for us, the probe limit suppresses the anomaly~\cite{Buchbinder:2007ar}. The one-loop diagram producing the anomaly is $\propto g_{YM}^2 N_7$, which vanishes in the probe limit. Similarly, the $U(N_7)$ WZW model exhibits a $U(N_7)$ anomaly, however the corresponding one-loop diagram is $\propto g_7^2 N_c$, which remains order one in the probe limit because $g_7^2 \propto 1/N_c$. As a result, in the holographic dual with probe D7-branes, only the $U(N_7)$ anomaly will be apparent.

Let us compare our $\psi_L$ to those of the CFT approach to the Kondo effect. Our $\psi_L$ do not come from an s-wave reduction, rather they are defect fermions in a genuinely relativistic theory. Nevertheless, they realize the necessary Kac-Moody current algebra. The role of $\N=4$ SYM is simply to provide a CFT with a well-understood holographic dual. At the moment we cannot say whether $SU(N_c)$ and $SU(N_7)$ are the ``spin'' and ``channel'' symmetries, or vice-versa. Our choice (in the next subsection) to represent the impurity as a Wilson line of $SU(N_c)$ will unambiguously identify $SU(N_c)$ as the spin group. The spin current will thus be the $\psi_L$'s gauge current, $J^a = \psi_L^{\dagger} T^a \psi_L$, and $N_7$ will be the number of channels.

Now let us turn to the gravity side. In the Maldacena and probe limits we find $N_7$ probe D7-branes extended along $AdS_3 \times S^5$ inside $AdS_5 \times S^5$~\cite{Skenderis:2002vf,Harvey:2007ab,Buchbinder:2007ar,Harvey:2008zz}. The $SU(4)\simeq SO(6)$ R-symmetry is dual to the isometry of the $S^5$, which the D7-branes trivially preserve. The $U(N_7)$ current is dual to the $U(N_7)$ gauge field living on the worldvolume of the D7-branes. A current obeying a Kac-Moody current algebra will be dual to a CS gauge field, where the rank and level of the algebra map to the rank and level of the gauge field~\cite{Gukov:2004id,Kraus:2006nb,Kraus:2006wn,Jensen:2010em,Andrade:2011sx}. More precisely, in Euclidean signature and using complex coordinates, the holomorphic component of the D7-brane gauge field is dual to the holomorphic Kac-Moody current. Our task is thus to locate a level-$N_c$ $U(N_7)$ CS term for the D7-brane worldvolume gauge field. The D7-brane action $S_{D7}$ includes a non-Abelian Dirac-Born-Infeld (DBI) term plus Wess-Zumino (WZ) terms. The latter indeed provides a CS term in the $AdS_3$ part of the worldvolume,
\bea
\label{eq:d7action}
S_{D7} & = & + \frac{1}{2} T_{D7} \left(2\pi\alpha'\right)^2 \int P[C_4] \wedge \textrm{tr} \left(F \wedge F \right)+ \ldots \nonumber \\ & = & - \frac{1}{2} T_{D7} \left(2\pi\alpha'\right)^2 \int P[F_5] \wedge \textrm{tr} \left(A\wedge dA + \frac{2}{3} A \wedge A \wedge A \right) + \ldots, \nonumber \\ & = & - \frac{N_c}{4\pi} \int_{AdS_3} \textrm{tr} \left(A\wedge dA + \frac{2}{3} A \wedge A \wedge A \right) + \ldots
\eea
where $T_{D7} = g_s^{-1} \alpha'^{-4}(2\pi)^{-7}$ is the D7-brane tension, the integration is over the D7-brane worldvolume (we used Lorentzian signature), $P[\ldots]$ is a pullback, $F=dA + A\wedge A$ is the worldvolume field strength, and $\ldots$ represents all other terms in the D7-brane action, including the DBI term and any boundary terms. In eq.~\eqref{eq:d7action}, we reached the second line via integration by parts and the third line using eq.~\eqref{eq:fiveform}, assuming that $A$ does not depend on the $S^5$ directions. The $U(N_7)_{N_c}$ current algebra is thus visible in the bulk in the probe limit as a level-$N_c$ $U(N_7)$ CS gauge field, as advertised.

The bulk CS action has a single derivative, so a special boundary term is required for a well-posed variational problem~\cite{Kraus:2006nb,Kraus:2006wn,Jensen:2010em,Andrade:2011sx}. Introducing $\gamma_{ij}$ as the induced metric on a surface at fixed radial position near the boundary, with indices $i,j$ running over $x^0,x^1$, the D7-brane action includes the boundary term
\beq
\label{eq:d7boundary}
S_{D7} = + \frac{N_c}{8\pi} \int \, d^2x \sqrt{-\gamma} \, \textrm{tr} \left(\gamma^{ij} A_i A_j\right) + \ldots.
\eeq
In a holographic calculation of the one-point function of the field theory's stress-energy tensor, the only contribution from the CS gauge field comes from the boundary term in eq.~\eqref{eq:d7boundary}, which produces a Hamiltonian of Sugawara form, eq.~\eqref{eq:kondoham4}~\cite{Kraus:2006nb,Kraus:2006wn,Jensen:2010em,Andrade:2011sx}. Neither the bulk CS action in eq.~\eqref{eq:d7action} nor the boundary term in eq.~\eqref{eq:d7boundary} are gauge-invariant, and the gauge variation of their sum indicates the existence of a WZW model at the $AdS_3$ boundary~\cite{Jensen:2010em,Witten:1988hf}, and hence of a $U(N_7)$ anomaly, as advertised.

%%%%%%%%%%%%%%%%%%%%%%%%%%%%%%%%%%%%%%%%%%%%%%%%%%%%%%%%%%%%%%%%%%%%%%%%%%%%%%%%%%%%%%%%%%%%%%%%%%%%%%%%%
\subsection{The D5-branes}
\label{d5branes}
%%%%%%%%%%%%%%%%%%%%%%%%%%%%%%%%%%%%%%%%%%%%%%%%%%%%%%%%%%%%%%%%%%%%%%%%%%%%%%%%%%%%%%%%%%%%%%%%%%%%%%%%%

The D3/D5 intersection above has been studied in detail in refs.~\cite{Pawelczyk:2000hy,Camino:2001at,Skenderis:2002vf,Yamaguchi:2006tq,Gomis:2006sb,Kachru:2009xf,Sachdev:2010um,Mueck:2010ja,Faraggi:2011bb,Jensen:2011su,Karaiskos:2011kf,Harrison:2011fs,Faraggi:2011ge} some of whose results we will now review. We will review the field theory side first, and then the gravity side.

The D3/D5 intersection above preserves eight real supercharges. The 3-5 and 5-3 strings will have eight ND directions, hence again the ground state is in the Ramond sector, and after the GSO projection we obtain $N_5$ fermions restricted to the (0+1)-dimensional intersection with the D3-branes. We will denote these fermions as $\chi$, which are bi-fundamental under $SU(N_c)\times U(N_5)$. The D5-branes clearly break the rotational symmetry in the directions transverse to the D3-branes from $SO(6)$ down to $SO(5)$. That reduced symmetry is reflected in the action for the $\chi$'s,
\beq
\label{d5flavoraction}
S_5 = \int dx^0 \, \chi^{\dagger} \left( i \partial_0 - A_0 - \Phi_9 \right) \chi,
\eeq
where $\Phi_9$ is the adjoint scalar of $\N=4$ SYM whose eigenvalues represent the positions of the D3-branes in $x^9$. In eq.~\eqref{d5flavoraction}, both $A_0$ and $\Phi_9$ are restricted to the location of the $\chi$'s. If we integrate out the $\chi$'s, then we obtain the insertion of a half-supersymmetric $SU(N_c)$ Wilson line in the $\N=4$ SYM path integral~\cite{Yamaguchi:2006tq,Gomis:2006sb}. More precisely, we obtain a Maldacena line, which involves both the $\N=4$ SYM gauge field and $\Phi_9$~\cite{Maldacena:1998im}, but for simplicity we will use the term ``Wilson line.'' The $SU(N_c)$ representation of the Wilson line is determined as explained around eq.~\eqref{eq:abrikosov}. For example, if $N_5=1$, then choosing some $U(N_5)=U(1)$ charge $\chi^{\dagger} \chi = q$ corresponds to inserting an $SU(N_c)$ Wilson line in a totally anti-symmetric representation whose Young tableau is a single column with $q$ boxes.

Clearly the $\chi$ will be our slave fermions. In particular, given that the $\chi$'s give rise to an $SU(N_c)$ Wilson line, we can unambiguously identify $SU(N_c)$ as ``spin'' in our system. The spin operator is therefore the $\chi$'s $SU(N_c)$ gauge current: $S^a=\chi^{\dagger} T^a \chi$. Indeed, $SU(N_c)$ is the only group under which our chiral fermions $\psi_L$ and our slave fermions $\chi$ are both charged, and hence must play the role of the spin group.

Let us now turn to the gravity side. After the Maldacena and probe limits we obtain $N_5$ probe D5-branes extended along $AdS_2 \times S^4$ inside $AdS_5 \times S^5$~\cite{Pawelczyk:2000hy,Camino:2001at,Skenderis:2002vf}. The D5-brane trivially preserves the $SO(5)$ isometry of the $S^4$ inside $S^5$, dual to the $SO(5)$ R-symmetry. The $U(N_5)$ currents are dual to the D5-brane $U(N_5)$ worldvolume gauge fields. Splitting $U(N_5) = U(1) \times SU(N_5)$, for simplicity we will present only the terms in the D5-brane action $S_{D5}$ involving the $U(1)$ gauge field,
\beq
\label{eq:d5action}
S_{D5} = - N_5 \, T_{D5} \int d^6\xi \sqrt{-\textrm{det}\left(P[g] + f \right)} \, + \, N_5 \, T_{D5} \int P[C_4] \wedge f + \ldots,
\eeq
where $T_{D5} = g_s^{-1} \alpha'^{-3}(2\pi)^{-5}$ is the D5-brane tension, $\xi$ denotes the worldvolume coordinates, $g$ is the background metric, $f=da$ is the worldvolume $U(1)$ field strength, and the $\ldots$ represents all other terms in the action.

For a detailed analysis of the equations of motion arising from $S_{D5}$, see refs.~\cite{Pawelczyk:2000hy,Camino:2001at}. We will emphasize two key facts about the solutions of the equations of motion. The first is that the $P[C_4]$ in the WZ term in eq.~\eqref{eq:d5action} acts as a source for $f$, producing some electric flux in the $AdS_2$ part of the worldvolume. That electric flux arises from the endpoints of fundamental strings dissolved into the D5-branes, whose number is quantized (in the full string theory, not just the supergravity approximation). Translating to the dual field theory, each string corresponds to a single box in the Young tableau specifying the representation of the $SU(N_c)$ Wilson line~\cite{Yamaguchi:2006tq,Gomis:2006sb}. A single D5-brane carrying string charge corresponds to a Young tableau with a single column, \textit{i.e.}\ a totally anti-symmetric representation. Describing an arbitrary representation requires multiple D5-branes, as explained in detail in refs.~\cite{Yamaguchi:2006tq,Gomis:2006sb}.

The Young tableau of a totally antisymmetric representation can have at most $N_c-1$ boxes. How does that constraint appear in the bulk? That brings us to the second key fact, concerning the embedding of the D5-branes. Let us write the metric of the $S^5$ as
\beq
ds^2_{S^5} = d\Theta^2 + \sin^2 \Theta \, ds^2_{S^4},
\eeq
where $\Theta \in [0,\pi]$ and $ds^2_{S^4}$ represents the metric of a round, unit-radius $S^4$. The D5-branes wrap that $S^4$, so the angle of latitude $\Theta$ appears as a worldvolume scalar field. In the equations of motion $\Theta$ is coupled to $f$, and so the quantization of electric flux in the $AdS_2$ part of the worldvolume is tied to the embedding of the D5-branes: a D5-brane carrying a fixed number of dissolved strings must sit at a fixed value of $\Theta$. When the number of strings is zero, $\Theta=0$, so the D5-brane sits at the ``north pole'' of the $S^5$. The square root factor in the DBI term of eq.~\eqref{eq:d5action} includes a factor of $\sin^4\Theta$, which vanishes at that point, and the D5-brane action degenerates. Such a non-existent D5-brane corresponds to a Wilson line in the trivial representation, \textit{i.e.}\ the identity operator. As the number of strings increases, $\Theta$ increases monotonically from zero to $\pi$. When the number of strings is $N_c$, $\Theta=\pi$, so the D5-brane sits at the ``south pole'' of the $S^5$, and its action again degenerates~\cite{Camino:2001at}. Such a D5-brane represents a Wilson line whose representation is a single column with $N_c$ boxes, \textit{i.e.}\ the anti-symmetric singlet. A D5-brane thus cannot carry more than $N_c-1$ units of string charge. For more details about these D5-brane embeddings, see refs.~\cite{Pawelczyk:2000hy,Camino:2001at,Skenderis:2002vf,Yamaguchi:2006tq,Gomis:2006sb,Kachru:2009xf,Karaiskos:2011kf}.

Half-supersymmetric Wilson lines in $\N=4$ SYM actually admit another, equivalent, holographic description as D3-branes extended along $AdS_2 \times S^2$ inside $AdS_5$, located at a point on the $S^5$, and carrying string charge~\cite{Rey:1998ik,Drukker:2005kx,Gomis:2006sb,Gomis:2006im}. Each D3-brane corresponds to a Wilson line in a totally symmetric representation, whose Young tableau is a single row. The number of rows is at most $N_c$; each D3-brane consumes one unit of $F_5$ flux, so at most $N_c$ D3-branes can appear in the bulk. In field theory terms, the D3-branes correspond to slave bosons rather than slave fermions, although the two are equivalent via bosonization~\cite{Gomis:2006sb,Gomis:2006im}. We prefer to work with the D5-branes, which have a simpler embedding.

Crucially, both the D3- and D5-brane descriptions of Wilson lines involve electric flux in $AdS_2$, the main differences being the tensions of the D-branes and the WZ terms. We will retain only the ``universal'' sector in our bottom-up model, namely electric flux of a YM gauge field in $AdS_2$, representing an approximation to a DBI term.

%%%%%%%%%%%%%%%%%%%%%%%%%%%%%%%%%%%%%%%%%%%%%%%%%%%%%%%%%%%%%%%%%%%%%%%%%%%%%%%%%%%%%%%%%%%%%%%%%%%%%%%%%
\subsection{The Kondo Coupling}
\label{kondocoupling}
%%%%%%%%%%%%%%%%%%%%%%%%%%%%%%%%%%%%%%%%%%%%%%%%%%%%%%%%%%%%%%%%%%%%%%%%%%%%%%%%%%%%%%%%%%%%%%%%%%%%%%%%%

The D5/D7 intersection above breaks all supersymmetry. The 7-5 and 5-7 strings have two ND directions and so give rise to a tachyon: the lightest mode of these open strings is a complex scalar with mass-squared $-1/(4\alpha')$. The tachyon potential has been computed up to quartic order~\cite{Gava:1997jt,Aganagic:2000mh}, and the endpoint of the instability is known~\cite{Gava:1997jt,Polchinski:1998rr,Aganagic:2000mh}: the D5-branes dissolve into the D7-branes, becoming magnetic flux in the $x^1$ and $x^9$ directions, which is possible thanks to the D7-brane WZ coupling $\int P[C_6] \wedge F$. The magnetic flux distribution with lowest energy is simply uniform. The initial configuration thus includes separate D5- and D7-branes while the final ground state includes only D7-branes with constant magnetic flux in $x^1$ and $x^9$. In the initial configuration the D5-branes break translational invariance in $x^1$ and $x^9$, but the final configuration is translationally invariant in the $(x^1,x^9)$-plane. Any Dq/D(q+2) intersection with two ND directions exhibits essentially this same instability~\cite{Gava:1997jt,Polchinski:1998rr}.

The tachyon is bi-fundamental under $U(N_7) \times U(N_5)$ and is a singlet of $SU(N_c)$. The corresponding operator is easily identified as ${\mathcal{O}} \equiv \psi_L^{\dagger} \chi$, precisely the operator that appears in the double-trace Kondo coupling in eq.~\eqref{eq:doubletrace} and that we want to condense at low temperature in the large-$N_c$ limit. Obviously the running of our $\lambda_K$ could be very different from that of the original Kondo problem, however, because our theory has ``extra'' massless degrees of freedom, which could dramatically change the IR physics. Indeed, starting from the original Kondo Hamiltonian we have, in effect, gauged the spin group, and in fact have introduced an entire (3+1)-dimensional $\N=4$ vector multiplet.

Luckily, the beta functions for double-trace couplings in theories nearly identical to ours have been computed in refs.~\cite{Pomoni:2008de,Pomoni:2010et}. Consider a (3+1)-dimensional field theory in which all single-trace couplings, collectively denoted $\lambda$, have beta functions that vanish to leading order at large $N_c$. Suppose such a theory has a complex single-trace scalar operator $\tilde{\mathcal{O}}$ such that $\tilde{\mathcal{O}}\tilde{\mathcal{O}}^{\dagger}$ is classically marginal, giving rise to a classically-marginal double-trace coupling $\lambda_{\textrm{DT}}$. The results of refs.~\cite{Pomoni:2008de,Pomoni:2010et} for the associated beta function, $\beta_{\textrm{DT}}$, as well as for $\tilde{\mathcal{O}}$'s anomalous dimension, are
\begin{subequations}
\beq
\label{eq:delta}
\langle \tilde{\mathcal{O}}(x)\tilde{\mathcal{O}}(0)^{\dagger}\rangle = \frac{v(\lambda)}{2\pi^2} \frac{1}{x^{2\Delta}}, \qquad \Delta = 2 + \gamma(\lambda) + \frac{v(\lambda)}{1+\gamma(\lambda)} \, \lambda_{\textrm{DT}},
\eeq
\beq
\label{eq:betadt}
\beta_{\textrm{DT}} = \frac{v(\lambda)}{1+\gamma(\lambda)} \, \lambda_{\textrm{DT}}^2 + 2 \gamma(\lambda) \, \lambda_{\textrm{DT}} + b(\lambda),
\eeq
\end{subequations}
where $v(\lambda)>0$ is the normalization of $\tilde{\mathcal{O}}$'s two-point function, $\gamma(\lambda)$ is $\tilde{\mathcal{O}}$'s anomalous dimension at $\lambda_{\textrm{DT}}=0$, and $b(\lambda)$ is the coefficient of the double-trace terms, induced by single-trace interactions, in $\tilde{{\mathcal{O}}}$'s Coleman-Weinberg potential. To leading order at large $N_c$, $v(\lambda)$, $\gamma(\lambda)$, and $b(\lambda)$ depend only on $\lambda$, and not on $\lambda_{\textrm{DT}}$, as indicated. The authors of refs.~\cite{Pomoni:2008de,Pomoni:2010et} also computed $\tilde{\mathcal{O}}$'s Coleman-Weinberg potential. The derivations in refs.~\cite{Pomoni:2008de,Pomoni:2010et} relied \textit{only} on large-$N_c$ diagrammatics, hence the results for $\Delta$, $\beta_{\textrm{DT}}$ and $\tilde{\mathcal{O}}$'s Coleman-Weinberg potential are \textit{exact} as functions of $\lambda$ and $\lambda_{\textrm{DT}}$, receiving only $1/N_c$ corrections.

The only difference between our theory and those studied in refs.~\cite{Pomoni:2008de,Pomoni:2010et} is that unlike $\tilde{\mathcal{O}}$, our scalar operator ${\mathcal{O}}$ is restricted to a (0+1)-dimensional defect. The results of refs.~\cite{Pomoni:2008de,Pomoni:2010et} are thus not immediately applicable to our case. Nevertheless, given that the results of refs.~\cite{Pomoni:2008de,Pomoni:2010et} rely only on large-$N_c$ diagrammatics, we expect that $\lambda_K$'s beta function, $\beta_K$, and ${\mathcal{O}}$'s anomalous dimension and Coleman-Weinberg potential will exhibit the same parametric dependence on $\lambda$ and $\lambda_K$ as found in refs.~\cite{Pomoni:2008de,Pomoni:2010et}. For example, we expect $\beta_K$ to be a quadratic polynomial in $\lambda_K$ with $\lambda$-dependent coefficients that can be written in terms of $v(\lambda)$, $\gamma(\lambda)$, and $b(\lambda)$, perhaps with numerical factors differing from those in eq.~\eqref{eq:betadt}.\footnote{We thank E.~Pomoni for discussions about how the results of refs.~\cite{Pomoni:2008de,Pomoni:2010et} may change for defect operators.} For the sake of argument, we will assume that is the case. (Even if that is not the case, we expect that what follows will remain a valid description of the physics, at least qualitatively.)

Consider fixing $\lambda$ and treating $\beta_K$ as a function of $\lambda_K$. The zeroes of $\beta_K$ are then roots of a quadratic equation, which come in three kinds: two real roots, a single double root, or two complex roots. In physical terms, the first case corresponds to distinct UV and IR fixed points and the second case corresponds to the merger of the UV and IR fixed points. In the third case no fixed points exist, rather Landau poles appear both in the UV and IR, and scale invariance is completely broken in the quantum theory~\cite{Pomoni:2008de,Pomoni:2010et}.

The result of refs.~\cite{Pomoni:2008de,Pomoni:2010et} for $\mathcal{O}$'s Coleman-Weinberg potential indicates that when $\beta_K$ has two real roots $\langle {\mathcal{O}} \rangle = 0$, whereas when $\beta_K$ has two complex roots, $\langle {\mathcal{O}} \rangle \neq 0$. In a transition where two real roots merge and move into the complex plane, the mass gap will exhibit exponential, or ``BKT'', scaling near the critical point~\cite{Kaplan:2009kr}. The authors of refs.~\cite{Pomoni:2008de,Pomoni:2010et} speculate that for a theory with a holographic dual the transition will occur when the mass of the scalar dual to ${\mathcal{O}}$ violates the Breitenlohner-Freedman (BF) stability bound, which indeed has been observed~\cite{Jensen:2010ga,Kutasov:2011fr,Iqbal:2011aj}. We will see the same mechanism in our bottom-up model.

When $\lambda \ll 1$ we could compute $\gamma(\lambda)$, $v(\lambda)$, and $b(\lambda)$ using perturbation theory, and determine the zeroes of $\beta_K$. In the $\lambda \gg 1$ limit, such a calculation is much more difficult. In the holographic description the tachyon appears as a complex scalar localized at the intersection of the probe D7- and D5-branes, $AdS_2 \times S^4$. The tachyon is dual to ${\mathcal{O}}$, hence the tachyon's mass and action normalization determine $\gamma(\lambda)$ and $v(\lambda)$. The crucial question is thus: can we compute the tachyon's effective action in $AdS_5 \times S^5$? For the D5/D7 intersection in flat $\mathbb{R}^{9,1}$, a disk-level worldsheet calculation of the tachyon four-point function gives the tachyon effective potential up to quartic order~\cite{Gava:1997jt,Aganagic:2000mh}. Such a worldsheet calculation in $AdS_5 \times S^5$ would be much more difficult, thanks to the non-zero curvature and RR five-form flux.

We can of course simply \textit{guess}\ the ground state solution for the D5/D7 intersection in $AdS_5 \times S^5$, using our intution for the same intersection in flat $\mathbb{R}^{(9,1)}$. What should the ground state solution look like? We expect a solution with D7-branes alone with worldvolume magnetic flux in the directions along the D7-branes but transverse to the D5-branes, $x^1$ and $\Theta$. Our D5-branes also carry string charge, as reviewed in subsection~\ref{d5branes}, so we also expect the D7-brane solution to carry string charge. At the AdS boundary the magnetic flux and string charge should be localized in $x^1$ and $\Theta$. Given that in $\mathbb{R}^{(9,1)}$ the magnetic flux ``wants'' to spread out and become uniform, in AdS we expect the magnetic flux to spread out as we descend into the bulk, eventually becoming uniform in the $x^1$ and $\Theta$ directions. Uniform magnetic flux deep in the bulk would signal restoration of translation invariance in $x^1$ in the IR, as in the original Kondo problem, where translation invariance is restored because the impurity is screened. A D7-brane solution with the above properties has the same asymptotics as separate D7- and D5-branes, making a comparison meaningful. Whichever solution has smaller on-shell action would be energetically preferred. If the solution with D7-branes alone was preferred, that would provide strong evidence that indeed the tachyon condenses.

Our system actually shares many features with the Sakai-Sugimoto system~\cite{Sakai:2004cn,Sakai:2005yt} (among others). In that system D4-branes provide the background spacetime and the probes are D8- and anti-D8-branes. The Sakai-Sugimoto model admits two descriptions. The first is the ``two-brane'' description, which employs distinct D8- and anti-D8-brane actions with a tachyon field bi-fundamental under the D8- and anti-D8-brane worldvolume gauge fields~\cite{Casero:2007ae,Bergman:2007pm,Dhar:2007bz}. The tachyon is dual to a quark mass operator, and tachyon condensation is dual to the formation of a non-zero chiral condensate. The tachyon's effective action is difficult to compute, for reasons similar to those in our system. The alternative is the ``one-brane'' description, which employs a single D8-brane action, with solutions that represent fused D8- and anti-D8-branes~\cite{Sakai:2004cn,Sakai:2005yt}. Intuitively, these one-brane solutions represent the endpoint of the tachyon condensation.

In what follows we consider only a two-brane description of our system, retaining an explicit tachyon field, leaving a one-brane description, \textit{i.e.}\ a D7-brane solution with the properties described above, for the future. Additionally, rather than attempting to calculate the tachyon's effective action in our top-down model, we will switch to a bottom-up model, retaining only the minimal ingredients for a holographic Kondo effect. Although strictly speaking we do not know whether our top-down model actually realizes a Kondo effect, we will present several pieces of evidence that our bottom up model does.

%%%%%%%%%%%%%%%%%%%%%%%%%%%%%%%%%%%%%%%%%%%%%%%%%%%%%%%%%%%%%%%%%%%%%%%%%%%%%%%%%%%%%%%%%%%%%%%%%%%%%%%%%
%%%%%%%%%%%%%%%%%%%%%%%%%%%%%%%%%%%%%%%%%%%%%%%%%%%%%%%%%%%%%%%%%%%%%%%%%%%%%%%%%%%%%%%%%%%%%%%%%%%%%%%%%
\section{Bottom-Up Holographic Model}
\label{bottomup}
%%%%%%%%%%%%%%%%%%%%%%%%%%%%%%%%%%%%%%%%%%%%%%%%%%%%%%%%%%%%%%%%%%%%%%%%%%%%%%%%%%%%%%%%%%%%%%%%%%%%%%%%%
%%%%%%%%%%%%%%%%%%%%%%%%%%%%%%%%%%%%%%%%%%%%%%%%%%%%%%%%%%%%%%%%%%%%%%%%%%%%%%%%%%%%%%%%%%%%%%%%%%%%%%%%%

Let us extract from our top-down model the essential ingredients for a holographic Kondo effect. Obviously we need $AdS_3$. In our top-down model the $AdS_3$ was a subspace of $AdS_5$, but in our bottom-up model we will treat $AdS_3$ as the entire spacetime. We need a level-$N$ $U(k)$ CS gauge field, dual to the charge $U(1)$ and channel $SU(k)_N$ currents. In an $AdS_2$ subspace, localized in the field theory spatial direction, we need a YM gauge field, whose electric flux at the boundary encodes the impurity's representation, and a complex scalar bi-fundamental under the two gauge fields, dual to an operator of the form ${\mathcal{O}}\equiv\psi^{\dagger}_L \chi$.

For simplicity, we will treat both of the gauge fields and the scalar as probes. The background metric is then fixed. We will be interested in non-zero $T$, so we will use the (2+1)-dimensional AdS-Schwarzschild (or BTZ) black hole metric,
\beq
\label{metric}
ds^2 = g_{\mu\nu} dx^{\mu} dx^{\nu} = \frac{1}{z^2} \left ( \frac{dz^2}{h(z)} - h(z) \, dt^2 + dx^2 \right), \qquad h(z) = 1 - z^2/z_H^2,
\eeq
where $z$ is the holographic radial coordinate, with the boundary at $z = 0$ and the horizon at $z=z_H$, $t$ and $x$ are the field theory time and space directions, and we have chosen units in which the $AdS_3$ radius of curvature is unity. The Hawking temperature is $T=1/(2\pi z_H)$.

The bulk action of our model is
\begin{subequations}
\label{eq:action}
\beq
S = S_{CS} + S_{AdS_2},
\eeq
\beq
S_{CS} = - \frac{N}{4\pi} \int \textrm{\tr} \left( A \wedge dA + \frac{2}{3} A \wedge A \wedge A \right),
\eeq
\beq
\label{eq:ads2action}
S_{AdS_2} = - N\int d^3x \, \delta(x) \sqrt{-g} \left [ \frac{1}{4} \tr f^{mn} f_{mn} + g^{mn} \left(D_m \Phi\right)^{\dagger} D_n \Phi + V(\Phi^{\dagger} \Phi )\right],
\eeq
\end{subequations}
where $A$ and $a$ are the CS and YM gauge fields with field strengths $F=dA+A\wedge A$ and $f=da + a \wedge a$, resepectively, our $AdS_2$ subspace is localized to $x=0$, where $m,n = z,t$ and $g$ is the determinant of the induced $AdS_2$ metric, $\Phi$ is our bi-fundamental scalar with gauge-covariant derivative $D_m$ and potential $V(\Phi^{\dagger} \Phi)$, and we have omitted boundary terms, which we discuss in detail in the next subsection. The symmetries determine all couplings in our model, except for those in $V(\Phi^{\dagger}\Phi)$, which we are free to choose.

In our top-down model, the type IIB supergravity and probe D-brane actions are in fact effective actions that will receive corrections in both $\alpha'$ and $g_s$. In particular, the probe D-brane actions will receive corrections of higher order in $A$ and $a$ suppressed by powers of $\alpha'$. In our model eq.~\eqref{eq:action} we include only the CS and YM terms, which we believe are universal in the sense of effective field theory, \textit{i.e.}\ any holographic model of the Kondo effect will admit some description in terms of an action of the form in eq.~\eqref{eq:action}. Moreover, we have chosen the CS and YM actions in eq.~\eqref{eq:action} to scale linearly with $N$ to mimic the probe D-brane actions, which scale linearly with the D-brane tensions $T_{Dq} \propto 1/g_s \propto N_c$. That is the natural scaling with $N$ for any bulk object dual to fields in the fundamental representation of the gauge group. In eq.~\eqref{eq:ads2action} we have also chosen the action for our ``tachyon'' $\Phi$ to scale linearly with $N$, even though in our top-down model the tachyon is dual to the gauge singlet operator $\mathcal{O} = \psi^{\dagger}_L \chi$, so that the natural scaling of the tachyon action is order one (order $N^0$). From the perspective of our top-down model, we are assuming that the tachyon is order $\sqrt{N}$, which after a re-scaling would give our order-one $\Phi$ with an action of order $N$.

Neither a $(2+1)$-dimensional CS gauge field nor a $(1+1)$-dimensional YM field describes any propagating bulk degrees of freedom. (The same is true also for $(2+1)$-dimensional gravity.) The only dynamical bulk degrees of freedon in our model are in the complex scalar field $\Phi$. Nevertheless, our model contains enough non-trivial dynamics to describe much of the basic physics of the Kondo effect, as we will show.

As reviewed in section~\ref{kondo}, a Kondo Hamiltonian is specified by three data: $N$, $k$, and the spin representation of the impurity. In our model $N$ and $k$ appear as the level and rank of the CS gauge field. As in our top-down model, in the field theory the $SU(N)_k$ spin symmetry is gauged and hence not directly visible in the bulk. Only the $SU(k)_N$ channel and $U(1)$ charge symmetries are visible, via the CS gauge field. The impurity's representation will be encoded in the electric flux of $f$, as we reviewed in subsection~\ref{d5branes} and as we discuss in detail below.

Starting now, for simplicity we will take both the CS and YM gauge groups to be $U(1)$. In field theory terms, choosing a $U(1)$ CS gauge field means choosing a single channel, $k=1$, and choosing a $U(1)$ $AdS_2$ gauge field means choosing an impurity in a representation whose Young tableau is a single column or a single row. In our top-down model these would appear in the bulk via a D5- or D3-brane, respectively, but in our bottom-up model we have discarded the terms that distinguish the two. In other words, we cannot distinguish the totally symmetric and totally anti-symmetric representations. Luckily, the distinction is probably not crucial. As reviewed at the end of subsection~\ref{cft}, for the standard (non-holographic) Kondo problem, for any $N$ and with $k=1$, with a symmetric representation either critical or under-screening will occur, while for an anti-symmetric representation critical screening will occur. In all cases, the main effect in the IR is merely a phase shift.

With $U(1)$ gauge fields, $\Phi$'s covariant derivative is simply
\beq
D_m \Phi = \partial_m \Phi + i A_m \Phi - i a_m \Phi,
\eeq
where we take $\Phi$'s charges under the two $U(1)$s to be simply $\pm 1$. Splitting $\Phi$ into a phase $\psi$ and a modulus $\phi$,
\beq
\Phi = e^{i\psi} \phi,
\eeq
the equations of motion for $A$, $a$, $\psi$, and $\phi$ are, respectively,\footnote{We choose $\epsilon^{ztx}=+1$.}
\begin{subequations}
\beq
\label{Aeom}
\epsilon^{n\mu\nu} F_{\mu\nu} =- 8 \pi \delta(x) \sqrt{-g} \, g^{nm} (A_m - a_m + \partial_m \psi) \phi^2,
\eeq
\beq
\label{aeom}
\partial_m \left( \sqrt{-g} \, g^{mp} g^{nq} f_{pq}\right) = - 2\sqrt{-g} \, g^{nm} (A_m - a_m + \partial_m \psi) \phi^2,
\eeq
\beq
\label{psieom}
\partial_n \left( \sqrt{-g} \, g^{nm} \left( A_m -a_m + \partial_m \psi \right) \phi^2 \right) = 0,
\eeq
\beq
\label{phieom}
\partial_m \left( \sqrt{-g} \, g^{mn} \partial_n \phi \right) = \sqrt{-g} \, g^{mn} (A_m - a_m + \partial_m \psi)(A_n - a_n + \partial_n \psi) \phi + \frac{1}{2}\sqrt{-g} \, \frac{\partial V}{\partial \phi}.
\eeq
\end{subequations}
If we define a current
\beq
\label{eq:current}
J^m \equiv 2\sqrt{-g} \, g^{mn} \left( A_n -a_n + \partial_n \psi \right) \phi^2,
\eeq
then the right-hand-sides of eqs.~\eqref{Aeom} and~\eqref{aeom} are both $\propto J^m$. A derivative $\partial_n$ of either eq.~\eqref{Aeom} or~\eqref{aeom} produces eq.~\eqref{psieom}, the equation of motion for the phase $\psi$, which expresses conservation of the current, $\partial_m J^m=0$.

We need an ansatz to solve the equations of motion. We want static solutions, in which all of the fields are independent of $t$. We want the $AdS_2$ gauge field to have some electric flux, so we need $f_{zt}(z)\neq 0$. Choosing a gauge with $a_z=0$, that means we need $a_t(z)$.\footnote{We thank D.~Dorigoni for extensive discussions about gauge fixing in our equations of motion.} A straightforward exercise then shows that the following ansatz is consistent: all fields are zero except for $A_x(z)$, $a_t(z)$, and $\phi(z)$. In other words, this subset of fields does not source any other fields. To obtain explicit solutions, we must commit to a specific form of $V(\Phi^{\dagger}\Phi)$. We will make the simplest choice: just a mass term,
\beq
\label{eq:scalarpot}
V(\Phi^{\dagger} \Phi) = M^2 \Phi^{\dagger} \Phi.
\eeq
We will discuss below our choice for the value of $M^2$. With our ansatz, $\psi$'s equations of motion is trivially satisfied, while the equations of motion for $A$, $a$, and $\phi$ reduce to, respectively,
\begin{subequations}
\beq
 \label{Aeom2}
\partial_z A_x = 4 \pi \delta(x) \sqrt{-g} \, g^{tt} \, a_t \, \phi^2,
\eeq
\beq
\label{aeom2}
\partial_z \left( \sqrt{-g} \, g^{zz} g^{tt} \, \partial_z a_t \right) = 2 \sqrt{-g} \, g^{tt} \, a_t \, \phi^2,
\eeq
\beq
\label{phieom2}
\partial_z \left( \sqrt{-g} \, g^{zz} \, \partial_z \phi \right) = \sqrt{-g} \, g^{tt} \, a_t^2 \, \phi + \sqrt{-g} \, M^2 \, \phi.
\eeq
\end{subequations}

With our ansatz, only the $t$ component of the current $J^m$ on the right-hand-sides of eqs.~\eqref{Aeom} and~\eqref{aeom} can be non-zero. Our ansatz thus admits only source \textit{charges}, not source \textit{currents}, for both $A$ and $a$. For a CS gauge field an electric charge induces a magnetic flux, as in eq.~\eqref{Aeom2}, thanks to the $\epsilon^{m\mu\nu}$ in eq.~\eqref{Aeom}. This fact will be of central importance when we discuss the phase shift at the IR fixed point of our model in subsection~\ref{fixedpoint}.

With our ansatz the CS gauge field does not appear in $a$ or $\phi$'s equation of motion, eqs.~\eqref{aeom2} and~\eqref{phieom2}. We thus only need to solve for $a_t(z)$ and $\phi(z)$, and then insert those solutions into eq.~\eqref{Aeom2} to obtain the solution for $A_x(z)$. Written explicitly, $a_t(z)$ and $\phi(z)$'s equations of motion are (primes denote $\partial_z$)
\begin{subequations}
\label{eq:aphieoms3}
\bea
a_t'' + \frac{2}{z} a_t' -  2\frac{\phi^2}{z^2 h} \, a_t =0, \label{aeom3} \\
\phi''  + \frac{h'}{h} \phi' + \frac{a_t^2}{h^2} \, \phi - \frac{M^2}{z^2 h} \phi = 0. \label{phieom3}
\eea
\end{subequations}
The behavior of $a_t(z)$ and $\phi(z)$ at the $AdS_2$ boundary will play a starring role in what follows, so let us now discuss the asymptotic forms of the solutions in detail.

%%%%%%%%%%%%%%%%%%%%%%%%%%%%%%%%%%%%%%%%%%%%%%%%%%%%%%%%%%%%%%%%%%%%%%%%%%%%%%%%%%%%%%%%%%%%%%%%%%%%%%%%%
\subsection{Asymptotics and Boundary Terms}
\label{asymptotics}
%%%%%%%%%%%%%%%%%%%%%%%%%%%%%%%%%%%%%%%%%%%%%%%%%%%%%%%%%%%%%%%%%%%%%%%%%%%%%%%%%%%%%%%%%%%%%%%%%%%%%%%%%

A non-zero $T$ does not affect the near-boundary behavior of the fields, so in this subsection we will set $T=0$ and hence $h=1$.

First we will consider solutions with $\phi(z)=0$. The solution for $a_t(z)$ is then simple,
\beq
\label{eq:atsol1}
a_t(z) = \frac{Q}{z} + \mu,
\eeq
with coefficients $Q$ and $\mu$ of dimension zero and one, respectively. For the $a_t(z)$ living on one of the D5-branes of subsection~\ref{d5branes}, $Q$ and $\mu$ would be proportional to the charge $q$ and chemical potential, respectively, associated with the slave fermion $\chi$'s $U(1)$ symmetry. In particular, the electric flux at the boundary is $\lim_{z\to 0} \sqrt{-g} f^{tz} = - Q$, which in the quantum theory (type IIB string theory) must obey a quantization condition, corresponding to the fact that in the bulk an integer number of strings is dissolved into the D5-brane and in the field theory the Young tableau (a single column) must have an integer number of boxes. We will ignore the quantization condition, and treat $Q$ as a continuous parameter. Given our choice of $N$ factors in the action eq.~\eqref{eq:action}, we expect $\langle \chi^{\dagger} \chi \rangle \propto N Q$, so that $Q$ of order one implies $q$ of order $N$, that is, $Q \sim q/N$.

The $a_t(z)$ in eq.~\eqref{eq:atsol1} diverges linearly at the boundary. That is not a surprise: in $AdS_D$ the solution would be $a_t(z) = \mu + Q z^{D-3}$. When $D>3$, $\lim_{z \to 0}a_t(z)$ is finite. In $AdS_4$, either Dirichlet (fixed $\mu$) or Neumann (fixed $Q$) boundary conditions are allowed~\cite{Marolf:2006nd}. In $AdS_3$, $a_t(z)$ diverges logarithmically at the boundary, and so only the Neumann boundary condition is allowed~\cite{Marolf:2006nd}. Similarly, in $AdS_2$ only a Neumann boundary condition is allowed for $a_t(z)$. The dual field theory statement is that the $U(1)$ charge $\chi^{\dagger} \chi = q$ is not allowed to fluctuate, as discussed below eq.~\eqref{eq:abrikosov}. Although in $AdS_2$ the $Q/z$ term is non-normalizable, we must still identify $Q$ as a charge, not a chemical potential. To see why, set $n=r$ in eq.~\eqref{aeom}: the resulting equation is the charge conservation equation, $\partial_t Q=0$.

Only one particular boundary term produces a well-posed variational problem for $a_t$ in which $Q$ is fixed and $\mu$ is free to fluctuate. If we introduce a cutoff surface at $z=\varepsilon$ and denote the metric induced on this surface in the $AdS_2$ subspace as $\gamma_{tt}$, then the required boundary term for $a_t$ is~\cite{Castro:2008ms}
\beq
\label{eq:sat}
S_{a_t} = + \frac{N}{2} \int dt \, \sqrt{-\gamma_{tt}} \, \gamma^{tt} \, a_t^2.
\eeq
Although $S_{a_t}$ does not look gauge-invariant, as shown in ref.~\cite{Castro:2008ms} $S_{a_t}$ is invariant under all gauge transformations that preserve the asymptotic form of the $AdS_2$ Maxwell field, \textit{i.e.}\ that preserve the value of $Q$ and keep $a_z=0$ at the $AdS_2$ boundary. Moreover, $S_{a_t}$ also renders $a_t$'s contribution to the on-shell bulk action finite.

Now let us consider solutions with $\phi(z) \neq 0$. We begin by treating $\phi(z)$ as a probe with respect to $a_t(z)$. We thus ignore the right-hand-side of eq.~\eqref{aeom2}, and insert the solution for $a_t(z)$ in eq.~\eqref{eq:atsol1} into $\phi(z)$'s equation of motion eq.~\eqref{phieom2}, with the result
\beq
\phi'' + \frac{-M^2 + Q^2}{z^2} \, \phi + \frac{2Q\mu}{z} \, \phi + \mu^2 \phi = 0.
\eeq
The electric flux shifts the scalar's mass squared, $M^2 \to M^2 - Q^2$, due to the $1/z$ term in the $a_t(z)$ of eq.~\eqref{eq:atsol1} (hence this effect occurs only in $AdS_2$). For sufficiently large $Q^2$, the scalar will violate the BF bound. That is actually good news for us, once we recall our discussion at the end of subsection~\ref{kondocoupling}: the violation of the BF bound in the bulk should coincide with condensation of the dual operator ${\mathcal{O}}$ in the field theory, triggered by the double-trace coupling, which is precisely what we want for the Kondo effect.

To mimic the CFT description of the Kondo effect as closely as possible, we will demand that ${\mathcal{O}}$'s dimension is $1/2$, so that ${\mathcal{O}}{\mathcal{O}}^{\dagger} $ is exactly marginal. In the bulk that means the scalar will sit precisely at the $AdS_2$ BF bound, which here means $M^2 - Q^2 = -1/4$ and hence $M^2 = - 1/4 + Q^2$. Our scalar will thus sit at the boundary of the space of $M^2-Q^2$ values that violate the BF bound and produce an instability. In field theory terms we are, quite artificially, adjusting ${\mathcal{O}}$'s anomalous dimension to make ${\mathcal{O}}$ dimension $1/2$ for any $Q$.

For our choice of $M^2$, the asymptotic form of $\phi(z)$ is
\beq
\phi(z) = \alpha \, z^{1/2} \ln \left( \Lambda z \right) + \beta \, z^{1/2} + {\mathcal{O}}\left(z^{3/2} \log \left(\Lambda z\right)\right),
\eeq
where $\Lambda$ is an arbitrary scale factor that must be included to define the logarithm, and where the coefficients $\alpha$ and $\beta$ have dimension $1/2$. Following refs.~\cite{Witten:2001ua,Berkooz:2002ug}, we introduce the double-trace coupling by imposing a linear relation between $\alpha$ and $\beta$,
\beq
\label{eq:phibc}
\alpha = \kappa \, \beta,
\eeq
where $\kappa$ is proportional to the double-trace coupling, \textit{i.e.}\ the Kondo coupling. We can determine the running of $\kappa$ holographically as follows~\cite{Witten:2001ua}. The value of $\phi(z)$ cannot depend on our choice of $\Lambda$. Suppose we begin with ``bare'' parameters $\beta_0$, $\kappa_0$, $\Lambda_0$, and then rescale $\Lambda_0 \to \Lambda$, obtaining new parameters $\beta$ and $\kappa$. By demanding that $\phi(z)$ remain unchanged in this process, we find $\beta_0 \kappa_0 = \beta \kappa$ and
\beq
\kappa = \frac{\kappa_0}{1+\kappa_0 \ln \left(\frac{\Lambda_0}{\Lambda}\right)}.
\eeq
If $\kappa<0$ then the theory exhibits asymptotic freedom: in the UV, meaning $\Lambda/\Lambda_0 \to +\infty$, $\kappa \to 0$, whereas in the IR, meaning $\Lambda/\Lambda_0 \to 0$, $\kappa$ diverges when $\Lambda = \Lambda_0 e^{1/\kappa_0}$. If $\kappa>0$ then these UV and IR behaviors of $\kappa$ are reversed. We thus identify $N \lambda_K \propto - \kappa$.

The boundary terms required for $\phi(z)$ are, in terms of the complex scalar $\Phi$,
\beq
\label{eq:sphi}
S_{\Phi} = - N\int dt\,\sqrt{-\gamma_{tt}} \left(\frac{1}{2} + \frac{1}{\ln \left(\Lambda \varepsilon\right)} - \frac{1}{\kappa} \frac{1}{\left(\ln \left(\Lambda \varepsilon\right)\right)^2}\right) \Phi^{\dagger} \Phi.
\eeq
The first two terms in eq.~\eqref{eq:sphi} are the standard counterterms for a scalar at the BF bound, and cancel all $\Phi$-related divergences of the on-shell action. The third term, $\propto 1/\kappa$, is finite, and is required to produce a well-posed variational problem in which $\alpha = \kappa \beta$~\cite{Papadimitriou:2007sj,Faulkner:2010gj}. A straightforward calculation shows that, with the boundary terms in eq.~\eqref{eq:sphi}, $\langle {\mathcal{O}} \rangle \propto N\alpha$, with a $\kappa$-independent proportionality constant.

Now let us consider the most general solutions, no longer treating $\phi(z)$ as a probe with respect to $a_t(z)$. In these cases, the asymptotics of the fields are (for clarity, in eq.~\eqref{eq:fullasymp} we choose $\Lambda$ to be the inverse $AdS_3$ radius of curvature, which we set to unity above)
\begin{subequations}
\label{eq:fullasymp}
\bea
\phi(z) & = & \alpha \, z^{1/2} \ln z + \beta \, z^{1/2} + {\mathcal{O}}\left(z^{3/2} \left(\ln z \right)^4 \right), \\ a_t(z) & = & \frac{Q}{z} + \mu + c_1 \ln z + c_2 \left(\ln z \right)^2 + c_3 \left(\ln z\right)^3+ {\mathcal O}\left(z \left(\ln z\right)^5 \right),
\eea
\beq
c_1 = 2Q \left( 2 \alpha^2 -2 \alpha \beta + \beta^2 \right), \qquad c_2 = 2Q \left( - \alpha^2 + \alpha \beta \right), \qquad c_3 = \frac{2}{3} Q \, \alpha^2.
\eeq
\end{subequations}
Despite the appearance of logarithmic divergences in $a_t(z)$'s asymptotics, the boundary terms in eqs.~\eqref{eq:sat} and~\eqref{eq:sphi} still suffice to render the on-shell action finite. Those boundary terms also produce a well-posed variational problem, bearing in mind the following important point. A variational problem is only well-posed with fixed $z\to 0$ asymptotics. In particular, once we fix $\phi(z)$'s asymptotics, we cannot allow $Q$ to vary: to do so would alter $\phi(z)$'s asymptotics, rendering the variational problem nonsensical.\footnote{We thank I.~Papadimitriou for many very useful correspondences about holographic renormalization and about the variational problem in our system.} The general lesson is that in $AdS_2$ a charged scalar drastically changes $a_t(z)$'s variational problem.

Given the asymptotic forms of $\phi(z)$ and $a_t(z)$ in eq.~\eqref{eq:fullasymp} we can determine $A_x(z)$'s asymptotics from eq.~\eqref{Aeom2}. As $z \to 0$, the derivative $A_x'(z)$ diverges logarithmically,
\beq
A_x'(z) = - 4\pi \, \delta(x) \, Q \left( \alpha \ln z + \beta \right)^2 +  {\mathcal{O}}(z \left(\ln z\right)^5).
\eeq
An integration gives $A_x(z) = {\mathcal{O}}(z \left(\ln z\right)^2)$, where we set an integration constant to zero to guarantee that $A_x(z) \to 0$ as $z \to 0$. In short, $A_x(z) \to 0$ infinitely steeply as $z \to 0$.

To summarize, the complete action for our system is the bulk action in eq.~\eqref{eq:action}, plus the boundary terms in eqs.~\eqref{eq:sat} and~\eqref{eq:sphi}, plus a boundary term for the CS gauge field $A$ of the form in eq.~\eqref{eq:d7boundary}.

%%%%%%%%%%%%%%%%%%%%%%%%%%%%%%%%%%%%%%%%%%%%%%%%%%%%%%%%%%%%%%%%%%%%%%%%%%%%%%%%%%%%%%%%%%%%%%%%%%%%%%%%%
\subsection{Stability Analysis}
\label{stability}
%%%%%%%%%%%%%%%%%%%%%%%%%%%%%%%%%%%%%%%%%%%%%%%%%%%%%%%%%%%%%%%%%%%%%%%%%%%%%%%%%%%%%%%%%%%%%%%%%%%%%%%%%

In this subsection we consider non-zero $T$ and $a_t(z)$, and we show analytically (without numerics) that the trivial solution for the scalar, $\phi(z)=0$, is unstable if $T/\mu$ is sufficiently small and $\kappa<0$, thus proving that a phase transition must occur as we reduce $T/\mu$. Our stability analysis does not reveal the transition temperature $T_c$, however. We will study the phase transition and determine $T_c$ in subsection~\ref{phasetransition}.

\begin{figure}[htbp]
\centering
\includegraphics[width=4in]{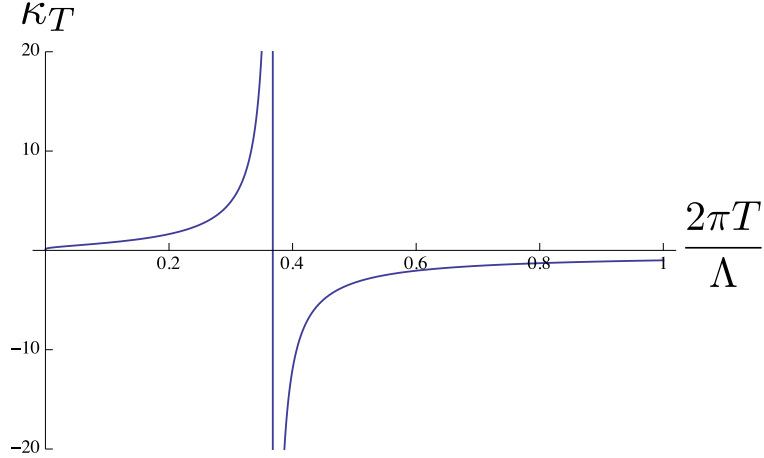}
\caption{The value of $\kappa_T$ in eq.~\protect\eqref{eq:betaTkappaT} as a function of $(2\pi T)/\Lambda$ for a representative negative (anti-ferromagnetic) value of the double-trace coupling $\kappa$. For the plot we chose $\kappa=-1$, leading to the divergence of $\kappa_T$ at $(2\pi T)/\Lambda = e^{-1} \approx 0.368$, which allows us to identify the Kondo temperature $T_K = \frac{1}{2\pi}\Lambda \, e^{1/\kappa} = \frac{1}{2 \pi} \Lambda \, e^{-1}$.}
\label{fig:kappaT}
\end{figure}

We begin by re-scaling to produce dimensionless coordinates,
\beq
\label{eq:rescaling}
\left(z/z_H,t/z_H,x/z_H\right) \to (z,t,x),
\eeq
which leaves the metric in eq.~\eqref{metric} invariant except for $h(z) = 1 - z^2/z_H^2 \to 1-z^2$. The $AdS_3$ boundary remains at $z=0$, but the horizon is now at $z=1$. To keep the one-form $a_t(z) dt$ invariant, we also take $a_t(z) z_H \to a_t(z)$, which is then dimensionless. After the re-scaling $\phi(z)$'s asymptotics become
\beq
\phi(z) = \alpha_T \, z^{1/2} \ln z + \beta_T \, z^{1/2}+ {\mathcal{O}}\left(z^{3/2} \left(\ln z\right)^4 \right),
\eeq
where we take $\alpha_T = \kappa_T \beta_T$, with $\beta_T$ and $\kappa_T$ related to $\beta$ and $\kappa$ from subsection~\ref{asymptotics} as
\beq
\label{eq:betaTkappaT}
\kappa_T \beta_T = \, z_H^{1/2} \,\kappa \, \beta, \qquad \kappa_T = \frac{\kappa}{1+\kappa \ln \left(\Lambda z_H\right)}.
\eeq
We want an anti-ferromagnetic coupling, so we choose $\kappa<0$, as explained in subsection~\ref{asymptotics}. A plot of $\kappa_T$ as a function of $1/(\Lambda z_H) = (2\pi T)/\Lambda$, for the representative choice $\kappa=-1$ appears in fig.~\ref{fig:kappaT}. The most prominent feature in fig.~\ref{fig:kappaT} is the divergence of $\kappa_T$ at a finite $T$, which we identify as the Kondo temperature $T_K = \frac{1}{2\pi}\Lambda \, e^{1/\kappa}$. Notice from fig.~\ref{fig:kappaT} that $\kappa_T < 0$ for $T>T_K$ while $\kappa_T > 0$ (\textit{opposite} to the sign of $\kappa$) for $T<T_K$. By plugging $\Lambda = (2 \pi T_K) e^{-1/\kappa}$ and $z_H = (2\pi T)^{-1}$ into eq.~\eqref{eq:betaTkappaT}, we find $T/T_K = e^{-1/\kappa_T}$.

%%%%%%%%%%%%%%%%%%%%%%%%%%%%%%%%%%%%%%%%%%%%%%%%%%%%%%%%%%%%%%%%%%%%%%%%%%%%%%%%%%%%%%%%%%%%%%%%%%%%%%%%%
%%%%%%%%%%%%%%%%%%%%%%%%%%%%%%%%%%%%%%%%%%%%%%%%%%%%%%%%%%%%%%%%%%%%%%%%%%%%%%%%%%%%%%%%%%%%%%%%%%%%%%%%%
%%%%%%%%%%%%%%%%%%%%%%%%%%%%%%%%%%%%%%%%%%%%%%%%%%%%%%%%%%%%%%%%%%%%%%%%%%%%%%%%%%%%%%%%%%%%%%%%%%%%%%%%%
%%%%%%%%%%%%%%%%%%%%%%%%%%%%%%%%%%%%%%%%%%%%%%%%%%%%%%%%%%%%%%%%%%%%%%%%%%%%%%%%%%%%%%%%%%%%%%%%%%%%%%%%%

To demonstrate that the trivial solution $\phi(z)=0$ is unstable at low $T$, we will perform a linearized stability analysis. We begin by writing $\phi = \phi_0 + \delta \phi$, with $\phi_0$ an arbitrary background solution and $\delta \phi$ the fluctuation, and similarly for all other fields. To write the linearized equations of motion of the fluctuations compactly, let us define
\beq
\Delta_m \equiv A_m - a_m + \partial_m \psi,
\eeq
which we also split into a background solution and a fluctuation, $\Delta_m = \Delta_m^0 + \delta \Delta_m$. The linearized equations for $\delta A_{\mu}$, $\delta a_m$, $\delta \psi$, and $\delta \phi$ are then, respectively,
\begin{subequations}
\label{fluceoms}
\beq
\epsilon^{m\mu\nu} \delta F_{\mu\nu} + 8\pi\delta(x) \sqrt{-g} \, g^{mn} \left[ \Delta_n^0 \, 2 \phi_0 \, \delta\phi + \phi_0^2 \, \delta \Delta_n \right] = 0,
\eeq
\beq
\partial_n \left(\sqrt{-g} \, g^{np} g^{mq} \, \delta f_{pq}\right) + 2 \sqrt{-g} \, g^{mn} \left[\Delta_n^0 \, 2 \phi_0 \, \delta \phi + \phi_0^2 \, \delta \Delta_n \right] = 0,
\eeq
\beq
\partial_m \left( \sqrt{-g} \, g^{mn} \left[ \Delta_n^0 \, 2 \phi_0 \, \delta \phi  + \phi_0^2 \, \delta \Delta_n \right]\right) = 0,
\eeq
\beq
\label{moduluseom}
\partial_m \left(\sqrt{-g} \, g^{mn} \partial_n \delta\phi\right) - \sqrt{-g} M^2 \delta \phi - \sqrt{-g} g^{mn} \left[ \Delta_m^0 \Delta_n^0 \delta \phi + 2 \phi_0 \Delta_m^0 \delta\Delta_n \right]=0.
\eeq
\end{subequations}
We will Fourier transform all fluctuations in field theory directions, for example
\beq
\delta\phi(z,t) = \int \frac{d\omega}{2\pi} \, e^{-i \omega t} \, \delta\phi(z,\omega),
\eeq
where $\omega$ is the re-scaled, dimensionless frequency.

Now let us consider the background solution with all fields zero except for $a_t^0(z)$. We solve eq.~\eqref{aeom3} for $a_t^0(z)$ with $\phi_0(z)=0$ and impose the regularity condition that $a_t^0(z)$ vanish at the horizon $z=1$, which fixes $\mu = -Q$, giving us
\beq
\label{eq:atsolbg}
a_t^0(z) = Q \left(\frac{1}{z} -1 \right).
\eeq
From eq.~\eqref{fluceoms} we see that when $\phi_0(z)=0$, all of the fluctuations decouple from one another at linear order. We will thus focus exclusively on the equation for $\delta\phi(z,\omega)$, eq.~\eqref{moduluseom},
\beq
\label{phifluceom}
\delta \phi '' + \frac{h'}{h} \delta \phi' + \left[ \frac{\omega^2}{h^2} + \frac{a_t^0(z)^2}{h^2} - \frac{M^2}{hz^2} \right] \delta \phi = 0.
\eeq
Eq.~\eqref{phifluceom} is of the form of Riemann's differential equation, whose solutions can be written in terms of hypergeometric functions. In particular, the two linearly-independent solutions of eq.~\eqref{phifluceom} are
\beq
\delta \phi_{\pm}(z,\omega) = \left(\frac{z}{1-z}\right)^{\mathfrak{a}_{\pm}} \left(\frac{1+z}{1-z}\right)^{\mathfrak{b}_{\pm}} {}_2 F_1\left(\mathfrak{a}_\pm + \mathfrak{b}_{\pm} - \frac{i\omega}{2},\mathfrak{a}_\pm + \mathfrak{b}_{\pm} + \frac{i\omega}{2},1+2\mathfrak{b}_\pm,\frac{1+z}{1-z}\right), \nonumber
\eeq
\beq
\mathfrak{a}_{\pm} = \frac{1}{2} \pm \sqrt{M^2 + \frac{1}{4} - Q^2}, \qquad \mathfrak{b}_{\pm} = \pm i \sqrt{Q^2 + \frac{1}{4} \omega^2}.
\eeq
The most general solution of eq.~\eqref{phifluceom} is a linear combination of the $\delta \phi_{\pm}(z,\omega)$ solutions,
\beq
\label{phiflucsolgeneral}
\delta \phi (z,\omega) = C_+(\omega) \, \delta \phi_+(z,\omega) + C_-(\omega) \, \delta \phi_-(z,\omega).
\eeq
As explained in subsection~\ref{asymptotics}, we want $M^2 = -1/4 + Q^2$, in which case $\mathfrak{a}_{\pm}=1/2$. Expanding eq.~\eqref{phiflucsolgeneral} about the horizon $z=1$, we find to leading non-trivial order
\beq
\delta \phi(z,\omega) = C_{\textrm{in}}(\omega) \left(1-z\right)^{- i \omega/2} + C_{\textrm{out}}(\omega)\left(1-z\right)^{+i\omega/2},
\eeq
which is a linear combination of in-going and out-going waves, with coefficients $C_{\textrm{in}}(\omega)$ and $C_{\textrm{out}}(\omega)$ that are themselves linear combinations of $C_{\pm}(\omega)$. In particular, demanding that $C_{\textrm{out}}(\omega)=0$, as appropriate for computing $\mathcal{O}$'s \textit{retarded} Green's function~\cite{Son:2002sd}, fixes
\beq
\frac{C_+(\omega)}{C_-(\omega)} = - e^{2\pi i \mathfrak{b}_+} \, \frac{\Gamma\left(1+2 \mathfrak{b}_-\right)}{\Gamma\left(1+2 \mathfrak{b}_+\right)} \frac{\Gamma\left(\frac{1}{2} - \frac{i\omega}{2} + \mathfrak{b}_+\right)^2}{\Gamma\left(\frac{1}{2}-\frac{i\omega}{2} + \mathfrak{b}_-\right)^2}.
\eeq
Expanding eq.~\eqref{phiflucsolgeneral} (with $M^2 = -1/4 + Q^2$) about the boundary $z=0$, at leading non-trivial order we find the expected form,
\beq
\delta \phi(z,\omega) = \alpha(\omega) \, z^{1/2} \ln z + \beta(\omega) \, z^{1/2},
\eeq
where because the ratio $C_+(\omega)/C_-(\omega)$ is fixed, the ratio $\alpha(\omega)/\beta(\omega)=\kappa_T(\omega)$ is fixed,
\beq
\label{kappatomega}
\kappa_T(\omega) = \frac{e^{2\pi i \mathfrak{b}_+}\Gamma\left(\frac{1}{2} - \frac{i\omega}{2} + \mathfrak{b}_+\right)\Gamma\left(\frac{1}{2} + \frac{i\omega}{2} + \mathfrak{b}_-\right) - \Gamma\left(\frac{1}{2} + \frac{i\omega}{2} + \mathfrak{b}_+\right)\Gamma\left(\frac{1}{2} - \frac{i\omega}{2} + \mathfrak{b}_-\right)}{h_+ e^{2\pi i \mathfrak{b}_+}\Gamma\left(\frac{1}{2} - \frac{i\omega}{2} + \mathfrak{b}_+\right)\Gamma\left(\frac{1}{2} + \frac{i\omega}{2} + \mathfrak{b}_-\right) - h_- \Gamma\left(\frac{1}{2} + \frac{i\omega}{2} + \mathfrak{b}_+\right)\Gamma\left(\frac{1}{2} - \frac{i\omega}{2} + \mathfrak{b}_-\right) },
\eeq
\beq
h_{\pm} = i \pi + \ln 2 + H_{- \frac{1}{2} + \frac{i\omega}{2} + \mathfrak{b}_{\pm}} + H_{- \frac{1}{2} - \frac{i\omega}{2} + \mathfrak{b}_{\pm}}, \nonumber
\eeq
where $H_n$ denotes the $n^{\textrm{th}}$ harmonic number.

\begin{figure}[t]
  \centering
  $\begin{array}{ccc}
  \includegraphics[width=0.33\textwidth]{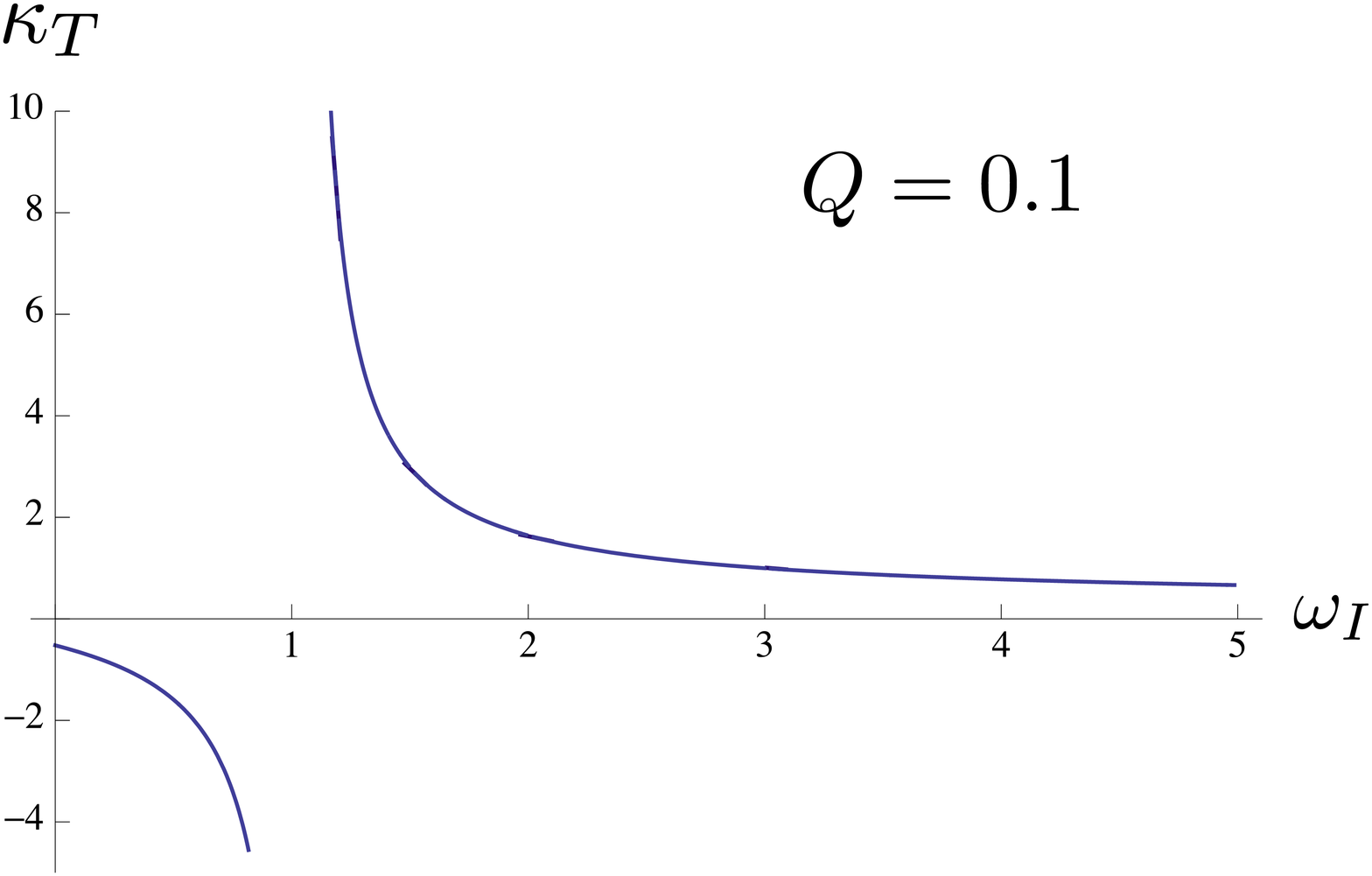} & \includegraphics[width=0.33\textwidth]{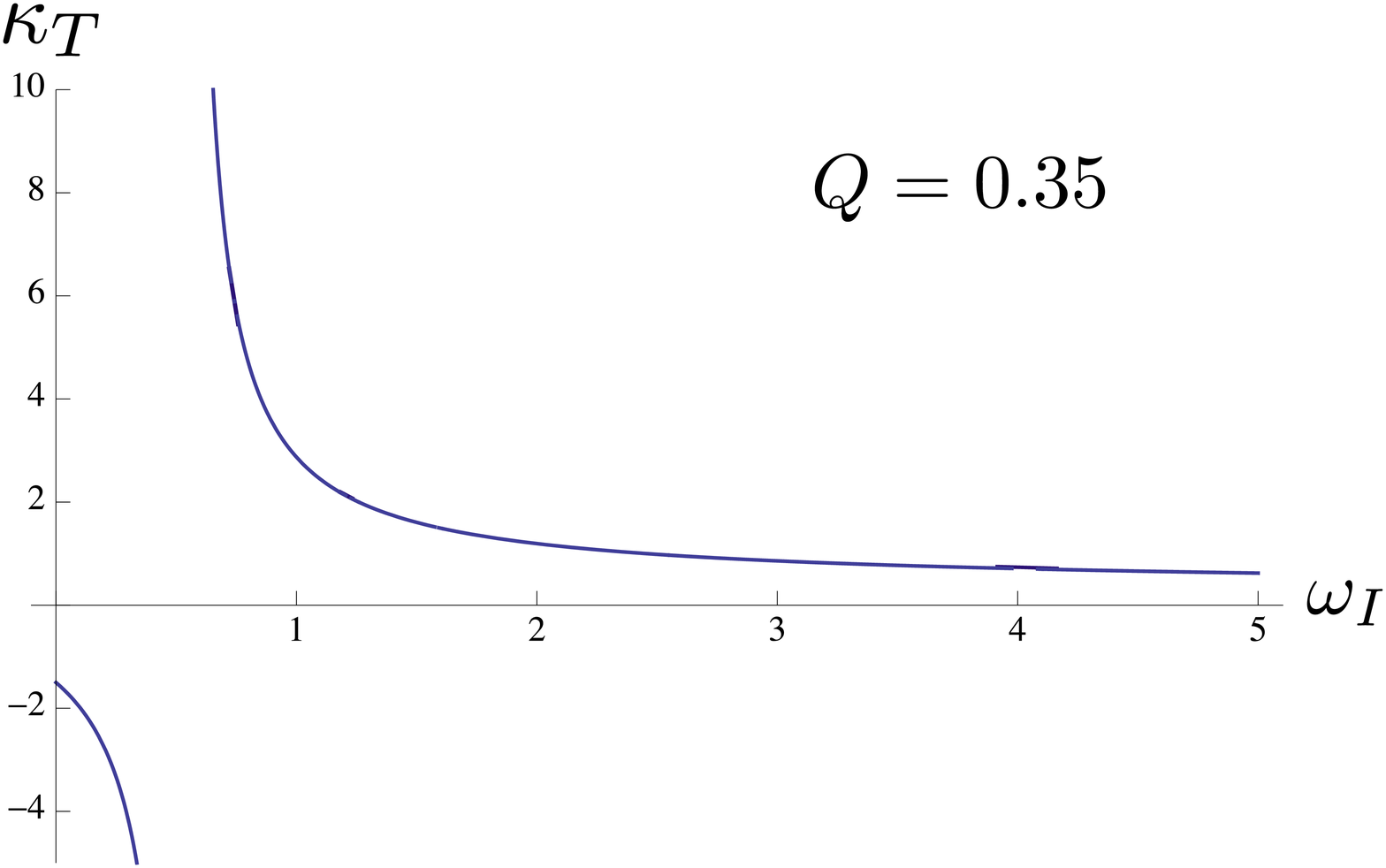} & \includegraphics[width=0.33\textwidth]{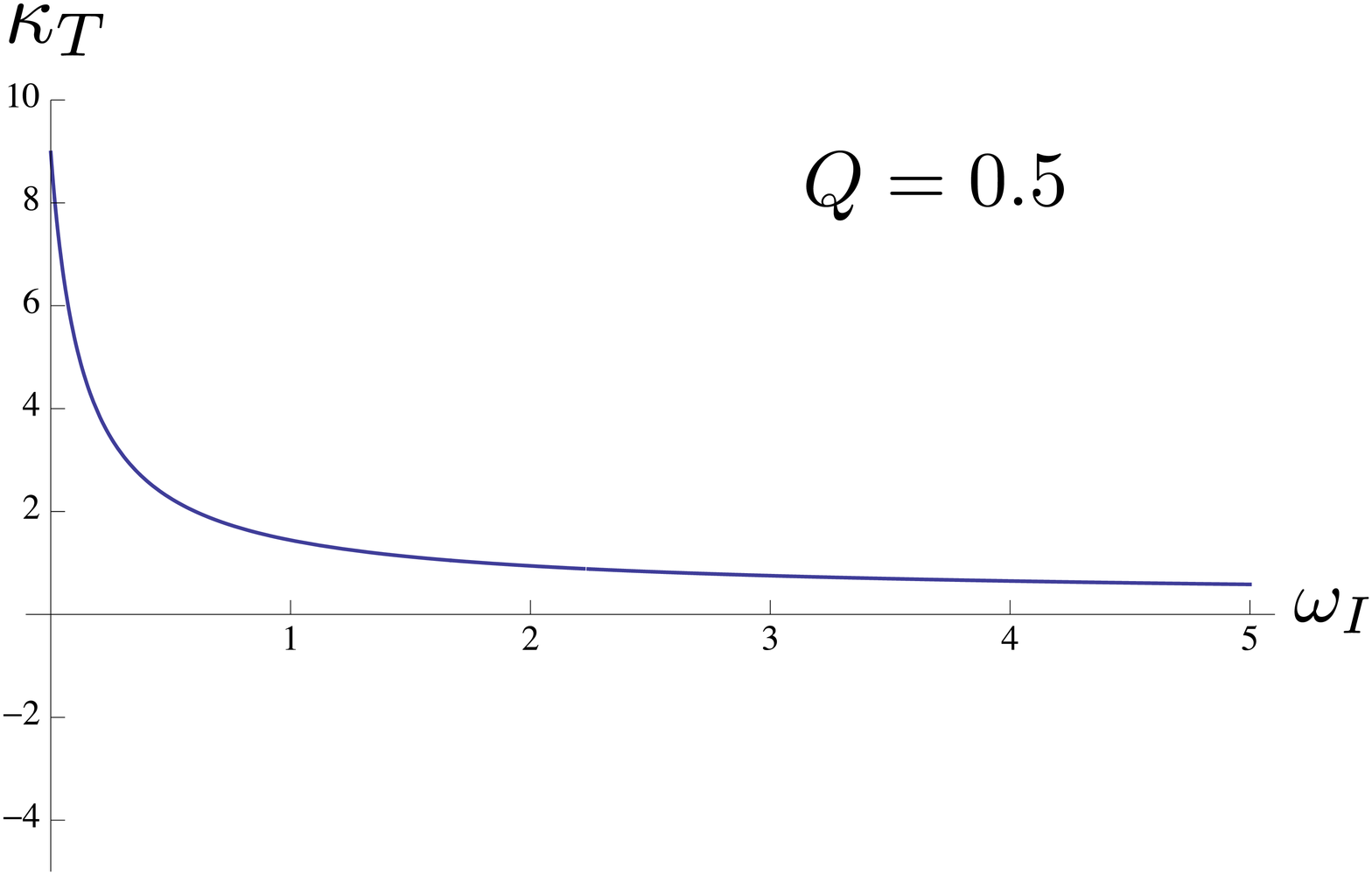} \\
  (a.) & (b.) & (c.)
  \end{array}$
    \caption{The value of $\kappa_T$ in eq.~\protect\eqref{kappatomega} as a function of the imaginary part of the frequency, $\omega_I \equiv \textrm{Im}\,\omega$, with vanishing real part, $\omega_R \equiv \textrm{Re}\,\omega=0$, for (a.) $Q=0.1$, (b.) $Q=0.35$, and (c.) $Q=0.5$.}
    \label{fig:kappaTomega}
\end{figure}

Let us split $\omega$ into real and imaginary parts, $\omega = \omega_R + i \omega_I$. A solution for $\delta \phi(z,\omega)$ with $C_{\textrm{out}}(\omega)=0$ and $\omega_I >0$, with any $\omega_R$, represents an in-going fluctuation growing exponentially in time, $e^{-i\omega t} \propto e^{+\omega_I t}$, and hence an instability. In the field theory such an unstable mode appears as a pole in ${\mathcal{O}}$'s retarded Green's function in the ``wrong'' half of the complex $\omega$-plane. We will search for unstable solutions with $\omega_R=0$. We do so only for simplicity: though not immediately obvious, a straightforward exercise shows that when $\omega_R=0$ the $\kappa_T(\omega)$ in eq.~\eqref{kappatomega} is purely real. Plots of $\kappa_T(\omega)$, with $\omega_R=0$ and $\omega_I>0$ appear in fig.~\ref{fig:kappaTomega}, for some representative values of $Q$. The $\kappa_T$ in eq.~\protect\eqref{kappatomega} depends only on $Q^2$, via $\mathfrak{b}_{\pm}$, hence in fig.~\ref{fig:kappaTomega} we take $Q>0$ without loss of generality. Every point on the plots of $\kappa_T(\omega)$ in fig.~\ref{fig:kappaTomega} represents an in-going solution with $\omega_R=0$ and $\omega_I>0$, \textit{i.e.}\ an unstable mode.

%%%%%%%%%%%%%%%%%%%%%%%%%%%%%%%%%%%%%%%%%%%%%%%%%%%%%%%%%%%%%%%%%%%%%%%%%%%%%%%%%%%%%%%%%%%%%%%%%%%%%%%%%
%%%%%%%%%%%%%%%%%%%%%%%%%%%%%%%%%%%%%%%%%%%%%%%%%%%%%%%%%%%%%%%%%%%%%%%%%%%%%%%%%%%%%%%%%%%%%%%%%%%%%%%%%
%%%%%%%%%%%%%%%%%%%%%%%%%%%%%%%%%%%%%%%%%%%%%%%%%%%%%%%%%%%%%%%%%%%%%%%%%%%%%%%%%%%%%%%%%%%%%%%%%%%%%%%%%
%%%%%%%%%%%%%%%%%%%%%%%%%%%%%%%%%%%%%%%%%%%%%%%%%%%%%%%%%%%%%%%%%%%%%%%%%%%%%%%%%%%%%%%%%%%%%%%%%%%%%%%%%

To understand fig.~\ref{fig:kappaTomega}, let us begin with $T \gg T_K$ and then cool the system, recalling from fig.~\ref{fig:kappaT} how $\kappa_T$ behaves as we reduce $T$. From fig.~\ref{fig:kappaT} we see that when $T\gg T_K$, $\kappa_T$ is small and negative, so in fig.~\ref{fig:kappaTomega} we should imagine a horizontal line just below the $\omega_I$ axis. That line never hits the curves for $\kappa_T(\omega)$ in fig.~\ref{fig:kappaTomega}: no unstable mode appears, for any $Q$, when $T \gg T_K$. As we decrease $T$, from fig.~\ref{fig:kappaT} we see that $\kappa_T$ becomes increasingly negative, so in fig.~\ref{fig:kappaTomega} we move our imaginary horizontal line down. In figs.~\ref{fig:kappaTomega} (a.) and (b.), where $Q=0.1$ and $Q=0.35$, respectively, our imaginary horizontal line will eventually hit the $\kappa_T(\omega)$ curve: an unstable mode appears. That does not happen in fig.~\ref{fig:kappaTomega} (c.), where $Q=0.5$. More generally, from eq.~\eqref{kappatomega} we find that the unstable mode appears for $\kappa_T(\omega)<0$, or equivalently when $T > T_K$, only when $Q \lesssim 0.47$.

As $T \to T_K^+$, fig.~\ref{fig:kappaT} shows that $\kappa_T \to -\infty$. Our imaginary horizontal line then drops to the lower boundary of fig.~\ref{fig:kappaTomega}, where the $Q \lesssim 0.47$ instability persists, but no instability appears yet for $Q \gtrsim 0.47$. As $T \to T_K^-$, fig.~\ref{fig:kappaT} shows that $\kappa_T \to \infty$, so when $T$ passes through $T_K$ our imaginary horizontal line jumps from the lower boundary to the upper boundary of fig.~\ref{fig:kappaTomega}, and, upon further cooling, descends towards the $\omega_I$ axis. In figs.~\ref{fig:kappaTomega} (a.) and (b.), our imaginary horizontal line then intersects the upper branches of the $\kappa_T(\omega)$ curves: when $Q\lesssim 0.47$, the unstable mode persists down to $T<T_K$. In fig.~\ref{fig:kappaTomega} (c.), however, our imaginary horizontal line intersects the $\kappa_T(\omega)$ curve for the first time at $\kappa_T \approx 8.98$: if $Q \gtrsim 0.47$, then the unstable mode first appears only when when $T< T_K$.

Increasing $Q$ increases the dimension of the impurity's representation. We have thus found that the larger we make $Q$, the lower we must make $T$ to trigger an instability. The intuitive lesson is that a ``big'' impurity is ``more stable'' than a ``small'' impurity.

In our model, when $Q=0$ the plot of $\kappa_T$ versus $\omega_I$ is similar to fig.~\ref{fig:kappaTomega} (a.), indicating an instability. Apparently in our system the Kondo effect can somehow occur even for an impurity in the \textit{trivial} representation. That is not a complete surprise: holographic superconductivity triggered by a double-trace coupling can indeed occur even with \textit{zero} charge density~\cite{Faulkner:2010gj}. Moreover, if $\kappa=0$, which via eq.~\eqref{eq:betaTkappaT} means $\kappa_T=0$, then for any $Q$ the instability has $\omega_I= \infty$, \textit{i.e.}\ the unstable mode decouples, as we see in the examples in fig.~\ref{fig:kappaTomega}. In other words, if $\kappa=0$, then the instability does not appear for any $Q$.

To summarize, for \textit{any} $Q$, including $Q=0$, and \textit{any non-zero} $\kappa$, an instability occurs in our model at sufficiently low $T$. As a result, a phase transition must occur, although we cannot yet determine the order of the transition or the transition temperature $T_c$. So far all we know is that a transition occurs somewhere near $T_K$, as defined via eq.~\eqref{eq:betaTkappaT}. We will determine $T_c$ directly from thermodynamics in the next subsection.

%%%%%%%%%%%%%%%%%%%%%%%%%%%%%%%%%%%%%%%%%%%%%%%%%%%%%%%%%%%%%%%%%%%%%%%%%%%%%%%%%%%%%%%%%%%%%%%%%%%%%%%%%
\subsection{The Phase Transition}
\label{phasetransition}
%%%%%%%%%%%%%%%%%%%%%%%%%%%%%%%%%%%%%%%%%%%%%%%%%%%%%%%%%%%%%%%%%%%%%%%%%%%%%%%%%%%%%%%%%%%%%%%%%%%%%%%%%

In this subsection we construct non-trivial solutions for $\phi(z)$ that have smaller Euclidean action than the trivial solution $\phi(z)=0$ when $T$ is below a $T_c$ that we will calculate. Such a non-trivial solution will represent the endpoint of the instability found in subsection~\ref{stability}. We will construct these non-trivial solutions using numerics. In field theory terms, we construct states with $\langle {\mathcal{O}} \rangle \neq 0$ that have lower free energy than the $\langle {\mathcal{O}} \rangle =0$ state when $T \leq T_c$, indicating spontaneous symmetry breaking.

Notice that $A_x(z)$ contributes nothing to the on-shell action: with our ansatz, the bulk CS term vanishes trivially, and $A_x(z)={\mathcal{O}}(z \left(\ln z\right)^2)$ vanishes sufficiently quickly when $z \to 0$ to guarantee that the boundary term of the form in eq.~\eqref{eq:d7boundary} vanishes when $\varepsilon \to 0$.

Let us Wick-rotate to Euclidean signature. In practical terms, we take $g_{tt} \to + h(z)/z^2$ and reverse the overall sign of $S_{AdS_2}$ in eq.~\eqref{eq:ads2action}, producing the Euclidean action $S^{\textrm{E}}_{AdS_2}$. The same statements apply to the boundary terms discussed in subsection~\ref{asymptotics}. We also compactify the Euclidean time direction into a circle of circumference $1/T$, which after the re-scaling in eq.~\eqref{eq:rescaling} becomes circumference $2\pi$. Notice that $a_t(z)$ also Wick-rotates such that the signs in the equations of motion eq.~\eqref{eq:aphieoms3} are unchanged.

To obtain non-trivial solutions of eq.~\eqref{eq:aphieoms3} numerically, we ``shoot from the boundary,'' as follows. The $z \to 0$ asymptotics of $a_t(z)$ and $\phi(z)$ involves four parameters, $Q$, $\mu$, $\beta_T$, and $\kappa_T$. In all of our numerics we take $Q=-1/2$. We next choose $\mu$, $\beta_T$, and $\kappa_T$, and numerically integrate the equations of motion up to $z$ near the horizon $z=1$. We then vary $\beta_T$ and $\kappa_T$ until we obtain a solution obeying the conditions for regularity at the horizon, $a_t(z=1)=0$ and $\phi'(z=1)=0$. Once we obtain an acceptable solution, we proceed to a new $\mu$ and repeat the process. If we un-do the re-scaling in eq.~\eqref{eq:rescaling}, then $\mu \to \mu/(2\pi T)$, so changing the dimensionless $\mu$ is equivalent to changing the dimensionful $\mu$ relative to $T$.

Given solutions for $a_t(z)$ and $\phi(z)$, numerical evaulation of $S^{\textrm{E}}_{AdS_2}$ is straightforward, with a finite result thanks to the counterterms of subsection~\ref{asymptotics}. The field theory's free energy is then $\mathcal{F} = T S^{\textrm{E}}_{AdS_2}$. For the trivial solution $\phi(z)=0$, where $a_t(z)$ is the solution in eq.~\eqref{eq:atsolbg}, the free energy is ${\mathcal{F}}= N\pi T Q\mu= -N\pi T Q^2$. For solutions with $\phi(z) \neq 0$, we will subtract $-N\pi T Q^2$ from $\mathcal{F}$ to obtain the free energy difference $\Delta {\mathcal{F}}$. If $\Delta {\mathcal{F}} <0$, then the $\phi(z)\neq 0$ solution has smaller ${\mathcal{F}}$ and hence is thermodynamically favored over the $\phi(z)=0$ solution.

In our numerics we consider only $\kappa_T>0$, as appropriate in the $T\leq T_K$ regime when $\kappa<0$, as explained below eq.~\eqref{eq:betaTkappaT}. The stability analysis of subsection~\ref{stability} suggests that $T_c \lesssim T_K$ when $\kappa_T>0$ and $T_c \gtrsim T_K$ when $\kappa_T<0$. We will indeed find from our numerical solutions that $T_c \lesssim T_K$ when $\kappa_T>0$.

\begin{figure}[htbp]
\centering
\includegraphics[width=0.5\textwidth]{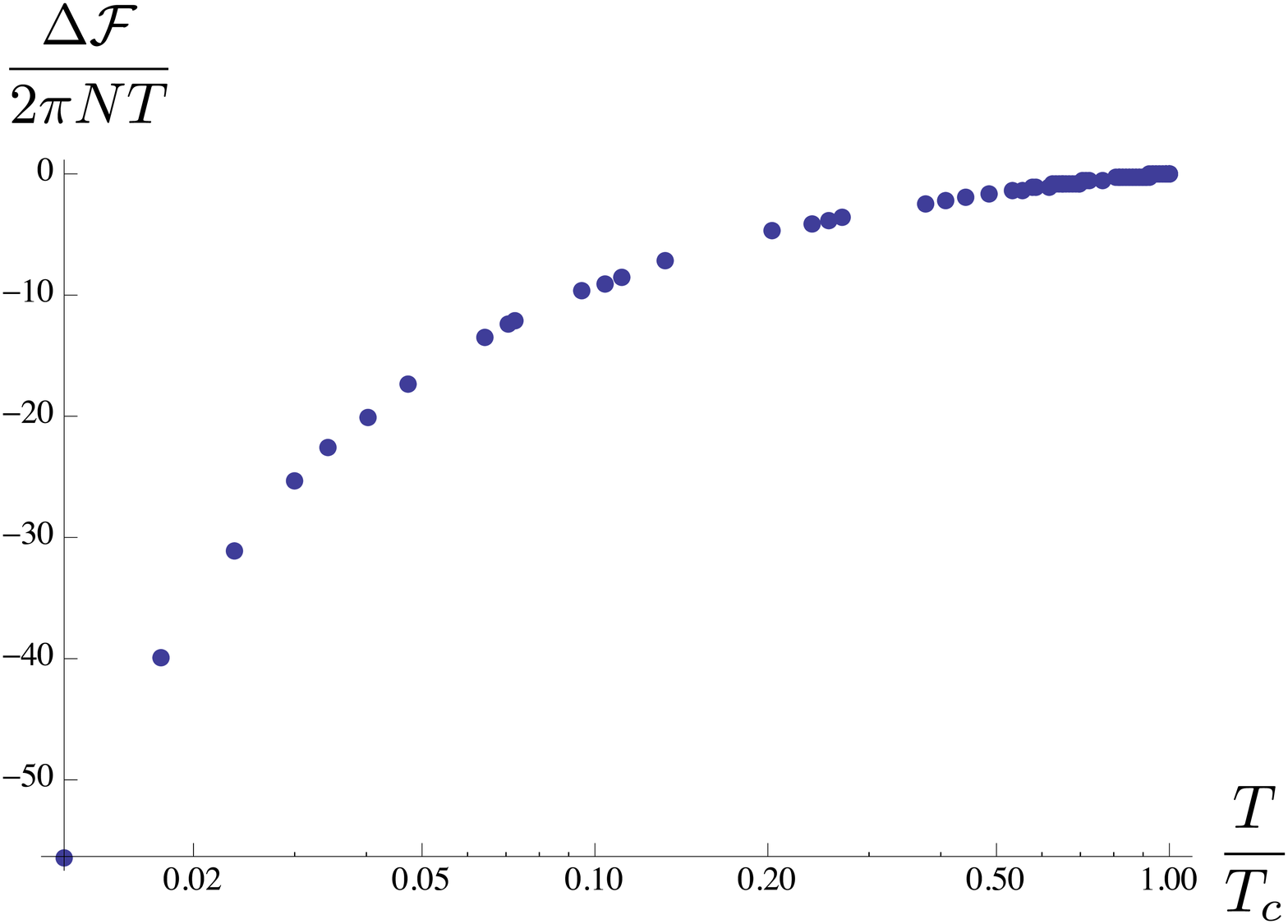}
\caption{Log-linear plot of our numerical results for the free energy difference $\Delta {\mathcal{F}}$ between the condensed ($\langle {\mathcal{O}}\rangle\neq0$) and uncondensed ($\langle {\mathcal{O}}\rangle=0$) phases, in units of $(2\pi NT)$, as a function of $T/T_c$. We find $\Delta {\mathcal{F}} <0$, indicating that the condensed phase is thermodynamically favored for $T\leq T_c$.}
\label{fig:F}
\end{figure}

We find that non-trivial solutions exist only for $\mu \geq 1/2  = -Q$ (the $\mu$ of the trivial solution), or using dimensionful quantities, $\mu \geq - 2 \pi Q T$. Moreover, the non-trivial solutions are always thermodynamically favored over the trivial solution, so we identify the critical temperature as $T_c = - \mu/(2\pi Q)$. Fig.~\ref{fig:F} shows our numerical results for $\Delta {\mathcal{F}}/(2\pi N T)$ versus $T/T_c$, clearly showing $\Delta {\mathcal{F}} <0$. We have thus demonstrated that a second-order phase transition occurs in our system at $T=T_c$.

%%%%%%%%%%%%%%%%%%%%%%%%%%%%%%%%%%%%%%%%%%%%%%%%%%%%%%%%%%%%%%%%%%%%%%%%%%%%%%%%%%%%%%%%%%%%%%%%%%%%%%%%%
%%%%%%%%%%%%%%%%%%%%%%%%%%%%%%%%%%%%%%%%%%%%%%%%%%%%%%%%%%%%%%%%%%%%%%%%%%%%%%%%%%%%%%%%%%%%%%%%%%%%%%%%%
%%%%%%%%%%%%%%%%%%%%%%%%%%%%%%%%%%%%%%%%%%%%%%%%%%%%%%%%%%%%%%%%%%%%%%%%%%%%%%%%%%%%%%%%%%%%%%%%%%%%%%%%%
%%%%%%%%%%%%%%%%%%%%%%%%%%%%%%%%%%%%%%%%%%%%%%%%%%%%%%%%%%%%%%%%%%%%%%%%%%%%%%%%%%%%%%%%%%%%%%%%%%%%%%%%%

For $Q=-1/2$ we find numerically that $\kappa_T \approx 9.04$ at $T_c$, which gives us $T_c/T_K = e^{-1/\kappa_T} \approx 0.895$, and so indeed $T_c \lesssim T_K$, as advertised. Reassuringly, $\kappa_T\approx 9.06$ is very close to $\kappa_T \approx 8.98$, the value of $\kappa_T$ in fig.~\ref{fig:kappaTomega} (c.) where an instability (with $\omega_R=0$) first appears as we reduce $T$. In other words, the onset of instability in the normal phase coincides with the second-order phase transition, as expected.

\begin{figure}[htbp]
  \centering
  $\begin{array}{cc}
  \includegraphics[width=0.45\textwidth]{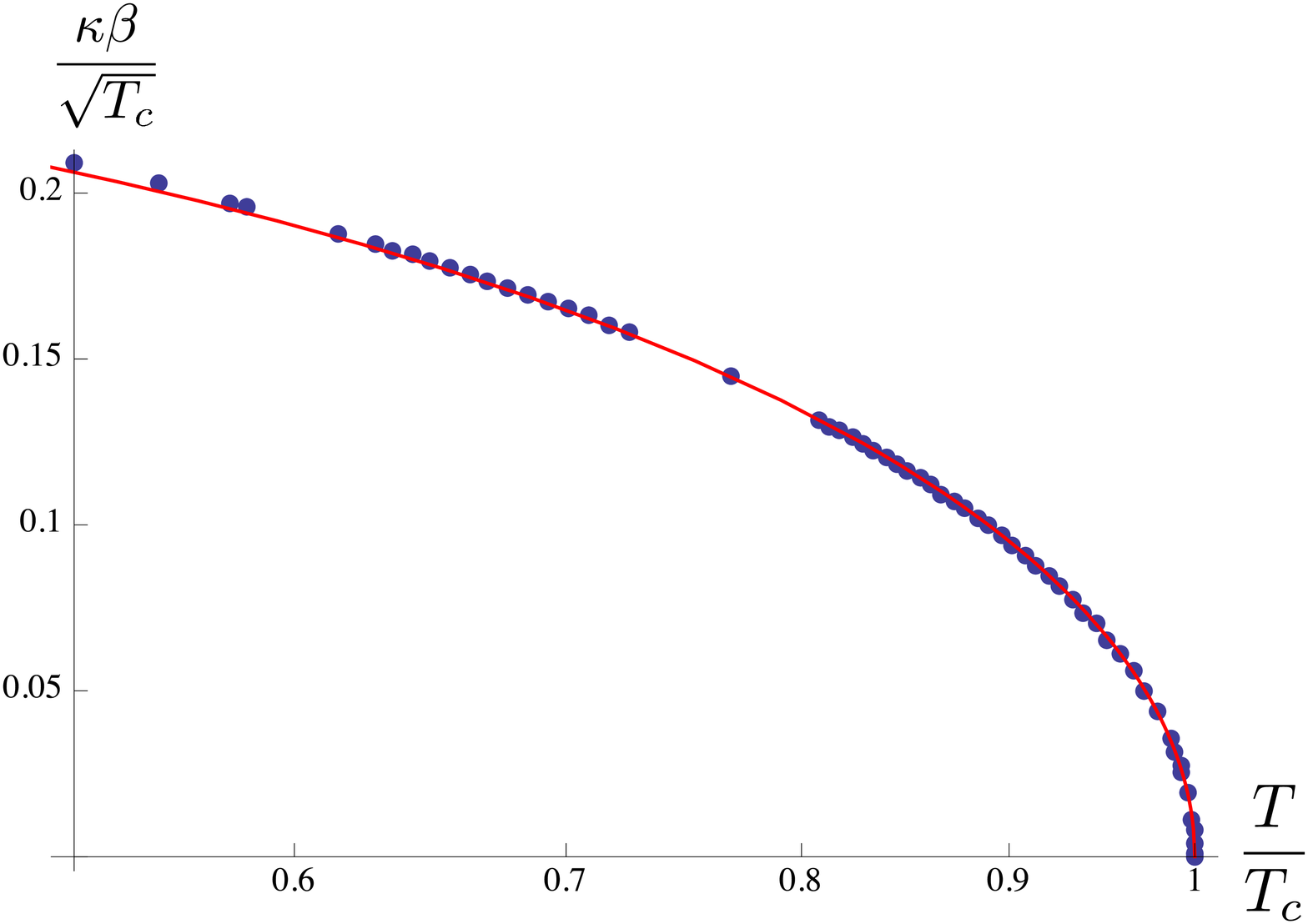} & \includegraphics[width=0.45\textwidth]{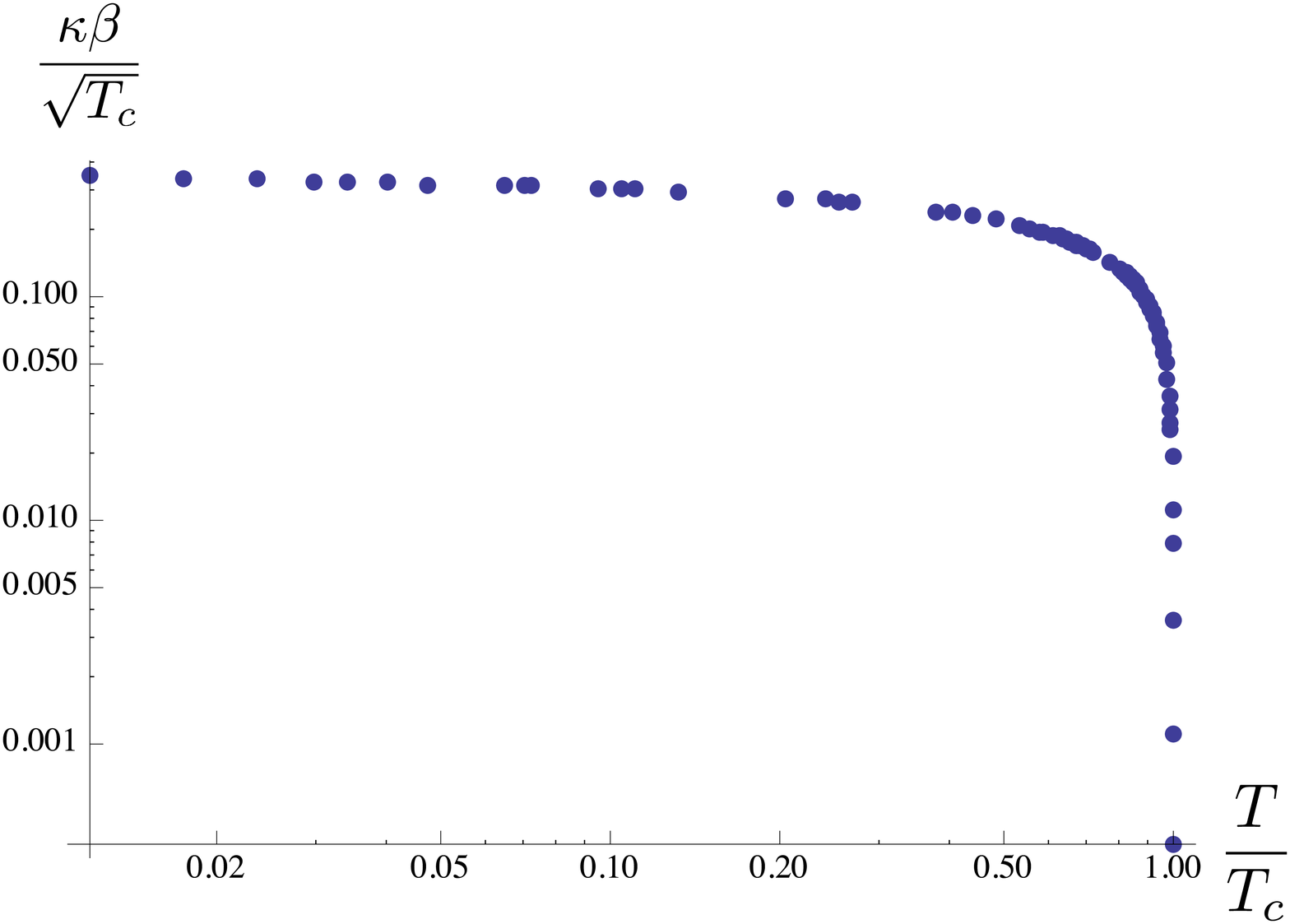} \\
  (a.) & (b.)
  \end{array}$
    \caption{Plots of our numerical results for $\kappa \beta/\sqrt{T_c} \propto \langle {\mathcal{O}}\rangle/(N\sqrt{T_c})$ as a function of $T/T_c$. (a.) Log-linear plot for $T$ just below $T_c$. The solid red curve is $0.30(1-T/T_c)^{1/2}$, where we obtained the number $0.30$ from a fit to the data. The exponent $1/2$ reveals a mean-field transition. (b.) Log-log plot over a larger range of $T/T_c$, revealing that $\langle {\mathcal{O}} \rangle$ approaches a finite constant as $T/T_c \to 0$.}
    \label{fig:alpha}
\end{figure}

As mentioned below eq.~\eqref{eq:sphi}, the condensate $\langle {\mathcal{O}} \rangle \propto N \alpha = N \kappa \beta$. Our numerical solutions give us $\kappa_T$ and $\beta_T$, so using eq.~\eqref{eq:betaTkappaT} we can obtain $\kappa \beta/\sqrt{T_c} \propto \langle {\mathcal{O}} \rangle/(N\sqrt{T_c})$, which we plot as a function of $T/T_c$ in fig.~\ref{fig:alpha}. In fig.~\ref{fig:alpha} (a.) we see the characteristic behavior of a mean-field transition, $\langle {\mathcal{O}} \rangle \propto (T_c-T)^{1/2}$ for $T\lesssim T_c$. In fig.~\ref{fig:alpha} (b.) we plot $\kappa \beta/\sqrt{T_c}$ over a larger range of $T/T_c$, revealing that $\kappa\beta/\sqrt{T_c}$ appears to approach a finite constant as $T /T_c\to 0$.

%%%%%%%%%%%%%%%%%%%%%%%%%%%%%%%%%%%%%%%%%%%%%%%%%%%%%%%%%%%%%%%%%%%%%%%%%%%%%%%%%%%%%%%%%%%%%%%%%%%%%%%%%
%%%%%%%%%%%%%%%%%%%%%%%%%%%%%%%%%%%%%%%%%%%%%%%%%%%%%%%%%%%%%%%%%%%%%%%%%%%%%%%%%%%%%%%%%%%%%%%%%%%%%%%%%
%%%%%%%%%%%%%%%%%%%%%%%%%%%%%%%%%%%%%%%%%%%%%%%%%%%%%%%%%%%%%%%%%%%%%%%%%%%%%%%%%%%%%%%%%%%%%%%%%%%%%%%%%
%%%%%%%%%%%%%%%%%%%%%%%%%%%%%%%%%%%%%%%%%%%%%%%%%%%%%%%%%%%%%%%%%%%%%%%%%%%%%%%%%%%%%%%%%%%%%%%%%%%%%%%%%

In summary, our system is essentially a holographic superconductor~\cite{Hartnoll:2008vx,Hartnoll:2008kx,Faulkner:2010gj} in $AdS_2$. In field theory terms, we have found a second-order, mean-field, symmetry-breaking phase transition at a critical temperature $T_c = - \mu/(2\pi Q)$ in which the $(0+1)$-dimensional charged operator $\mathcal{O} = \psi_L^{\dagger} \chi$ condenses due to the marginally-relevant double-trace coupling ${\mathcal{O}}{\mathcal{O}}^{\dagger}$. Such a transition is extremely similar to that in the large-$N$ approach to the Kondo effect~\cite{0022-3719-19-17-017,PhysRevB.35.5072,2003PhRvL..90u6403S,2004PhRvB..69c5111S}, as reviewed in subsection~\ref{largen}, giving us confidence in our model. In the next subsection we will find many more similarities that will make us even more confident.

%%%%%%%%%%%%%%%%%%%%%%%%%%%%%%%%%%%%%%%%%%%%%%%%%%%%%%%%%%%%%%%%%%%%%%%%%%%%%%%%%%%%%%%%%%%%%%%%%%%%%%%%%
\subsection{The IR Fixed Point}
\label{fixedpoint}
%%%%%%%%%%%%%%%%%%%%%%%%%%%%%%%%%%%%%%%%%%%%%%%%%%%%%%%%%%%%%%%%%%%%%%%%%%%%%%%%%%%%%%%%%%%%%%%%%%%%%%%%%

What is the IR fixed point of our holographic Kondo model? Does our model exhibit under-, critical, or over-screening? Does a phase shift occur? What is the leading irrelevant operator when we deform about the IR fixed point?

Generically in holographic superconductors, to see an IR fixed point we must take $T \to 0$, and include the back-reaction of the gauge field and scalar on the metric. Recalling that the coordinate $z$ is dual to the field theory RG scale, we expect an IR fixed point to appear deep in the bulk, $z \to \infty$, where the full solution will approach a solution with some scaling isometry, dual to some scaling symmetry. For example, in the $z \to \infty$ limit the solution may exhibit Lifshitz scaling with a dynamical exponent fixed by the scalar's mass and charge~\cite{Gubser:2009cg,Horowitz:2009ij}. Indeed, in general as $T \to 0$ the behavior of solutions deep in the bulk depends sensitively on the scalar's potential~\cite{Gubser:2009cg,Horowitz:2009ij}. If the scalar approaches a finite constant deep in the bulk then a scale-invariant solution may emerge there. If the scalar diverges as $z \to \infty$, then the system may never settle into a scale-invariant solution.

In this subsection we will show how some characteristic features of the Kondo effect appear at the IR fixed point of our model. We will work exclusively in the probe limit. In particular, we will assume that the probe limit remains reliable all the way down to $T=0$. The probe limit will not suffice to characterize the IR fixed point completely. For example, we will not be able to determine what scaling symmetry emerges in the IR, since that requires calculating the back-reaction on the metric and then identifying the isometry group that emerges deep in the bulk as $T \to 0$. Nevertheless, the probe limit will suffice to illustrate how the leading irrelevant operator, the screening of the impurity, and the phase shift appear in the holographic dual. We will discuss each of these in turn.

%%%%%%%%%%%%%%%%%%%%%%%%%%%%%%%%%%%%%%%%%%%%%%%%%%%%%%%%%%%%%%%%%%%%%%%%%%%%%%%%%%%%%%%%%%%%%%%%%%%%%%%%%
\subsubsection{The Leading Irrelevant Operator}
%%%%%%%%%%%%%%%%%%%%%%%%%%%%%%%%%%%%%%%%%%%%%%%%%%%%%%%%%%%%%%%%%%%%%%%%%%%%%%%%%%%%%%%%%%%%%%%%%%%%%%%%%

We begin by studying the approach to the IR fixed point, which is controlled by the leading irrelevant operator $\oirr$, of dimension $\dirr>1$, as discussed in subsection~\ref{irrop}.

Roughly speaking, taking $T \to 0$ means $z_H \to \infty$, so the horizon recedes to infinity, and the blackening factor in the metric $h(z) = 1 - z^2/z_H^2  \to 1$. Plugging $h(z)=1$ into $a_t(z)$ and $\phi(z)$'s equations of motion, eqs.~\eqref{aeom3} and~\eqref{phieom3}, respectively, we find
\begin{subequations}
\label{eq:aphieomszeroT}
\bea
a_t'' + \frac{2}{z} a_t' -  2\frac{\phi^2}{z^2} \, a_t =0, \label{aeomzeroT} \\
\phi'' + a_t^2 \, \phi - \frac{M^2}{z^2} \phi = 0. \label{phieomzeroT}
\eea
\end{subequations}

Suppose we know the solutions for $a_t(z)$ and $\phi(z)$ in the $z \to \infty$ limit. The bulk theory in that background is dual to the IR fixed point. In particular, we can invoke the AdS/CFT dictionary in that background: every fluctuation of that background is dual to an operator of the IR fixed point theory, where the mass of a fluctuation maps to the dimension of the dual operator. The leading irrelevant operator $\oirr$ will be dual to a fluctuation that vanishes as $z \to \infty$ and either diverges or approaches a constant as $z \to 0$, as appropriate for an operator that is only important at high energy. Indeed, if we perturb the $z \to \infty$ solutions by that fluctuation, then we can match onto a domain wall solution that approaches the $z \to 0$ solutions described in subsection~\ref{asymptotics}~\cite{Gubser:2009cg}. In field theory language, we can perturb the IR fixed point by $\oirr$ and go ``up the RG flow'' to reach the UV fixed point.

Obviously, the leading deformation about the $z \to \infty$ solutions must be a fluctuation of either $a_t(z)$ or $\phi(z)$, so $\oirr$ will be either the operator dual to $a_t(z)$, $\Oa$, or the operator dual to $\phi(z)$, $\Ophi$, or one of the double-trace operators $\Oa^2$ or $\Ophi^2$. In the UV, $a_t(z)$ is dual to $\chi^{\dagger}\chi$ and $\phi(z)$ is dual to $\psi_L^{\dagger}\chi$. Eqs.~\eqref{eq:aphieomszeroT} represents an RG flow under which these operators may mix, so the operators $\Oa$ and $\Ophi$ of the IR fixed point theory are not necessarily $\chi^{\dagger}\chi$ and $\psi^{\dagger}_L\chi$. The CS gauge field $A_x(z)$ does not enter eq.~\eqref{eq:aphieomszeroT}, indicating that the dual operator $\psi^{\dagger}_L \psi_L$ does not directly participate in the RG flow. In particular, $A_x(z)$ contributes nothing to the on-shell action, as explained in subsection~\ref{phasetransition}, hence we expect that $A_x(z)$ will have no influence on the low-$T$ scalings of thermodynamic oberservables. We will therefore assume that $\oirr$ will not be $\psi_L^{\dagger} \psi_L$. We will discuss $A_x(z)$'s role in our model in subsection~\ref{phaseshift}.

At finite $T$ we expect both $\langle \Oa \rangle \neq 0$ and $\langle \Ophi \rangle \neq 0$, so in our model at finite $T$ we expect $\langle \oirr \rangle \neq 0$ always. As explained in subsection~\ref{cft}, we then expect the impurity's leading contribution to any thermodynamic quantity at low $T$ to be linear in $\lirr$. For example, we expect the impurity's contribution to the entropy at low $T$ to be $\imps \propto \lirr T^{\dirr-1}$. We also expect that generically the impurity's leading contributions to transport properties at low $T$ will be linear in $\lirr$. For example, we expect the impurity's leading contribution to $\rho$ to be $\propto \lirr T^{\dirr}$. The upshot is that to determine the low-$T$ scalings of thermodynamic and transport properties in our model, we just need to compute $\dirr$.

Our goal in this subsection is to identify the possible $\oirr$ and $\dirr$ in our model as a function of $M^2=-1/4+Q^2$. Doing so requires two steps. First, for a given value of $M^2$, we must solve eq.~\eqref{eq:aphieomszeroT} in the $z \to \infty$ limit. Second, we must find the leading deformation about the $z \to \infty$ solution.

In the $z \to \infty$ limit, $a_t(z)$ and $\phi(z)$ have only two options: diverge or approach a constant, possibly including zero. As mentioned in subsections~\ref{stability} and~\ref{phasetransition}, when $T>0$ we demand that $a_t(z_H)=0$. Here we will demand continuity of the $T\to0$ or $z_H \to \infty$ limit: we will impose $\lim_{z \to \infty} a_t(z) = 0$. In our model, $\phi(z)$'s potential includes only a mass term $\propto M^2 \phi(z)^2$. If $-1/4\leq M^2<0$ then $\phi(z)$'s potential is unbounded from below, and we expect $\phi(z)$ to diverge as $z \to \infty$. In that case, the system may never settle into a scale-invariant solution, and moreover $\phi(z)$'s stress-energy tensor will almost certainly diverge as $z \to \infty$, invalidating the probe limit. We will thus demand that $\lim_{z \to \infty} \phi(z)$ is a constant, or equivalently that $M^2\geq0$. Given our choice $M^2 = -1/4 + Q^2$, that means $Q^2 \geq 1/2$. If $M^2>0$ then we expect $\lim_{z \to \infty} \phi(z)=0$, whereas if $M^2=0$ then $\lim_{z \to \infty}\phi(z)$ has no preferred value \textit{a priori}, but may be pushed to some value by $\phi(z)$'s coupling to $a_t(z)$. We can only determine that value by solving eq.~\eqref{eq:aphieomszeroT} for all $z$ and extracting $\lim_{z \to \infty} \phi(z)$. With these boundary conditions, the bulk solution representing the IR fixed point will be $a_t(z)=0$ with constant $\phi(z)$.

We expect a scaling symmetry to emerge when $z \to \infty$ only if $a_t(z)$ and $\phi(z)$ approach powers of $z$ as $z \to \infty$, since in general any more complicated function will introduce one or more scales. We will thus look for power-law solutions of eq.~\eqref{eq:aphieomszeroT} in the $z \to \infty$ limit,
\beq
\label{eq:powersols}
\phi(z) = \phinf \, z^X, \qquad a_t(z) = \ainf \, z^Y,
\eeq
where $\phinf$, $\ainf$, $X$, and $Y$ are $z$-independent constants. To enfore our boundary conditions we require $X \leq 0$ and $Y \leq 0$. In particular, if $Y=0$ then we must demand $\ainf=0$. Plugging eq.~\eqref{eq:powersols} into eq.~\eqref{eq:aphieomszeroT}, we find
\begin{subequations}
\label{eq:powereqs}
\bea
\label{eq:powereqs1}
Y(Y+1) & - & 2 \phinf^2 \, z^{2X} = 0, \\
\label{eq:powereqs2}
X(X-1) & - & M^2 + \ainf^2 \, z^{2Y+2}=0.
\eea
\end{subequations}
The table below summarizes the solutions of eqs.~\eqref{eq:powereqs1} and~\eqref{eq:powereqs2} that obey our boundary conditions.
\begin{table}[h!]
\begin{center}
\begin{tabular}{|c|c|c|c|c|}
  \hline
  $X$ & $Y$ & $M^2\geq0$ & $\oirr$ & $\dirr$ \\ \hline \hline
  $\frac{1}{2} - \sqrt{\frac{1}{4} + M^2}$ & $0$ & $>0$ & $\Ophi$ & $\frac{1}{2} + \sqrt{\frac{1}{4} + M^2}$  \\\hline
  $\frac{1}{2} - \sqrt{\frac{1}{4} + M^2 - \ainf^2}$ & $-1$ & $<\ainf^2 + 2$ & $\Ophi$ & $\frac{1}{2} + \sqrt{\frac{1}{4} + M^2 - \ainf^2}$ \\\hline
  $\frac{1}{2} - \sqrt{\frac{1}{4} + M^2 - \ainf^2}$ & $-1$ & $>\ainf^2 + 2$ & $(\Oa)^2$ & $2$ \\\hline
  $0$ & $-\frac{1}{2} - \sqrt{\frac{1}{4} + 2 \phinf^2}$ & $0$ & $\Oa$ & $\frac{1}{2} + \sqrt{\frac{1}{4} + 2 \phinf^2}$ \\\hline
\end{tabular}
\end{center}
\end{table}

We will now explain in detail each entry of the table above. The cases $X<0$, where $\lim_{z \to \infty} \phi(z)=0$, and $X=0$, where $\lim_{z \to \infty} \phi(z)=\phinf$ may be non-zero, lead to qualitatively different classes of solutions.

If $X<0$ then in eq.~\eqref{eq:powereqs1} as $z \to \infty$ the term $\phinf^2 z^{2X}$ is suppressed and the equation becomes $Y(Y+1)=0$, with solutions $Y=0$ and $-1$. Consider first the case $Y=0$. In that case, in eq.~\eqref{eq:powereqs2} the term $\ainf^2 \, z^{2Y+2} = \ainf^2 z^2$ dominates as $z \to \infty$, forcing us to set $\ainf=0$, as expected when $Y=0$. Eq.~\eqref{eq:powereqs2} is then the same as for a scalar of mass $M$ in $AdS_2$, so that $X = 1/2 \pm \sqrt{1/4 + M^2}$. Crucially, these powers are different from those in $\phi(z)$'s expansion near the $AdS_2$ boundary, which are fixed not by $M^2$ but by $M^2 - Q^2$, as discussed in subsection~\ref{asymptotics}. To obtain $X<0$ we must choose the minus sign and demand $M^2 >0$. In short, we find a solution with $X = 1/2 - \sqrt{1/4 + M^2}$ and $Y=0$ with $M^2>0$. The solution representing the IR fixed point is $a_t(z)=0$ and $\phi(z)=0$, and the leading deformation is $\phi(z)= \phinf z^{1/2 - \sqrt{1/4+M^2}}$. Invoking the AdS/CFT dictionary, we interpret that deformation as a source for $\oirr$, and so identify $\oirr=\Ophi$, where $\lirr \propto \phinf$ and $\dirr = 1/2 + \sqrt{1/4 + M^2}$, with $M^2>0$. These results appear in the first row of the table.

Now consider $X<0$ and $Y=-1$. From eq.~\eqref{eq:powereqs2} we find $X=1/2 \pm \sqrt{1/4 + M^2 - \ainf^2}$. To obtain $X<0$ we choose the minus sign and demand $M^2 - \ainf^2>0$. The solution representing the IR fixed point is $a_t(z)=0$ and $\phi(z)=0$. Whether $a_t(z)$ or $\phi(z)$ is the leading deformation then depends on how $X$ compares to $Y=-1$. If $X>Y$, meaning $1/2 - \sqrt{1/4 + M^2 - \ainf^2}> -1$ or equivalently $M^2 - \ainf^2 <2$, then the leading deformation is $\phi(z) = \phinf z^{1/2 - \sqrt{1/4 + M^2 - \ainf^2}}$. We thus again identify $\oirr=\Ophi$, now with $\dirr = 1/2 + \sqrt{1/4 + M^2 - \ainf^2}$, as listed in the second row of the table. If $X<Y$, meaning $1/2 - \sqrt{1/4 + M^2 - \ainf^2}<-1$ or equivalently $M^2 - \ainf^2 > 2$, then the leading deformation is $a_t(z) = \ainf/z$, suggesting that $\Oa$ is exactly marginal. In that case, $\oirr$ is likely the double-trace operator $(\Oa)^2$, with $\dirr=2$, as listed in the third row of the table.

Finally, when $X=0$, eq.~\eqref{eq:powereqs1} is the same as for a vector field in $AdS_2$ with mass squared $2 \phinf^2$. In other words, $\phi(z)$ gives $a_t(z)$ a mass via the Higgs mechanism, as expected. We thus find $Y = -1/2 \pm \sqrt{1/4 + 2 \phinf^2}$. To obtain $Y<0$ we must choose the minus sign. When $Y<0$ in eq.~\eqref{eq:powereqs2}, the term $\ainf^2 \, z^{2Y+2}$ is suppressed as $z \to \infty$, and the equation reduces to $M^2=0$, as expected when $X=0$. In other words, when $M^2=0$, $\phi(z)$ approaches the constant $\phinf$ as $z \to \infty$, which gives a constant mass to $a_t(z)$, so that $Y=-1/2 - \sqrt{1/4 + 2 \phinf^2}$. The solution representing the IR fixed point is $a_t(z)=0$ and $\phi(z) = \phinf$. The leading deformation is $a_t(z) = \ainf z^{-1/2 - \sqrt{1/4 + 2 \phinf^2}}$, so we identify $\oirr=\Oa$, with $\dirr = 1/2 + \sqrt{1/4 + 2 \phinf^2}$, as listed in the fourth row of the table.

\begin{figure}[t]
  \centering
  \includegraphics[width=0.5\textwidth]{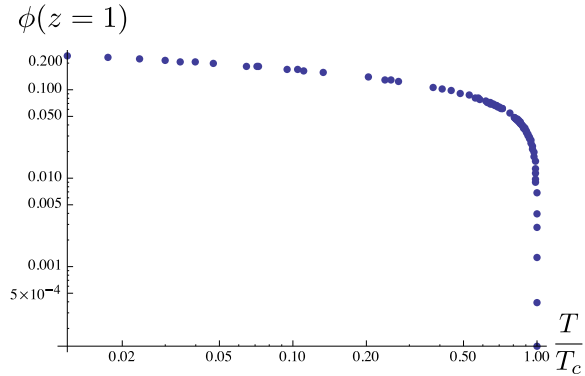}
    \caption{Log-log plot of our numerical results for the value of the scalar at the horizon, $\phi(z=1)$, as a function of $T/T_c$, down to $T/T_c = 0.012$, for $Q=-1/2$. We find that $\phi(z=1)$ appears to approach a non-zero constant as $T/T_c \to 0$, namely $\phi(z=1)\approx 0.2$.}
    \label{fig:phihor}
\end{figure}

As mentioned above, when $M^2=0$ we can only determine $\phinf$ by solving eq.~\eqref{eq:aphieomszeroT} for all $z$ and then extracting $\lim_{z \to \infty} \phi(z)$. Equivalently, with $T>0$ we can calculate the value of $\phi(z)$ at the horizon, which after the re-scalings in eq.~\eqref{eq:rescaling} is $\phi(z=1)$, and then take the limit $T \to 0$, where the horizon recedes to infinity. Fig.~\ref{fig:phihor} shows our numerical results for $\phi(z=1)$ with $Q=-1/2$ and hence $M^2 = -1/4 + Q^2 = 0$, as a function of $T/T_c$, down to $T/T_c=0.012$. Our numerical results suggest that $\phinf \approx 0.2$, so that $\dirr \approx 1.07$. Apparently, when $M^2=0$ in our model, $\oirr = \Oa$ is only weakly irrelevant.

Under the assumption that the probe limit remains valid as $T \to 0$, we have thus enumerated all possible $\oirr$ in our model, as summarized in the table above. When $M^2>0$, multiple options for $\oirr$ exist. For a given choice of $M^2>0$, to determine which $\oirr$ is actually realized requires solving eq.~\eqref{eq:aphieomszeroT} for all $z$ and then studying the $z \to \infty$ asymptotics of the solutions. We will leave that for future research.

The table above shows that in our model generically $\dirr$ is not an integer, and hence the IR fixed point cannot be described by free fields. That is not surprising: in the dual field theory, the 't Hooft coupling is always large, regardless of how our double-trace Kondo coupling runs, and hence all fixed points in our model are strongly-coupled.

Our results are dramatically different from those of the standard Kondo system with an impurity in a totally anti-symmetric representation of $SU(N)$, reviewed in section~\ref{kondo}. When $k=1$, which leads to critical screening and a trivial IR CFT, $\oirr=\mathcal{J}^a \mathcal{J}^a$ with $\dirr=2$, leading to the $\imps$ and $\rho$ in eq.~\eqref{eq:criticalscalings}. Recall that $\mathcal{J}^a$ the spin current of the IR CFT, \textit{i.e.}\ after absorbing the spin, eq.~\eqref{absorb}, which is a linear combination of $J^a J^a$, $S^a J^a$, and $S^aS^a$, or equivalently $(\psi_L^{\dagger}\psi_L)^2$, $(\psi_L^{\dagger} \chi)^2$, and $(\chi^{\dagger}\chi)^2$. For that to appear as $\oirr$ in our model the leading deformation about the $z \to \infty$ solution would have to be a linear combination of the CS gauge field, the scalar, and the $AdS_2$ Maxwell field that is constant in $z$. When $k>1$, which leads to over-screening and a non-trivial IR CFT, $\oirr$ is obtained by contracting the spin current with the adjoint primary of $SU(N)$, and $\dirr = 1 + \frac{N}{N+k}$~\cite{1998PhRvB..58.3794P}, leading to the $\imps$ and $\rho$ in eq.~\eqref{eq:overscreenedscalings}. In fact, in terms of the Abrikosov pseudo-fermions $\chi$, and at large $N$, $\oirr=(\psi_L^{\dagger} \chi)^2$~\cite{1998PhRvB..58.3794P}. If $N \to \infty$ with $k \ll N$, the analogue of our probe limit, then $\dirr \to 2$. For that to occur in our model, the leading deformation about the $z \to \infty$ solution would have to be a constant $\phi(z) = \phinf$.

%%%%%%%%%%%%%%%%%%%%%%%%%%%%%%%%%%%%%%%%%%%%%%%%%%%%%%%%%%%%%%%%%%%%%%%%%%%%%%%%%%%%%%%%%%%%%%%%%%%%%%%%%
%%%%%%%%%%%%%%%%%%%%%%%%%%%%%%%%%%%%%%%%%%%%%%%%%%%%%%%%%%%%%%%%%%%%%%%%%%%%%%%%%%%%%%%%%%%%%%%%%%%%%%%%%
%%%%%%%%%%%%%%%%%%%%%%%%%%%%%%%%%%%%%%%%%%%%%%%%%%%%%%%%%%%%%%%%%%%%%%%%%%%%%%%%%%%%%%%%%%%%%%%%%%%%%%%%%
%%%%%%%%%%%%%%%%%%%%%%%%%%%%%%%%%%%%%%%%%%%%%%%%%%%%%%%%%%%%%%%%%%%%%%%%%%%%%%%%%%%%%%%%%%%%%%%%%%%%%%%%%

%%%%%%%%%%%%%%%%%%%%%%%%%%%%%%%%%%%%%%%%%%%%%%%%%%%%%%%%%%%%%%%%%%%%%%%%%%%%%%%%%%%%%%%%%%%%%%%%%%%%%%%%%
\subsubsection{Screening of the Impurity}
%%%%%%%%%%%%%%%%%%%%%%%%%%%%%%%%%%%%%%%%%%%%%%%%%%%%%%%%%%%%%%%%%%%%%%%%%%%%%%%%%%%%%%%%%%%%%%%%%%%%%%%%%

As discussed in subsection~\ref{asymptotics}, $a_t(z)$'s electric flux at the boundary, $\lim_{z\to 0} \sqrt{-g} f^{tz} = - Q$, encodes the impurity's representation in the UV. When $T > T_c$ and $\phi(z)=0$, that electric flux is constant from the boundary to the horizon. When $T \leq T_c$, the non-trivial $\phi(z)$ draws electric charge away from $a_t(z)$, reducing the electric flux at the horizon. At $T/T_c=0$, if $\phi(z)$ does not draw all the charge away from $a_t(z)$, then we may interpret the remaining non-zero flux $\lim_{z \to \infty} \sqrt{-g} f^{tz}$ as an impurity in the IR in a representation with smaller dimension than that in the UV, \textit{i.e.}\ the Young tableau has fewer boxes. This is under-screening. If $\phi(z)$ draws all the charge away from $a_t(z)$, so that $\lim_{z \to \infty} \sqrt{-g} f^{tz}=0$, then no impurity survives in the IR, as occurs in critical and over-screening.

\begin{figure}[t]
  \centering
  $\begin{array}{cc}
  \includegraphics[width=0.45\textwidth]{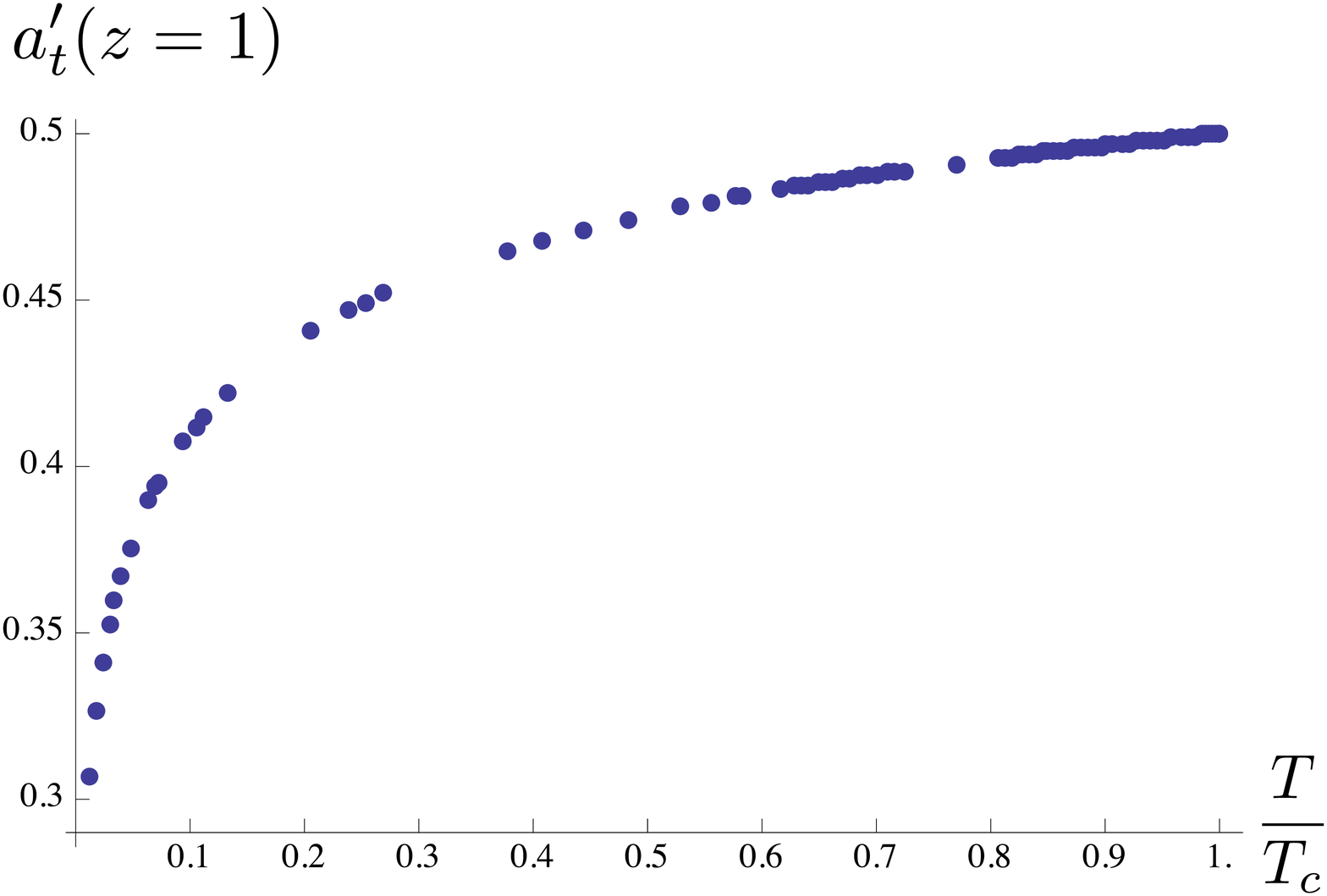} & \includegraphics[width=0.45\textwidth]{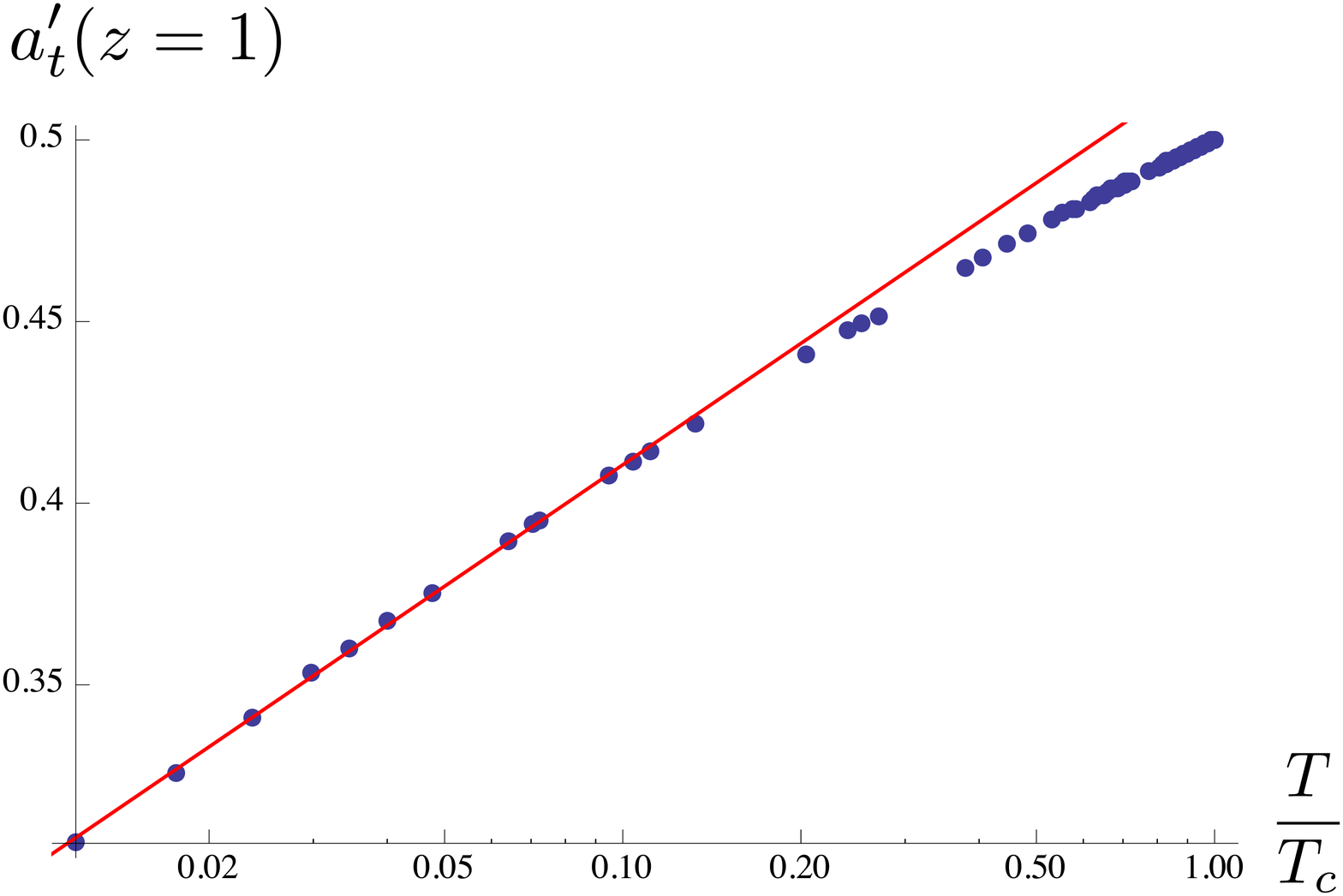} \\
  (a.) & (b.)
  \end{array}$
    \caption{Plots of our numerical results for the electric flux at the horizon, $\left . \sqrt{-g} f^{tz} \right |_{z=1}= \left . z^2 a_t'(z) \right |_{z=1}= a_t'(z=1)$, as a function of $T/T_c$, for $Q=-1/2$. (a.) Between $T/T_c=1$ and $T/T_c=0.012$, $a_t'(z=1)$ decreases by only about $40\%$, from $-Q=1/2$ to about $0.30$. (b.) The same as (a.) but a log-linear plot, revealing that $a_t'(z=1)$ decreases only logarithmically for $T/T_c \lesssim 0.20$: the solid red line is $0.522 + 0.048 \ln \left(T/T_c\right)$, obtained from a fit to the data.}
    \label{fig:fluxhor}
\end{figure}

For $Q=-1/2$ and hence $M^2 = -1/4 + Q^2 =0$, fig.~\ref{fig:fluxhor} shows our numerical results for the electric flux at the horizon, which after the re-scaling in eq.~\eqref{eq:rescaling} is simply $\left . \sqrt{-g} f^{tz} \right |_{z=1}= \left . z^2 a_t'(z) \right |_{z=1}= a_t'(z=1)$, as a function of $T/T_c$. Fig.~\ref{fig:fluxhor} (a.) shows that $a_t'(z=1)$ indeed decreases as $T/T_c$ decreases, although between $T/T_c=1$ and $T/T_c = 0.012$ the decrease is only $\approx 40\%$, from $-Q=1/2$ to about $0.30$. Fig.~\ref{fig:fluxhor} (b.) shows that the decrease is only logarithmic for $T/T_c\lesssim 0.20$. Our numerical results for $a_t'(z=1)$ are insufficient to extrapolate reliably to $T/T_c=0$, so we will leave the fate of $a_t'(z=1)$ at $T/T_c=0$ for future research.

%%%%%%%%%%%%%%%%%%%%%%%%%%%%%%%%%%%%%%%%%%%%%%%%%%%%%%%%%%%%%%%%%%%%%%%%%%%%%%%%%%%%%%%%%%%%%%%%%%%%%%%%%
%%%%%%%%%%%%%%%%%%%%%%%%%%%%%%%%%%%%%%%%%%%%%%%%%%%%%%%%%%%%%%%%%%%%%%%%%%%%%%%%%%%%%%%%%%%%%%%%%%%%%%%%%
%%%%%%%%%%%%%%%%%%%%%%%%%%%%%%%%%%%%%%%%%%%%%%%%%%%%%%%%%%%%%%%%%%%%%%%%%%%%%%%%%%%%%%%%%%%%%%%%%%%%%%%%%
%%%%%%%%%%%%%%%%%%%%%%%%%%%%%%%%%%%%%%%%%%%%%%%%%%%%%%%%%%%%%%%%%%%%%%%%%%%%%%%%%%%%%%%%%%%%%%%%%%%%%%%%%

As a word of caution, in the Kondo systems reviewed in section~\ref{kondo}, ``under-screening'' and ``critical screening'' are often equated with ``free IR fixed point,'' while ``over-screening'' is often equated with ``strongly-coupled IR fixed point.'' In our holographic Kondo model, however, ``under-screening'' and `critical-screening'' do not immediately imply a free IR fixed point: in our model the 't Hooft coupling is always large, and so we expect a strongly-coupled IR fixed point with under-, critical, or over-screening. In other words, in our model under- or critical screening may occur, but the IR fixed point will always be strongly-coupled.

%%%%%%%%%%%%%%%%%%%%%%%%%%%%%%%%%%%%%%%%%%%%%%%%%%%%%%%%%%%%%%%%%%%%%%%%%%%%%%%%%%%%%%%%%%%%%%%%%%%%%%%%%
\subsubsection{The Phase Shift}
\label{phaseshift}
%%%%%%%%%%%%%%%%%%%%%%%%%%%%%%%%%%%%%%%%%%%%%%%%%%%%%%%%%%%%%%%%%%%%%%%%%%%%%%%%%%%%%%%%%%%%%%%%%%%%%%%%%

How does a phase shift at the IR fixed point appear in our model? Here the CS gauge field $A_x(z)$ plays a starring role~\cite{Kraus:2006nb,Kraus:2006wn}. From $A_x(z)$'s equation of motion, eq.~\eqref{Aeom2}, we have
\beq
\label{axsol}
A_x(z) = 4\pi \, \delta(x) \, \int_0^z d\hat{z} \, \sqrt{-g} \, g^{tt} \, a_t \, \phi^2 = -2\pi \, \delta(x) \, \int_0^z d\hat{z} \, J^t(\hat{z}) \equiv -2\pi \, \delta(x) \, c(z),
\eeq
where $\hat{z}$ is a dummy variable, $J^t(z)$ is the electric charge density defined from eq.~\eqref{eq:current}, and $c(z) \equiv \int_0^z d\hat{z} \, J^t(\hat{z})$ is the net electric charge between the boundary and $z$. As discussed in subsection~\ref{asymptotics}, $A_x(z)\to 0$ as $z \to 0$. As $z \to z_H$ (not performing the re-scaling in eq.~\eqref{eq:rescaling}), $\phi(z)$ approaches a non-zero constant whereas to leading order $a_t(z) \propto g_{tt} \to 0$, so $A_x(z)$ approaches $\delta(x)$ times a non-zero constant, $-2\pi q(z_H)$. If we compactify $x$ into a circle, then we can consider the Wilson loop of $A_x(z)$ around $x$ at a fixed $z$,
\beq
W(z) \equiv \oint dx \, A_x(z).
\eeq
For the $A_x(z)$ in eq.~\eqref{axsol}, the $\delta(x)$ makes the $\oint dx$ trivial, so we find $W(z) = -2\pi c(z)$. The electric charge between the boundary and $z$ completely determines $W(z)$.

If we introduce a particle charged under the CS gauge field, with unit charge, and move the particle around the $x$ circle at a fixed $z$, then the particle will acquire a phase shift $e^{i W(z)}$. If $T>T_c$ then $\phi(z)=0$ and hence $c(z)=0$ and $W(z)=0$. If $T\leq T_c$, then $\phi(z)\neq0$ and hence $c(z)\neq 0$ and $W(z)\neq0$. When $T\leq T_c$, if we descend from $z=0$ down into the bulk, then the particle accumulates a larger and larger phase shift. When $T/T_c=0$, the phase shift between $z=0$ and $z \to \infty$, which is determined by the \textit{total} charge in the bulk, $\lim_{z \to \infty} c(z)$, is dual to the phase shift of the IR fixed point. The phase shift represents a spectral flow, as discussed in detail in refs.~\cite{Kraus:2006nb,Kraus:2006wn}, precisely as expected in the Kondo effect.

As mentioned in section~\ref{cft}, in the WZW description of the Kondo effect the phase shift appears as a shift in periodicity in $x$ of the $U(1)$ charge boson. To see the shift from the bulk perspective, let us recall that the $U(1)$ charge boson, which we will denote $\Psi(t,x)$, is related to a Wilson line of $A_z$. Specifically, if we define a bulk scalar field
\beq
\label{azwilson}
\Psi(t,x,z) \equiv \int_0^z d\hat{z} \, A_{\hat{z}}(t,x,\hat{z}),
\eeq
then the $U(1)$ charge bosons in the UV and IR are $\Psi(t,x,z=0)$ and $\Psi(t,x,z\to\infty)$, respectively~\cite{Davis:2007ka,Nickel:2010pr}. If we choose $A_x = 0$ gauge, then
\beq
\partial_x \Psi(t,x,z) = \int_0^z d\hat{z} \, \partial_x A_{\hat{z}}(t,x,\hat{z}) = \int_0^z d\hat{z} \, F_{x\hat{z}}.
\eeq
The periodicity of $\Psi(t,x,z)$ in $x$ at a fixed $z$ is then given by
\beq
\oint dx\, \partial_x \Psi(x,t,z) = \oint \, dx \, \int_0^z d\hat{z} \, F_{x\hat{z}}.
\eeq
For our solution, which is in $A_z=0$ gauge with $A_x(z)\to 0$ as $z \to 0$,
\beq
\oint dx\, \partial_x \Psi(x,t,z) = \oint \, dx \, \int_0^z d\hat{z} \, F_{x\hat{z}} = - \oint dx \, \int_0^z d\hat{z} \, \partial_{\hat{z}} A_x = 2 \pi c(z).
\eeq
The electric charge between the boundary and $z$ determines the periodicity of $\Psi(t,x,z)$ in $x$ at that $z$. When $T/T_c=0$, the total electric charge $\lim_{z\to\infty} c(z)$ will determine the difference in periodicity between $\Psi(t,x,z=0)$ and $\Psi(t,x,z\to\infty)$, and will thus determine the shift in periodicity of the $U(1)$ charge boson $\Psi(t,x)$ between the UV and IR.

In Kondo systems, one version of Friedel's sum rule relates the IR phase shift to the impurity's representation, or equivalently the number $q$ of Abrikosov pseudo-fermions $\chi$~\cite{1998PhRvB..58.3794P}. In our holographic Kondo system, that Friedel sum rule is precisely the relation between the phase shift and the bulk electric charge $c(z)$ that we have found. Consider for example critical screening, where $\phi(z)$ drains $a_t(z)$ of electric flux completely. In that case, upon integrating $a_t(z)$'s equation of motion, eq.~\eqref{aeom2}, we find $\lim_{z\to\infty} c(z) = -Q$, producing a phase shift of $2 \pi Q $. The key point is that $\phi(z)$ is bi-fundamental, so in eqs.~\eqref{Aeom2} and~\eqref{aeom2} the same charge density $J^t(z)$ appears as a source for both $A_x(z)$ and $a_t(z)$. As a result, any charge drained from $a_t(z)$ is transferred over to $A_x(z)$, directly tying the reduction of the dimension of the impurity's representation to the growth of the phase shift.

We can also identify a holographic version of ``absorbing the spin,'' eq.~\eqref{absorb}, as follows. We can re-write a (1+1)-dimensional conserved current $J^m$ in terms of a scalar,
\beq
J^m \equiv \epsilon^{mn} \, \partial_n \theta,
\eeq
and so re-write the coupling between the CS gauge field and $J^m$ as a $(1+1)$-dimensional $\theta$-angle,
\beq
-N \int d^3x \, \delta(x) \, A_m J^m = - N\int d^3x \, \delta(x) \, A_m \epsilon^{mn} \, \partial_n \theta = N\int d^3x \, \delta(x) \, \theta \frac{1}{2} \, \epsilon^{mn} F_{mn},
\eeq
where in the second equality we performed an integration by parts, ignoring the boundary terms. In our solution $J^z=0$ while $J^t(z)$ is non-zero, leading to non-zero $\theta(z)$, which is simply minus the net electric charge between the boundary and $z$:
\beq
c(z) = \int_0^z d\hat{z} \, J^t(\hat{z}) = \int_0^z d\hat{z} \, \epsilon^{t\hat{z}} \, \partial_{\hat{z}} \theta(\hat{z}) = - \theta(z).
\eeq
In particular, when $T=0$ we expect that $\thetainf \equiv \lim_{z \to \infty} \theta(z) = - \lim_{z\to\infty} c(z)$ will be minus the total electric charge in the bulk. In the $z \to \infty$ region, we can thus write an effective action, $S_{\theta}$, including only the CS gauge field and a constant $\theta$-angle,
\beq
\label{eq:CSeff}
S_{\theta} = - \frac{N}{4\pi} \int A \wedge F + N\int d^3x \, \delta(x) \, \thetainf \, \frac{1}{2} \epsilon^{mn} F_{mn},
\eeq
where we also require $\oint dx A_x = - 2 \pi \lim_{z \to \infty} c(z) = 2 \pi \thetainf$. Written explicitly, eq.~\eqref{eq:CSeff} is
\beq
\label{eq:stheta}
S_{\theta} = - \frac{N}{4\pi} \int d^3x \left[ A_z F_{tx} - A_t F_{zx} + A_x F_{zt} \right] + N\int d^3x \, \delta(x) \, \thetainf \, F_{zt}.
\eeq
The two terms in eq.~\eqref{eq:stheta} that are $\propto F_{zt}$ prompt us to define a new gauge field $\tilde{A}_{\mu}$, where $\tilde{A}_z = A_z$ and $\tilde{A}_t = A_t$, but
\beq
\label{eq:axredef}
\tilde{A}_x \equiv A_x - 4\pi \thetainf \, \delta(x).
\eeq
Equivalently, we can perform a singular gauge transformation to send $A_x \to \tilde{A}_x$. Since $\thetainf$ is a constant, the field strength is unchanged, $\tilde{F}_{\mu\nu} = F_{\mu\nu}$, but the action becomes
\beq
\label{eq:stheta2}
S_{\theta} = - \frac{N}{4\pi} \int \tilde{A} \wedge \tilde{F},
\eeq
and $\oint dx \, \tilde{A}_x = - 2\pi \thetainf$. In the $z \to \infty$ region, then, the CS gauge field absorbs the $\theta$-angle, leading to the effective bulk action eq.~\eqref{eq:stheta2}, consisting only of a CS term for $\tilde{A}_{\mu}$, where $\tilde{A}_x$ has a non-zero Wilson loop. In the field theory we interpret that as the current absorbing the spin in the IR, producing a current algebra with the same rank and level as in the UV, but now with a phase shift. In short, we propose that the CS gauge field absorbing a $\theta$-angle, as in eq.~\eqref{eq:axredef}, is the holographic version of ``absorbing the spin.''

Crucially, however, $A_{\mu}$ is dual to the \textit{charge} current, whereas in the CFT description of the Kondo effect the impurity is absorbed by the \textit{spin} current, as in eq.~\eqref{absorb}. In the solution of the original Kondo problem reviewed in section~\ref{cft}, $N=2$, $k=1$, and $s_{\textrm{imp}}=1/2$, the key difference between the UV and IR fixed points was the exchange of the spin conformal towers, such that integer and half-integer spin states had odd and even $U(1)$ charges in the UV but even and odd $U(1)$ charges in the IR. That is clearly equivalent to the converse, \textit{i.e.}\ fixing the spin conformal towers and exchanging the $U(1)$ charge conformal towers: the essential physics is the \textit{relative} exchange of spin and charge towers. Our eq.~\eqref{eq:axredef} appears to be a holographic realization of the exchange of $U(1)$ charge towers.

A ``rolling $\theta$-angle'' $\theta(z)$ in $AdS_2$ coupled to a CS gauge field in $AdS_3$ could actually provide an alternative holographic model of the Kondo effect. Indeed, at the level of effective field theory in the bulk we are free to write an action including only those fields, with all couplings allowed by symmetries. What is $\theta$ dual to in the field theory? We suspect that using $\theta$ is dual to introducing $S^a$ without introducing Abrikosov pseudo-fermions $\chi$. When we write $S^a = \chi^{\dagger} T^a \chi$, we introduce an additional $U(1)$ symmetry in the field theory, namely phase shifts of $\chi$. The corresponding $U(1)$ current $\chi^{\dagger} \chi$ is dual to the $U(1)$ gauge field $a_t$. Crucially, however, $\theta$ is not charged under any bulk gauge groups, and so a bulk effective theory for $\theta$ requires no $AdS_2$ gauge field. The dual field theory thus requires no additional $U(1)$ symmetry, suggesting that the $\chi$ are absent. In fact, we suspect that $\theta$ is dual to the double-trace operator $S^aJ^a$ (recall $J^a = \psi_L^{\dagger} T^a \psi_L$), for two reasons. First, in terms of our top-down model, $\theta$ exists only when both D7- and D5-branes are present, \textit{i.e.}\ $\theta$ arises from the 7-5 and 5-7 strings, and should thus be dual to an operator that only exists when both $J^a$ and $S^a$ are present. Second, in holography we expect a multi-trace operator to be dual to a multi-particle state, and indeed $\epsilon^{mn} \partial_n \theta = J^m \propto a^m \phi^2$ is a multi-particle operator. Alternatively, we could trade $\theta$ for $J^m$, and in fact a proposal to use external currents such as $J^m$ to represent impurities in a bulk effective theory appears already in refs.~\cite{Hashimoto:2012pb,Ishii:2012hw}.

In summary, the basic physics of our holographic Kondo model is as follows. At any $T$, the $AdS_2$ gauge field has non-zero electric flux at the boundary, representing the impurity in the UV. When we reduce $T$ through $T_c$, the $AdS_2$ scalar condenses. If we then descend from the boundary into the bulk, then we will see the scalar transfer electric charge away from the $AdS_2$ gauge field and over to the CS gauge field, which is possible because the scalar is bi-fundamental, \textit{i.e.}\ essentially the same current appears for both the $AdS_2$ and CS gauge fields in eqs.~\eqref{Aeom} and~\eqref{aeom}. For a CS gauge field, electric charge induces \textit{magnetic}\ flux, here meaning non-zero $F_{zx}$, which produces the expected phase shift, as explained above. Whether the impurity is screened completely in the IR becomes the question of whether the scalar transfers \textit{all} of the $AdS_2$ electric flux to the CS gauge field. In short, the UV fixed point appears holographically as a CS gauge field with no Wilson loop for $A_x$ and some non-zero electric flux in $AdS_2$, while the IR fixed point appears as a CS gauge field with a Wilson loop, possibly with some reduced electric flux in $AdS_2$.

%%%%%%%%%%%%%%%%%%%%%%%%%%%%%%%%%%%%%%%%%%%%%%%%%%%%%%%%%%%%%%%%%%%%%%%%%%%%%%%%%%%%%%%%%%%%%%%%%%%%%%%%%
%%%%%%%%%%%%%%%%%%%%%%%%%%%%%%%%%%%%%%%%%%%%%%%%%%%%%%%%%%%%%%%%%%%%%%%%%%%%%%%%%%%%%%%%%%%%%%%%%%%%%%%%%
\section{Summary and Outlook}
\label{summary}
%%%%%%%%%%%%%%%%%%%%%%%%%%%%%%%%%%%%%%%%%%%%%%%%%%%%%%%%%%%%%%%%%%%%%%%%%%%%%%%%%%%%%%%%%%%%%%%%%%%%%%%%%
%%%%%%%%%%%%%%%%%%%%%%%%%%%%%%%%%%%%%%%%%%%%%%%%%%%%%%%%%%%%%%%%%%%%%%%%%%%%%%%%%%%%%%%%%%%%%%%%%%%%%%%%%

We proposed a holographic model of the Kondo effect as a holographic superconductor in $AdS_2$ coupled as a defect to a CS gauge field in $AdS_3$. The parameters that define a Kondo model, $N$, $k$, and the impurity representation, map to the level $N$ and rank $k$ of the CS gauge field and to the electric flux of the $AdS_2$ gauge field at the boundary. In field theory language, our model includes two couplings, a single-trace 't Hooft coupling, which is always large, and a double-trace Kondo coupling, whose running is very similar to that of the original Kondo system. Indeed, our model is essentially a Kondo Hamiltonian coupled to a strongly-interacting sector holographically dual to classical Einstein gravity. Our model exhibits an RG flow with several signatures of the large-$N$, single-impurity Kondo effect: a dynamically-generated scale $T_K$, a second-order mean field phase transition, observables with power-law scalings in $T$ at low $T$, screening of the impurity, and a phase shift.

Many open questions remain for both our top-down and bottom-up models. In our top-down model, we can describe the impurity using either D5- or D3-branes, dual to either slave fermions or bosons, respectively~\cite{Rey:1998ik,Drukker:2005kx,Gomis:2006sb,Gomis:2006im}. If we chose D3-branes instead of D5-branes, what kind of tachyon appears? For either D5- or D3-branes, can we find a ``single-brane'' solution, involving the D7-brane only and representing the endpoint of the tachyon condensation, as described at the end of subsection~\ref{kondocoupling}?

In our bottom-up model, what happens with multiple channels, corresponding to a non-Abelian CS gauge field, and/or with impurities in different representations, corresponding to various arrangements of electric flux of a non-Abelian $AdS_2$ gauge field? Can we clarify what under-, critical, and over-screening look like in the bulk? What is the Wilson number in our model, \textit{i.e.}\ the ratio of the change in the heat capacity due to the impurity to the change in magnetic susceptibility, which characterizes the universality class~\cite{Wilson:1974mb,springerlink:10.1007/BF00654541,Nozieres:1975}? When $T \leq T_c$, can we identify the Kondo resonance, which appears as a peak in the electronic spectral function at the Fermi energy, and whose height is fixed by the Friedel sum rule?

Does our model obey the $g$-theorem, the analogue of the $c$-theorem for RG flows triggered by impurities~\cite{Affleck:1991tk}? In holography, a $c$-theorem typically translates into a null energy condition on bulk matter fields~\cite{Freedman:1999gp}. We thus expect that any holographic model of the Kondo effect with reasonably well-behaved matter fields (such as our model) will obey the $g$-theorem.

What other holographic models of the Kondo effect are possible? In all holographic Kondo models to date, the spin group was the gauge group $SU(N_c)$. Can we construct models in which spin is a global symmetry? This question is crucial: gauging the spin group introduces ``extra'' massless degrees of freedom, which may dramatically alter low-energy physics.

What can we learn about the Kondo effect from holography? For example, with multiple impurities, what is the bulk dual of the RKKY interaction? Does a lattice of holographic Kondo impurities with RKKY interactions exhibit a quantum phase transition, producing $\rho \propto T$, as conjectured for heavy fermion compounds? What can holography teach us about entanglement entropy and far-from-equilibrium physics in Kondo systems?

What can we learn about holography from the Kondo effect? In the CFT description of the original Kondo problem, $N=2$, $k=1$, and $s_{\textrm{imp}}=1/2$, the UV CFT describes free chiral fermions while the IR CFT describes free chiral fermions with a phase shift. A theory of free chiral fermions has an infinite number of conserved currents of arbitrarily high spin. At large $N$, such a theory may be dual to some version of Vasiliev's higher-spin gravity theory in $AdS_3$~\cite{Vasiliev:1995dn}. If so, then the existence of an impurity-driven RG flow between the two CFTs implies the existence of a solution to Vasiliev's theory in which a localized source, dual to the impurity, triggers the growth of a Wilson loop deep in $AdS_3$. In other words, Vasiliev's higher-spin gravity in $AdS_3$ may provide the dual of \textit{precisely} the large-$N$ Kondo Hamiltonian, with no additional degrees of freedom. As emphasized recently in ref.~\cite{doi:10.1080/00107514.2013.779477}, the holographic dual of a familiar condensed matter system could be very valuable for many reasons, one being to improve our understanding of holography.

%%%%%%%%%%%%%%%%%%%%%%%%%%%%%%%%%%%%%%%%%%%%%%%%%%%%%%%%%%%%%%%%%%%%%%%%%%%%%%%%%%%%%%%%%%%%%%%%%%%%%%%%%
%%%%%%%%%%%%%%%%%%%%%%%%%%%%%%%%%%%%%%%%%%%%%%%%%%%%%%%%%%%%%%%%%%%%%%%%%%%%%%%%%%%%%%%%%%%%%%%%%%%%%%%%%
\section*{Acknowledgements}
%%%%%%%%%%%%%%%%%%%%%%%%%%%%%%%%%%%%%%%%%%%%%%%%%%%%%%%%%%%%%%%%%%%%%%%%%%%%%%%%%%%%%%%%%%%%%%%%%%%%%%%%%
%%%%%%%%%%%%%%%%%%%%%%%%%%%%%%%%%%%%%%%%%%%%%%%%%%%%%%%%%%%%%%%%%%%%%%%%%%%%%%%%%%%%%%%%%%%%%%%%%%%%%%%%%

We thank I.~Affleck, S.~Bolognesi, S.-P.~Chao, A.~Cherman, M.~Chernicoff, R.~de Sousa, F.~Essler, M.~Gaberdiel, S.~Hartnoll, J.~Harvey, C.~Herzog, D.~Hofman, N.~Iqbal, K.~Jensen, S.~Kachru, A.~Karch, E.~Kiritsis, S.-J.~Lee, S.-S.~Lee, M.~Lippert, H.~Liu, A.~Ludwig, J.~McGreevy, R.~Meyer, V.~Niarchos, C.~Nu{\~n}ez, E.~Perlmutter, A.~Ramallo, L.~Rastelli, D.~T.~Son, D.~Tong, L.~Yaffe, S.~Yaida, J.~Zaanen, and P.~Zhao for useful conversations and correspondence. J.E. also especially thanks S. Sachdev for a discussion about refs.~\cite{2003PhRvL..90u6403S,2004PhRvB..69c5111S}. A.O'B. also especially thanks D.~Dorigoni for discussions about gauge fixing in our system, I.~Papadimitriou for help with holographic renormalization, E.~Pomoni for discussions about refs.~\cite{Pomoni:2008de,Pomoni:2010et}, and the Crete Center for Theoretical Physics for hospitality while this work was in progress. The work of  C.H.  is partially supported by the Israel Science Foundation (grant 1665/10). The work of A.O'B. was supported in part by the European Union grant FP7-REGPOT-2008-1-CreteHEPCosmo-228644. The research leading to these results has received funding from the European Research Council under the European Community's Seventh Framework Programme (FP7/2007-2013) / ERC grant agreement no. 247252. This work was supported in part by the Cluster of Excellence ``Origin and Structure of the Universe.''

\bibliographystyle{JHEP}
\bibliography{kondoholo}

\providecommand{\href}[2]{#2}\begingroup\raggedright\begin{thebibliography}{100}

\bibitem{PTP.32.37}
J.~Kondo, {\it {Resistance Minimum in Dilute Magnetic Alloys}},  {\em Prog.
  Theo. Phys.} {\bf 32} (1964), no.~1 37--49.

\bibitem{0034-4885-37-2-001}
C.~Rizzuto, {\it {Formation of Localized Moments in Metals: Experimental Bulk
  Properties}},  {\em Rep. Prog. Phys.} {\bf 37} (1974), no.~2 147.

\bibitem{GrŸner1978591}
G.~Gr{\"u}ner and A.~Zawadowski, {\it {Low Temperature Properties of Kondo
  Alloys}},  in {\em Progress in Low Temperature Physics} (D.~Brewer, ed.),
  vol.~7, Part B, pp.~591 -- 647.
\newblock Elsevier, 1978.

\bibitem{Goldhaber1998}
D.~Goldhaber-Gordon, H.~Shtrikman, D.~Mahalu, D.~Abusch-Magder, U.~Meirav, and
  M.~A. Kastner, {\it {Kondo Effect in a Single-electron Transistor}},  {\em
  Nature} {\bf 391} (1998) 156--159.

\bibitem{Cronenwett24071998}
S.~Cronenwett, T.~Oosterkamp, and L.~Kouwenhoven, {\it {A Tunable Kondo Effect
  in Quantum Dots}},  {\em Science} {\bf 281} (1998), no.~5376 540--544.

\bibitem{vanderWiel22092000}
W.~G. van~der Wiel, S.~D. Franceschi, T.~Fujisawa, J.~M. Elzerman, S.~Tarucha,
  and L.~P. Kouwenhoven, {\it {The Kondo Effect in the Unitary Limit}},  {\em
  Science} {\bf 289} (2000), no.~5487 2105--2108.

\bibitem{Wilson:1974mb}
K.~G. Wilson, {\it {The Renormalization Group: Critical Phenomena and the Kondo
  Problem}},  {\em Rev.Mod.Phys.} {\bf 47} (1975) 773.

\bibitem{springerlink:10.1007/BF00654541}
P.~Nozi{\`e}res, {\it {A ``Fermi-liquid'' Description of the Kondo Problem at
  Low Temperatures}},  {\em Jour. Low Temp. Phys.} {\bf 17} (1974) 31--42.
  10.1007/BF00654541.

\bibitem{Nozieres:1975}
P.~Nozi{\`e}res in {\em Low Temperature Physics Conference Proceedings}
  (Krusius and Vuorio, eds.), vol.~14, p.~339.
\newblock Elsevier, 1975.

\bibitem{PhysRevLett.45.379}
N.~Andrei, {\it {Diagonalization of the Kondo Hamiltonian}},  {\em Phys. Rev.
  Lett.} {\bf 45} (Aug, 1980) 379--382.

\bibitem{Wiegmann:1980}
P.~Wiegmann, {\it {Exact Solution of s-d Exchange Model at T=0}},  {\em Sov.
  Phys. JETP Lett.} {\bf 31} (1980) 364.

\bibitem{RevModPhys.55.331}
N.~Andrei, K.~Furuya, and J.~H. Lowenstein, {\it {Solution of the Kondo
  problem}},  {\em Rev. Mod. Phys.} {\bf 55} (Apr, 1983) 331--402.

\bibitem{doi:10.1080/00018738300101581}
A.~Tsvelick and P.~Wiegmann, {\it {Exact Results in the Theory of Magnetic
  Alloys}},  {\em Advances in Physics} {\bf 32} (1983), no.~4 453--713.

\bibitem{0022-3719-19-17-017}
P.~Coleman and N.~Andrei, {\it {Diagonalisation of the Generalised Anderson
  Model}},  {\em Jour. Phys.} {\bf C19} (1986) 3211--3233.

\bibitem{PhysRevB.35.5072}
P.~Coleman, {\it {Mixed Valence as an Almost Broken Symmetry}},  {\em Phys.
  Rev.} {\bf B35} (Apr, 1987) 5072--5116.

\bibitem{RevModPhys.59.845}
{Bickers, N.}, {\it {Review of Techniques in the Large-N Expansion for Dilute
  Magnetic Alloys}},  {\em Rev. Mod. Phys.} {\bf 59} (Oct, 1987) 845--939.

\bibitem{1998PhRvB..58.3794P}
O.~Parcollet, A.~Georges, G.~Kotliar, and A.~Sengupta, {\it {Overscreened
  Multi-channel SU(N) Kondo Model: Large-N Solution and Conformal Field
  Theory}},  {\em Phys. Rev.} {\bf B58} (Aug., 1998) 3794--3813,
  [\href{http://xxx.lanl.gov/abs/{arXiv:cond-mat/9711192}}{{\tt
  {arXiv:cond-mat/9711192}}}].

\bibitem{Affleck:1990zd}
I.~Affleck, {\it {A Current Algebra Approach To The Kondo Effect}},  {\em Nucl.
  Phys.} {\bf B336} (1990) 517.

\bibitem{Affleck:1990by}
I.~Affleck and A.~Ludwig, {\it {The Kondo Effect, Conformal Field Theory and
  Fusion Rules}},  {\em Nucl.Phys.} {\bf B352} (1991) 849--862.

\bibitem{Affleck:1990iv}
I.~Affleck and A.~Ludwig, {\it {Critical Theory of Overscreened Kondo Fixed
  Points}},  {\em Nucl.Phys.} {\bf B360} (1991) 641--696.

\bibitem{Affleck:1991tk}
I.~Affleck and A.~Ludwig, {\it {Universal Non-integer 'Ground State Degeneracy'
  in Critical Quantum Systems}},  {\em Phys.Rev.Lett.} {\bf 67} (1991)
  161--164.

\bibitem{PhysRevB.48.7297}
I.~Affleck and A.~Ludwig, {\it {Exact Conformal-field-theory Results on the
  Multichannel Kondo Effect: Single-fermion Green's function, Self-energy, and
  Resistivity}},  {\em Phys.Rev.} {\bf B48} (1993) 7297--7321.

\bibitem{Affleck:1995ge}
I.~Affleck, {\it {Conformal Field Theory Approach to the Kondo Effect}},  {\em
  Acta Phys. Polon.} {\bf B26} (1995) 1869--1932,
  [\href{http://xxx.lanl.gov/abs/cond-mat/9512099}{{\tt cond-mat/9512099}}].

\bibitem{Hewson:1993}
A.~Hewson, {\it {The Kondo Model to Heavy Fermions}},  {\em {Cambridge
  University Press}} (1993).

\bibitem{doi:10.1080/000187398243500}
D.~L. Cox and A.~Zawadowski, {\it {Exotic Kondo Effects in Metals: Magnetic
  Ions in a Crystalline Electric Field and Tunnelling Centres}},  {\em Advances
  in Physics} {\bf 47} (1998), no.~5 599--942,
  [\href{http://xxx.lanl.gov/abs/{arxiv:cond-mat/9704103}}{{\tt
  {arxiv:cond-mat/9704103}}}].

\bibitem{Doniach1977231}
S.~Doniach, {\it {The Kondo Lattice and Weak Anti-ferromagnetism}},  {\em
  Physica B+C} {\bf 91} (1977), no.~0 231 -- 234.

\bibitem{RevModPhys.69.809}
H.~Tsunetsugu, M.~Sigrist, and K.~Ueda, {\it {The Ground-state Phase Diagram of
  the One-dimensional Kondo Lattice Model}},  {\em Rev. Mod. Phys.} {\bf 69}
  (Jul, 1997) 809--864.

\bibitem{2006cond.mat.12006C}
P.~Coleman, {\it {Heavy Fermions: Electrons at the Edge of Magnetism}},  in
  {\em Handbook of Magnetism and Advanced Magnetic Materials: Fundamentals and
  Theory} (Kronmuller and Parkin, eds.), vol.~1, pp.~95--148.
\newblock John Wiley and Sons, 2007.
\newblock \href{http://xxx.lanl.gov/abs/[arxiv:cond-mat/0612006]}{{\tt
  [arxiv:cond-mat/0612006]}}.

\bibitem{2010uqpt.book..193S}
Q.~Si, {\it {Quantum Criticality and the Kondo Lattice}},  {\em Understanding
  Quantum Phase Transitions.~Series: Condensed Matter Physics, CRC Press}
  (Nov., 2010) 193--216, [\href{http://xxx.lanl.gov/abs/1012.5440}{{\tt
  arXiv:1012.5440}}].

\bibitem{2008NatPh...4..186G}
P.~Gegenwart, Q.~Si, and F.~Steglich, {\it {Quantum Criticality in
  Heavy-fermion Metals}},  {\em Nature Physics} {\bf 4} (Mar., 2008) 186--197,
  [\href{http://xxx.lanl.gov/abs/0712.2045}{{\tt arXiv:0712.2045}}].

\bibitem{AffleckEE}
I.~Affleck, N.~Laflorencie, and E.~Sorensen, {\it {Entanglement Entropy in
  Quantum Impurity Systems and Systems with Boundaries}},  {\em Journal of
  Physics A Mathematical General} {\bf 42} (Dec., 2009) 4009,
  [\href{http://xxx.lanl.gov/abs/0906.1809}{{\tt arXiv:0906.1809}}].

\bibitem{2006PhRvB..74x5113A}
F.~{Anders} and A.~{Schiller}, {\it {Spin Precession and Real-time Dynamics in
  the Kondo model: Time-dependent Numerical Renormalization-group Study}},
  {\em Phys. Rev.} {\bf B74} (Dec., 2006) 245113,
  [\href{http://xxx.lanl.gov/abs/{arXiv:cond-mat/0604517}}{{\tt
  {arXiv:cond-mat/0604517}}}].

\bibitem{2011arXiv1102.3982L}
C.~{Latta}, F.~{Haupt}, M.~{Hanl}, A.~{Weichselbaum}, M.~{Claassen},
  W.~{Wuester}, P.~{Fallahi}, S.~{Faelt}, L.~{Glazman}, J.~{von Delft}, H.~E.
  {T{\"u}reci}, and A.~{Imamoglu}, {\it {Quantum Quench of Kondo Correlations
  in Optical Absorption}},  {\em Nature} {\bf 474} (2011) 627--630,
  [\href{http://xxx.lanl.gov/abs/1102.3982}{{\tt arXiv:1102.3982}}].

\bibitem{Maldacena:1997re}
J.~M. Maldacena, {\it {The Large N Limit of Superconformal Field Theories and
  Supergravity}},  {\em Adv. Theor. Math. Phys.} {\bf 2} (1998) 231--252,
  [\href{http://xxx.lanl.gov/abs/hep-th/9711200}{{\tt hep-th/9711200}}].

\bibitem{Gubser:1998bc}
S.~S. Gubser, I.~R. Klebanov, and A.~M. Polyakov, {\it {Gauge Theory
  Correlators from Non-critical String Theory}},  {\em Phys. Lett.} {\bf B428}
  (1998) 105--114, [\href{http://xxx.lanl.gov/abs/hep-th/9802109}{{\tt
  hep-th/9802109}}].

\bibitem{Witten:1998qj}
E.~Witten, {\it {Anti-de Sitter Space and Holography}},  {\em Adv. Theor. Math.
  Phys.} {\bf 2} (1998) 253--291,
  [\href{http://xxx.lanl.gov/abs/hep-th/9802150}{{\tt hep-th/9802150}}].

\bibitem{Kachru:2009xf}
S.~Kachru, A.~Karch, and S.~Yaida, {\it {Holographic Lattices, Dimers, and
  Glasses}},  {\em Phys.Rev.} {\bf D81} (2010) 026007,
  [\href{http://xxx.lanl.gov/abs/0909.2639}{{\tt arXiv:0909.2639}}].

\bibitem{Sachdev:2010um}
S.~Sachdev, {\it {Holographic Metals and the Fractionalized Fermi Liquid}},
  {\em Phys.Rev.Lett.} {\bf 105} (2010) 151602,
  [\href{http://xxx.lanl.gov/abs/1006.3794}{{\tt arXiv:1006.3794}}].

\bibitem{Kachru:2010dk}
S.~Kachru, A.~Karch, and S.~Yaida, {\it {Adventures in Holographic Dimer
  Models}},  {\em New J.Phys.} {\bf 13} (2011) 035004,
  [\href{http://xxx.lanl.gov/abs/1009.3268}{{\tt arXiv:1009.3268}}].

\bibitem{Sachdev:2010uj}
S.~Sachdev, {\it {Strange Metals and the AdS/CFT Correspondence}},  {\em
  J.Stat.Mech.} {\bf 1011} (2010) P11022,
  [\href{http://xxx.lanl.gov/abs/1010.0682}{{\tt arXiv:1010.0682}}].

\bibitem{Mueck:2010ja}
W.~M{\"u}ck, {\it {The Polyakov Loop of Anti-symmetric Representations as a
  Quantum Impurity Model}},  {\em Phys.Rev.} {\bf D83} (2011) 066006,
  [\href{http://xxx.lanl.gov/abs/1012.1973}{{\tt arXiv:1012.1973}}].

\bibitem{Faraggi:2011bb}
A.~Faraggi and L.~Pando~Zayas, {\it {The Spectrum of Excitations of Holographic
  Wilson Loops}},  {\em JHEP} {\bf 1105} (2011) 018,
  [\href{http://xxx.lanl.gov/abs/1101.5145}{{\tt arXiv:1101.5145}}].

\bibitem{Jensen:2011su}
K.~Jensen, S.~Kachru, A.~Karch, J.~Polchinski, and E.~Silverstein, {\it
  {Towards a Holographic Marginal Fermi Liquid}},  {\em Phys.Rev.} {\bf D84}
  (2011) 126002, [\href{http://xxx.lanl.gov/abs/1105.1772}{{\tt
  arXiv:1105.1772}}].

\bibitem{Karaiskos:2011kf}
N.~Karaiskos, K.~Sfetsos, and E.~Tsatis, {\it {Brane Embeddings in Sphere
  Submanifolds}},  {\em Class.Quant.Grav.} {\bf 29} (2012) 025011,
  [\href{http://xxx.lanl.gov/abs/1106.1200}{{\tt arXiv:1106.1200}}].

\bibitem{Harrison:2011fs}
S.~Harrison, S.~Kachru, and G.~Torroba, {\it {A Maximally Supersymmetric Kondo
  Model}},  {\em Class.Quant.Grav.} {\bf 29} (2012) 194005,
  [\href{http://xxx.lanl.gov/abs/1110.5325}{{\tt arXiv:1110.5325}}].

\bibitem{Benincasa:2011zu}
P.~Benincasa and A.~Ramallo, {\it {Fermionic Impurities in Chern-Simons-Matter
  Theories}},  {\em JHEP} {\bf 1202} (2012) 076,
  [\href{http://xxx.lanl.gov/abs/1112.4669}{{\tt arXiv:1112.4669}}].

\bibitem{Faraggi:2011ge}
A.~Faraggi, W.~M{\"u}ck, and L.~Pando~Zayas, {\it {One-loop Effective Action of
  the Holographic Antisymmetric Wilson Loop}},  {\em Phys.Rev.} {\bf D85}
  (2012) 106015, [\href{http://xxx.lanl.gov/abs/1112.5028}{{\tt
  arXiv:1112.5028}}].

\bibitem{Benincasa:2012wu}
P.~Benincasa and A.~Ramallo, {\it {Holographic Kondo Model in Various
  Dimensions}},  {\em JHEP} {\bf 1206} (2012) 133,
  [\href{http://xxx.lanl.gov/abs/1204.6290}{{\tt arXiv:1204.6290}}].

\bibitem{Itsios:2012ev}
G.~Itsios, K.~Sfetsos, and D.~Zoakos, {\it {Fermionic Impurities in the
  Unquenched ABJM}},  {\em JHEP} {\bf 1301} (2013) 038,
  [\href{http://xxx.lanl.gov/abs/1209.6617}{{\tt arXiv:1209.6617}}].

\bibitem{Matsueda:2012gc}
H.~Matsueda, {\it {Multiscale Entanglement Renormalization Ansatz for Kondo
  Problem}},  \href{http://xxx.lanl.gov/abs/1208.2872}{{\tt arXiv:1208.2872}}.

\bibitem{Swingle:2009bg}
B.~Swingle, {\it {Entanglement Renormalization and Holography}},  {\em
  Phys.Rev.} {\bf D86} (2012) 065007,
  [\href{http://xxx.lanl.gov/abs/0905.1317}{{\tt arXiv:0905.1317}}].

\bibitem{Swingle:2012wq}
B.~Swingle, {\it {Constructing Holographic Spacetimes Using Entanglement
  Renormalization}},  \href{http://xxx.lanl.gov/abs/1209.3304}{{\tt
  arXiv:1209.3304}}.

\bibitem{2003PhRvL..90u6403S}
T.~Senthil, S.~Sachdev, and M.~Vojta, {\it {Fractionalized Fermi Liquids}},
  {\em Phys. Rev. Lett.} {\bf 90} (May, 2003) 216403,
  [\href{http://xxx.lanl.gov/abs/{arXiv:cond-mat/0209144}}{{\tt
  {arXiv:cond-mat/0209144}}}].

\bibitem{2004PhRvB..69c5111S}
T.~Senthil, M.~Vojta, and S.~Sachdev, {\it {Weak Magnetism and Non-Fermi
  liquids Near Heavy-fermion Critical Points}},  {\em Phys. Rev.} {\bf B69}
  (Jan., 2004) 035111,
  [\href{http://xxx.lanl.gov/abs/{arxiv:cond-mat/0305193}}{{\tt
  {arxiv:cond-mat/0305193}}}].

\bibitem{Skenderis:2002vf}
K.~Skenderis and M.~Taylor, {\it {Branes in AdS and pp-wave Spacetimes}},  {\em
  JHEP} {\bf 06} (2002) 025,
  [\href{http://xxx.lanl.gov/abs/hep-th/0204054}{{\tt hep-th/0204054}}].

\bibitem{Harvey:2007ab}
J.~A. Harvey and A.~B. Royston, {\it {Localized Modes at a D-brane--O-plane
  Intersection and Heterotic Alice Strings}},  {\em JHEP} {\bf 04} (2008) 018,
  [\href{http://xxx.lanl.gov/abs/0709.1482}{{\tt arXiv:0709.1482}}].

\bibitem{Buchbinder:2007ar}
E.~I. Buchbinder, J.~Gomis, and F.~Passerini, {\it {Holographic Gauge Theories
  in Background Fields and Surface Operators}},  {\em JHEP} {\bf 12} (2007)
  101, [\href{http://xxx.lanl.gov/abs/0710.5170}{{\tt arXiv:0710.5170}}].

\bibitem{Harvey:2008zz}
J.~A. Harvey and A.~B. Royston, {\it {Gauge/Gravity Duality with a Chiral
  N=(0,8) String Defect}},  {\em JHEP} {\bf 08} (2008) 006,
  [\href{http://xxx.lanl.gov/abs/0804.2854}{{\tt arXiv:0804.2854}}].

\bibitem{Pawelczyk:2000hy}
J.~Pawelczyk and S.-J. Rey, {\it {Ramond-Ramond Flux Stabilization of
  D-branes}},  {\em Phys.Lett.} {\bf B493} (2000) 395--401,
  [\href{http://xxx.lanl.gov/abs/hep-th/0007154}{{\tt hep-th/0007154}}].

\bibitem{Camino:2001at}
J.~Camino, A.~Paredes, and A.~Ramallo, {\it {Stable Wrapped Branes}},  {\em
  JHEP} {\bf 05} (2001) 011,
  [\href{http://xxx.lanl.gov/abs/hep-th/0104082}{{\tt hep-th/0104082}}].

\bibitem{Yamaguchi:2006tq}
S.~Yamaguchi, {\it {Wilson Loops of Anti-symmetric Representation and
  D5-branes}},  {\em JHEP} {\bf 0605} (2006) 037,
  [\href{http://xxx.lanl.gov/abs/hep-th/0603208}{{\tt hep-th/0603208}}].

\bibitem{Gomis:2006sb}
J.~Gomis and F.~Passerini, {\it {Holographic Wilson Loops}},  {\em JHEP} {\bf
  08} (2006) 074, [\href{http://xxx.lanl.gov/abs/hep-th/0604007}{{\tt
  hep-th/0604007}}].

\bibitem{Gukov:2004id}
S.~Gukov, E.~Martinec, G.~W. Moore, and A.~Strominger, {\it {Chern-Simons Gauge
  Theory and the AdS(3) / CFT(2) Correspondence}},
  \href{http://xxx.lanl.gov/abs/hep-th/0403225}{{\tt hep-th/0403225}}.

\bibitem{Kraus:2006nb}
P.~Kraus and F.~Larsen, {\it {Partition Functions and Elliptic Genera from
  Supergravity}},  {\em JHEP} {\bf 0701} (2007) 002,
  [\href{http://xxx.lanl.gov/abs/hep-th/0607138}{{\tt hep-th/0607138}}].

\bibitem{Kraus:2006wn}
P.~Kraus, {\it {Lectures on Black Holes and the AdS(3)/CFT(2) Correspondence}},
   {\em Lect. Notes Phys.} {\bf 755} (2008) 193--247,
  [\href{http://xxx.lanl.gov/abs/hep-th/0609074}{{\tt hep-th/0609074}}].

\bibitem{Jensen:2010em}
K.~Jensen, {\it {Chiral Anomalies and AdS/CMT in Two Dimensions}},  {\em JHEP}
  {\bf 1101} (2011) 109, [\href{http://xxx.lanl.gov/abs/1012.4831}{{\tt
  arXiv:1012.4831}}].

\bibitem{Andrade:2011sx}
T.~Andrade, J.~Jottar, and R.~Leigh, {\it {Boundary Conditions and Unitarity:
  the Maxwell-Chern-Simons System in $AdS_3/CFT_2$}},  {\em JHEP} {\bf 1205}
  (2012) 071, [\href{http://xxx.lanl.gov/abs/1111.5054}{{\tt
  arXiv:1111.5054}}].

\bibitem{Witten:2001ua}
E.~Witten, {\it {Multi-trace Operators, Boundary Conditions, and AdS/CFT
  Correspondence}},  \href{http://xxx.lanl.gov/abs/hep-th/0112258}{{\tt
  hep-th/0112258}}.

\bibitem{Berkooz:2002ug}
M.~Berkooz, A.~Sever, and A.~Shomer, {\it {'Double Trace' Deformations,
  Boundary Conditions and Space-time Singularities}},  {\em JHEP} {\bf 0205}
  (2002) 034, [\href{http://xxx.lanl.gov/abs/hep-th/0112264}{{\tt
  hep-th/0112264}}].

\bibitem{Faulkner:2010gj}
T.~Faulkner, G.~Horowitz, and M.~Roberts, {\it {Holographic Quantum Criticality
  from Multi-trace Deformations}},  {\em JHEP} {\bf 1104} (2011) 051,
  [\href{http://xxx.lanl.gov/abs/1008.1581}{{\tt arXiv:1008.1581}}].

\bibitem{Hartnoll:2008vx}
S.~A. Hartnoll, C.~P. Herzog, and G.~T. Horowitz, {\it {Building a Holographic
  Superconductor}},  {\em Phys. Rev. Lett.} {\bf 101} (2008) 031601,
  [\href{http://xxx.lanl.gov/abs/0803.3295}{{\tt arXiv:0803.3295}}].

\bibitem{Hartnoll:2008kx}
S.~A. Hartnoll, C.~P. Herzog, and G.~T. Horowitz, {\it {Holographic
  Superconductors}},  {\em JHEP} {\bf 12} (2008) 015,
  [\href{http://xxx.lanl.gov/abs/0810.1563}{{\tt arXiv:0810.1563}}].

\bibitem{DiFrancesco:1997nk}
P.~Di~Francesco, P.~Mathieu, and D.~Senechal, {\it {Conformal Field Theory}}, .
  Springer-Verlag New York Inc., (1997).

\bibitem{Felder:1999cv}
G.~Felder, J.~Fr{\"o}hlich, J.~Fuchs, and C.~Schweigert, {\it {Conformal
  Boundary Conditions and Three-dimensional Topological Field Theory}},  {\em
  Phys. Rev. Lett.} {\bf 84} (2000) 1659--1662,
  [\href{http://xxx.lanl.gov/abs/hep-th/9909140}{{\tt hep-th/9909140}}].

\bibitem{Bachas:2004sy}
C.~Bachas and M.~Gaberdiel, {\it {Loop Operators and the Kondo Problem}},  {\em
  JHEP} {\bf 11} (2004) 065,
  [\href{http://xxx.lanl.gov/abs/hep-th/0411067}{{\tt hep-th/0411067}}].

\bibitem{Alekseev:2007in}
A.~Alekseev and S.~Monnier, {\it {Quantization of Wilson Loops in
  Wess-Zumino-Witten Models}},  {\em JHEP} {\bf 08} (2007) 039,
  [\href{http://xxx.lanl.gov/abs/hep-th/0702174}{{\tt hep-th/0702174}}].

\bibitem{Monnier:2008jj}
S.~Monnier, {\it {Kondo Flow Invariants, Twisted K-theory and Ramond-Ramond
  Charges}},  {\em JHEP} {\bf 06} (2008) 022,
  [\href{http://xxx.lanl.gov/abs/0803.1565}{{\tt arXiv:0803.1565}}].

\bibitem{ZinnJustin1998}
P.~Zinn-Justin and N.~Andrei, {\it {The Generalized Multi-channel Kondo Model:
  Thermodynamics and Fusion Equations}},  {\em Nucl. Phys.} {\bf B528} (1998)
  648--682, [\href{http://xxx.lanl.gov/abs/cond-mat/9801158}{{\tt
  cond-mat/9801158}}].

\bibitem{PhysRevB.58.3814}
A.~Jerez, N.~Andrei, and G.~Zar\'and, {\it {Solution of the Multichannel
  Coqblin-Schrieffer Impurity Model and Application to Multilevel Systems}},
  {\em Phys. Rev.} {\bf B58} (1998) 3814--3841,
  [\href{http://xxx.lanl.gov/abs/cond-mat/9803137}{{\tt cond-mat/9803137}}].

\bibitem{PhysRevB.73.224445}
D.~Bensimon, A.~Jerez, and M.~Lavagna, {\it {Intermediate Coupling Fixed Point
  Study in the Overscreened Regime of Generalized Multichannel $\mathrm{SU}(N)$
  Kondo Models}},  {\em Phys. Rev.} {\bf B73} (2006) 224445.

\bibitem{Witten:1988hf}
E.~Witten, {\it {Quantum Field Theory and the Jones Polynomial}},  {\em
  Commun.Math.Phys.} {\bf 121} (1989) 351.

\bibitem{Maldacena:1998im}
J.~M. Maldacena, {\it {Wilson Loops in Large N Field Theories}},  {\em
  Phys.Rev.Lett.} {\bf 80} (1998) 4859--4862,
  [\href{http://xxx.lanl.gov/abs/hep-th/9803002}{{\tt hep-th/9803002}}].

\bibitem{Rey:1998ik}
S.-J. Rey and J.-T. Yee, {\it {Macroscopic Strings as Heavy Quarks in Large N
  Gauge Theory and Anti-de Sitter Supergravity}},  {\em Eur.Phys.J.} {\bf C22}
  (2001) 379--394, [\href{http://xxx.lanl.gov/abs/hep-th/9803001}{{\tt
  hep-th/9803001}}].

\bibitem{Drukker:2005kx}
N.~Drukker and B.~Fiol, {\it {All-genus Calculation of Wilson Loops Using
  D-branes}},  {\em JHEP} {\bf 0502} (2005) 010,
  [\href{http://xxx.lanl.gov/abs/hep-th/0501109}{{\tt hep-th/0501109}}].

\bibitem{Gomis:2006im}
J.~Gomis and F.~Passerini, {\it {Wilson Loops as D3-Branes}},  {\em JHEP} {\bf
  0701} (2007) 097, [\href{http://xxx.lanl.gov/abs/hep-th/0612022}{{\tt
  hep-th/0612022}}].

\bibitem{Gava:1997jt}
E.~Gava, K.~S. Narain, and M.~H. Sarmadi, {\it {On the Bound States of p- and
  (p+2)-branes}},  {\em Nucl. Phys.} {\bf B504} (1997) 214--238,
  [\href{http://xxx.lanl.gov/abs/hep-th/9704006}{{\tt hep-th/9704006}}].

\bibitem{Aganagic:2000mh}
M.~Aganagic, R.~Gopakumar, S.~Minwalla, and A.~Strominger, {\it {Unstable
  Solitons in Noncommutative Gauge Theory}},  {\em JHEP} {\bf 04} (2001) 001,
  [\href{http://xxx.lanl.gov/abs/hep-th/0009142}{{\tt hep-th/0009142}}].

\bibitem{Polchinski:1998rr}
J.~Polchinski, {\it {String Theory. Vol. 2: Superstring Theory and Beyond}}, .
  Cambridge Univ. Press (1998).

\bibitem{Pomoni:2008de}
E.~Pomoni and L.~Rastelli, {\it {Large N Field Theory and AdS Tachyons}},  {\em
  JHEP} {\bf 0904} (2009) 020, [\href{http://xxx.lanl.gov/abs/0805.2261}{{\tt
  arXiv:0805.2261}}].

\bibitem{Pomoni:2010et}
E.~Pomoni and L.~Rastelli, {\it {Intersecting Flavor Branes}},  {\em JHEP} {\bf
  1210} (2012) 171, [\href{http://xxx.lanl.gov/abs/1002.0006}{{\tt
  arXiv:1002.0006}}].

\bibitem{Kaplan:2009kr}
D.~Kaplan, J.-W. Lee, D.-T. Son, and M.~Stephanov, {\it {Conformality Lost}},
  {\em Phys.Rev.} {\bf D80} (2009) 125005,
  [\href{http://xxx.lanl.gov/abs/0905.4752}{{\tt arXiv:0905.4752}}].

\bibitem{Jensen:2010ga}
K.~Jensen, A.~Karch, D.~T. Son, and E.~G. Thompson, {\it {Holographic
  Berezinskii-Kosterlitz-Thouless Transitions}},  {\em Phys.Rev.Lett.} {\bf
  105} (2010) 041601, [\href{http://xxx.lanl.gov/abs/1002.3159}{{\tt
  arXiv:1002.3159}}].

\bibitem{Kutasov:2011fr}
D.~Kutasov, J.~Lin, and A.~Parnachev, {\it {Conformal Phase Transitions at Weak
  and Strong Coupling}},  {\em Nucl.Phys.} {\bf B858} (2012) 155--195,
  [\href{http://xxx.lanl.gov/abs/1107.2324}{{\tt arXiv:1107.2324}}].

\bibitem{Iqbal:2011aj}
N.~Iqbal, H.~Liu, and M.~Mezei, {\it {Quantum Phase Transitions in Semi-local
  Quantum Liquids}},  \href{http://xxx.lanl.gov/abs/1108.0425}{{\tt
  arXiv:1108.0425}}.

\bibitem{Sakai:2004cn}
T.~Sakai and S.~Sugimoto, {\it {Low Energy Hadron Physics in Holographic QCD}},
   {\em Prog. Theor. Phys.} {\bf 113} (2005) 843--882,
  [\href{http://xxx.lanl.gov/abs/hep-th/0412141}{{\tt hep-th/0412141}}].

\bibitem{Sakai:2005yt}
T.~Sakai and S.~Sugimoto, {\it {More on a Holographic Dual of QCD}},  {\em
  Prog.Theor.Phys.} {\bf 114} (2005) 1083--1118,
  [\href{http://xxx.lanl.gov/abs/hep-th/0507073}{{\tt hep-th/0507073}}].

\bibitem{Casero:2007ae}
R.~Casero, E.~Kiritsis, and A.~Paredes, {\it {Chiral Symmetry Breaking as Open
  String Tachyon Condensation}},  {\em Nucl.Phys.} {\bf B787} (2007) 98--134,
  [\href{http://xxx.lanl.gov/abs/hep-th/0702155}{{\tt hep-th/0702155}}].

\bibitem{Bergman:2007pm}
O.~Bergman, S.~Seki, and J.~Sonnenschein, {\it {Quark Mass and Condensate in
  HQCD}},  {\em JHEP} {\bf 0712} (2007) 037,
  [\href{http://xxx.lanl.gov/abs/0708.2839}{{\tt arXiv:0708.2839}}].

\bibitem{Dhar:2007bz}
A.~Dhar and P.~Nag, {\it {Sakai-Sugimoto Model, Tachyon Condensation and Chiral
  Symmetry Breaking}},  {\em JHEP} {\bf 0801} (2008) 055,
  [\href{http://xxx.lanl.gov/abs/0708.3233}{{\tt arXiv:0708.3233}}].

\bibitem{Marolf:2006nd}
D.~Marolf and S.~F. Ross, {\it {Boundary Conditions and New Dualities: Vector
  Fields in AdS/CFT}},  {\em JHEP} {\bf 11} (2006) 085,
  [\href{http://xxx.lanl.gov/abs/hep-th/0606113}{{\tt hep-th/0606113}}].

\bibitem{Castro:2008ms}
A.~Castro, D.~Grumiller, F.~Larsen, and R.~McNees, {\it {Holographic
  Description of AdS(2) Black Holes}},  {\em JHEP} {\bf 0811} (2008) 052,
  [\href{http://xxx.lanl.gov/abs/0809.4264}{{\tt arXiv:0809.4264}}].

\bibitem{Papadimitriou:2007sj}
I.~Papadimitriou, {\it {Multi-Trace Deformations in AdS/CFT: Exploring the
  Vacuum Structure of the Deformed CFT}},  {\em JHEP} {\bf 0705} (2007) 075,
  [\href{http://xxx.lanl.gov/abs/hep-th/0703152}{{\tt hep-th/0703152}}].

\bibitem{Son:2002sd}
D.~T. Son and A.~O. Starinets, {\it {Minkowski-space Correlators in AdS/CFT
  Correspondence: Recipe and Applications}},  {\em JHEP} {\bf 09} (2002) 042,
  [\href{http://xxx.lanl.gov/abs/hep-th/0205051}{{\tt hep-th/0205051}}].

\bibitem{Gubser:2009cg}
S.~S. Gubser and A.~Nellore, {\it {Ground States of Holographic
  Superconductors}},  {\em Phys. Rev.} {\bf D80} (2009) 105007,
  [\href{http://xxx.lanl.gov/abs/0908.1972}{{\tt arXiv:0908.1972}}].

\bibitem{Horowitz:2009ij}
G.~T. Horowitz and M.~M. Roberts, {\it {Zero Temperature Limit of Holographic
  Superconductors}},  {\em JHEP} {\bf 11} (2009) 015,
  [\href{http://xxx.lanl.gov/abs/0908.3677}{{\tt arXiv:0908.3677}}].

\bibitem{Davis:2007ka}
J.~L. Davis, M.~Gutperle, P.~Kraus, and I.~Sachs, {\it {Stringy NJL and
  Gross-Neveu Models at Finite Density and Temperature}},  {\em JHEP} {\bf 10}
  (2007) 049, [\href{http://xxx.lanl.gov/abs/0708.0589}{{\tt
  arXiv:0708.0589}}].

\bibitem{Nickel:2010pr}
D.~Nickel and D.~Son, {\it {Deconstructing Holographic Liquids}},  {\em New
  J.Phys.} {\bf 13} (2011) 075010,
  [\href{http://xxx.lanl.gov/abs/1009.3094}{{\tt arXiv:1009.3094}}].

\bibitem{Hashimoto:2012pb}
K.~Hashimoto and N.~Iizuka, {\it {Impurities in Holography and Transport
  Coefficients}},  \href{http://xxx.lanl.gov/abs/1207.4643}{{\tt
  arXiv:1207.4643}}.

\bibitem{Ishii:2012hw}
T.~Ishii and S.-J. Sin, {\it {Impurity Effect in a Holographic
  Superconductor}},  {\em JHEP} {\bf 1304} (2013) 128,
  [\href{http://xxx.lanl.gov/abs/1211.1798}{{\tt arXiv:1211.1798}}].

\bibitem{Freedman:1999gp}
D.~Freedman, S.~Gubser, K.~Pilch, and N.~Warner, {\it {Renormalization Group
  Flows from Holography Supersymmetry and a c theorem}},  {\em
  Adv.Theor.Math.Phys.} {\bf 3} (1999) 363--417,
  [\href{http://xxx.lanl.gov/abs/hep-th/9904017}{{\tt hep-th/9904017}}].

\bibitem{Vasiliev:1995dn}
M.~A. Vasiliev, {\it {Higher Spin Gauge Theories in Four-dimensions,
  Three-dimensions, and Two-dimensions}},  {\em Int.J.Mod.Phys.} {\bf D5}
  (1996) 763--797, [\href{http://xxx.lanl.gov/abs/hep-th/9611024}{{\tt
  hep-th/9611024}}].

\bibitem{doi:10.1080/00107514.2013.779477}
A.~Green, {\it {An Introduction to Gauge-gravity Duality and its Application in
  Condensed Matter}},  {\em Contemporary Physics} {\bf 54} (2013), no.~1
  33--48, [\href{http://xxx.lanl.gov/abs/arxiv:1304.5908}{{\tt
  arxiv:1304.5908}}].

\end{thebibliography}\endgroup

\end{document}